\documentclass[twocolumn]{aastex63}

\usepackage{graphicx, amsmath}
\usepackage{amsfonts}
\usepackage[caption=false]{subfig}
\usepackage{hyperref}
\usepackage{cleveref}
\usepackage{xspace}
\usepackage{rotating}
\usepackage{multirow}
\usepackage{relsize}
\usepackage{comment}

\newcommand{\lya}{Ly$\alpha$\xspace}
\newcommand{\notlya}{not-Ly$\alpha$}
\newcommand{\plya}{P(Ly$\alpha$)}

\newcommand{\Hb}{H$\beta$\xspace}
\newcommand{\angstrom}{\mbox{\normalfont\AA\xspace}}
\newcommand{\OII}{[\ion{O}{2}]\xspace}
\newcommand{\OIII}{[\ion{O}{3}]\xspace}
\newcommand{\HII}{\ion{H}{2}\xspace}

\newcommand{\MgII}{\ion{Mg}{2}\xspace}

\newcommand{\CIII}{\ion{C}{3}]\xspace}
\newcommand{\CIV}{\ion{C}{4}\xspace}

\newcommand{\civ}{C\,{\sc iv}}

\newcommand{\ciii}{C\,{\sc iii}]}

\newcommand{\oiii}{[O\,{\sc iii}]}
\newcommand{\oii}{[O\,{\sc ii}]}  

\newcommand{\caii}{Ca\,{\sc ii}}

\newcommand{\sqdeg}{deg$^{2}$\xspace}

\newcommand{\kms}{km s$^{-1}$}

\def\etal{{et~al.\null}}
\def\eg{e.g.,}

\usepackage{enumitem,amssymb}
\usepackage{float}
\usepackage[utf8]{inputenc}
\usepackage{color}
\usepackage{graphicx} 
\usepackage{amsmath} 
\usepackage{amssymb} 
\usepackage{wasysym}
\usepackage{mathtools}
\usepackage{newtxtext,newtxmath}
\def\arcs{\hbox{$^{\prime\prime}$}}

\newcommand{\cgsa}{erg s$^{-1}$ cm$^{-2}$ $\angstrom^{-1}$}
\newcommand{\cgs}{erg s$^{-1}$ cm$^{-2}$}

\graphicspath{{./}{figures/}}
\usepackage[title]{appendix}

\newcommand{\rv}[1]{#1}

\shorttitle{ELiXer}
\shortauthors{Davis, et al.}

\begin{document}

\title{The HETDEX Survey: Emission Line Exploration and Source Classification \footnote{Based on observations obtained with the Hobby-Eberly Telescope, which is a joint project of the University of Texas at Austin, the Pennsylvania State University, Ludwig-Maximilians-Universit\"at M\"unchen, and Georg-August-Universit\"at G\"ottingen.  The HET is named in honor of its principal benefactors, William P.~Hobby and Robert E.~Eberly.}}

\author[0000-0002-8925-9769]{Dustin Davis}
\affiliation{Department of Astronomy, The University of Texas at Austin, 2515 Speedway Boulevard, Austin, TX 78712, USA}

\author[0000-0002-8433-8185]{Karl Gebhardt}
\affiliation{Department of Astronomy, The University of Texas at Austin, 2515 Speedway Boulevard, Austin, TX 78712, USA}

\author[0000-0002-2307-0146]{Erin Mentuch Cooper}
\affiliation{Department of Astronomy, The University of Texas at Austin, 2515 Speedway Boulevard, Austin, TX 78712, USA}
\affiliation{McDonald Observatory, The University of Texas at Austin, 2515 Speedway Boulevard, Austin, TX 78712, USA}

\author[0000-0002-1328-0211]{Robin Ciardullo} \affiliation{Department of Astronomy \& Astrophysics, The Pennsylvania
State University, University Park, PA 16802, USA}
\affiliation{Institute for Gravitation and the Cosmos, The Pennsylvania State University, University Park, PA 16802, USA}

\author[0000-0002-7025-6058]{Maximilian Fabricius}
\affiliation{Max-Planck Institut f\"ur extraterrestrische Physik, Giessenbachstrasse 1, 85748 Garching, Germany}
\affiliation{Universit{\"a}ts-Sternwarte, Fakult{\"a}t f{\"u}r Physik, Ludwig-Maximilians Universit{\"a}t M{\"u}nchen, Scheinerstr. 1, 81679\ M{\"u}nchen, Germany}

\author[0000-0003-2575-0652]{Daniel J. Farrow}
\affiliation{Universit{\"a}ts-Sternwarte, Fakult{\"a}t f{\"u}r Physik, Ludwig-Maximilians Universit{\"a}t M{\"u}nchen, Scheinerstr. 1, 81679\ M{\"u}nchen, Germany}
\affiliation{Max-Planck Institut f\"ur extraterrestrische Physik, Giessenbachstrasse 1, 85748 Garching, Germany}

\author[0000-0003-2908-2620] {John J. Feldmeier}
\affiliation{Department of Physics, Astronomy, Geology, and Environmental Sciences, Youngstown State University
Youngstown, OH 44555}

\author[0000-0001-8519-1130]{Steven L. Finkelstein}
\affiliation{Department of Astronomy, The University of Texas at Austin, 2515 Speedway Boulevard, Austin, TX 78712, USA}

\author[0000-0003-1530-8713]{Eric Gawiser}
\affiliation{Department of Physics and Astronomy, Rutgers, The State University of New Jersey, Piscataway, NJ 08854, USA}

\author[0000-0001-6842-2371]{Caryl Gronwall}
\affiliation{Department of Astronomy \& Astrophysics, The Pennsylvania State University, University Park, PA 16802, USA}
\affiliation{Institute for Gravitation and the Cosmos, The Pennsylvania State University, University Park, PA 16802, USA}

\author[0000-0001-6717-7685]{Gary J. Hill}
\affiliation{McDonald Observatory, The University of Texas at Austin, 2515 Speedway Boulevard, Austin, TX 78712, USA}
\affiliation{Department of Astronomy, The University of Texas at Austin, 2515 Speedway Boulevard, Austin, TX 78712, USA}

\author[0000-0003-1008-225X]{Ulrich Hopp}
\affiliation{Universit{\"a}ts-Sternwarte, Fakult{\"a}t f{\"u}r Physik, Ludwig-Maximilians Universit{\"a}t M{\"u}nchen, Scheinerstr. 1, 81679\ M{\"u}nchen, Germany}
\affiliation{Max-Planck Institut f\"ur extraterrestrische Physik, Giessenbachstrasse 1, 85748 Garching, Germany}

\author[0000-0002-1496-6514]{Lindsay R. House}
\affiliation{Department of Astronomy, The University of Texas at Austin, 2515 Speedway Boulevard, Austin, TX 78712, USA}

\author[0000-0002-8434-979X]{Donghui Jeong}
\affiliation{Department of Astronomy \& Astrophysics, The Pennsylvania
State University, University Park, PA 16802, USA}
\affiliation{Institute for Gravitation and the Cosmos, The Pennsylvania State University, University Park, PA 16802, USA}

\author[0000-0002-0417-1494]{Wolfram Kollatschny}
\affiliation{Institut f\"{u}r Astrophysik, Universit\"{a}t G\"{o}ttingen, Friedrich-Hund-Platz 1, D-37077 G\"{o}ttingen, Germany}

\author[0000-0002-0136-2404]{Eiichiro Komatsu}
\affiliation{Max-Planck-Institut f\"{u}r Astrophysik, Karl-Schwarzschild-Str. 1, 85741 Garching, Germany}
\affiliation{Kavli Institute for the Physics and Mathematics of the Universe (WPI), Todai Institutes for Advanced Study, the University of Tokyo, Kashiwanoha, Kashiwa, Chiba 277-8583, Japan}

\author[0000-0003-1838-8528]{Martin Landriau}
\affiliation{Lawrence Berkeley National Laboratory, 1 Cyclotron Road, Berkeley, CA 94720, USA}

\author[0000-0001-5561-2010]{Chenxu Liu}
\affiliation{Department of Astronomy, The University of Texas at Austin, 2515 Speedway Boulevard, Austin, TX 78712, USA}

\author[0000-0002-6186-5476]{Shun Saito}
\affiliation{Institute for Multi-messenger Astrophysics and Cosmology, Department of Physics, Missouri University of Science and Technology, 1315 N Pine St, Rolla, MO 65409}
\affiliation{Kavli Institute for the Physics and Mathematics of the Universe (WPI), Todai Institutes for Advanced Study, the University of Tokyo, Kashiwanoha, Kashiwa, Chiba 277-8583, Japan}

\author[0000-0002-7327-565X]{Sarah Tuttle}
\affiliation{Department of Astronomy, University of Washington, Seattle, 3910 15th Ave NE, Room C319, Seattle WA 98195-0002}

\author[0000-0002-0784-1852]{Isak G. B. Wold}
\affil{Astrophysics Science Division, Goddard Space Flight Center, Greenbelt, MD 20771, USA}
\affil{Department of Physics, The Catholic University of America, Washington, DC 20064, USA }
\affil{Center for Research and Exploration in Space Science and Technology, NASA/GSFC, Greenbelt, MD 20771}

\author[0000-0003-2307-0629]{Gregory R. Zeimann}
\affiliation{Hobby-Eberly Telescope, University of Texas, Austin, Austin, TX, 78712, USA}

\author[0000-0003-3817-8739]{Yechi Zhang}
\affiliation{Institute for Cosmic Ray Research, The University of Tokyo, 5-1-5 Kashiwanoha, Kashiwa, Chiba 277-8582, Japan}
\affiliation{Department of Astronomy, Graduate School of Science, the University of Tokyo, 7-3-1 Hongo, Bunkyo, Tokyo 113-0033, Japan}

\defcitealias{Gebhardt+2021}{KG21}

\begin{abstract}

The Hobby-Eberly Telescope Dark Energy Experiment (HETDEX) is an untargeted spectroscopic survey that aims to measure the expansion rate of the Universe at $z \sim 2.4$ to 1\% precision for both $H(z)$ and $D_A(z)$. HETDEX is in the process of mapping in excess of one million Lyman-$\alpha$ emitting (LAE) galaxies and a similar number of lower-z galaxies as a tracer of the large-scale structure. The success of the measurement is predicated on the post-observation separation of galaxies with \lya\ emission from the lower-$z$ interloping galaxies, primarily \OII, with low contamination and high recovery rates. The Emission Line eXplorer (ELiXer) is the principal classification tool for HETDEX, providing a tunable balance between contamination and completeness as dictated by science needs. By combining multiple selection criteria, ELiXer improves upon the 20~\AA\ rest-frame equivalent width cut commonly used to distinguish LAEs from lower-$z$ \OII\ emitting galaxies. Despite a spectral resolving power, R $\sim800$, that cannot resolve the \oii\ doublet, we demonstrate the ability to distinguish LAEs from foreground galaxies with 98.1\% accuracy. We estimate a contamination rate of \lya\ by \OII\ of 1.2\% and a \lya\ recovery rate of 99.1\% using the default ELiXer configuration. These rates meet the HETDEX science requirements. 
\end{abstract}


\section{Introduction}\label{sec:intro}

It is generally acknowledged that the universe is expanding and that the expansion is accelerating. Though surprising at the time, the accelerated expansion has come to be the consensus understanding since the early work of \cite{Perlmutter_1999} and \cite{Riess_1998}. Since then, many observations have confirmed and refined the measures of this expansion with such increased precision that a possible tension may have emerged in the results from the various broad measurement camps \citep[][among others]{divalentino_2021,aloni_2021}. Regardless, whether this tension is a consequence of real physics, as yet unidentified systematics, or some combination, we are essentially limited to only two anchor points, one from the recent past \citep[$\sim 72~ \mathrm{km~s^{-1}~Mpc^{-1};}$][and others]{Riess_2009,Dhawan_2018,shoes_2021,Mortsell_2021} and one from the Epoch of Recombination \citep[$\sim 67~ \mathrm{km~s^{-1}~Mpc^{-1};}$][and others]{Alam_2017,planck_2018_overview,Aiola_2020}, from which to constrain descriptions of dark energy. Further understanding requires additional data points from different epochs in the expansion history of the Universe. Multiple efforts are in progress to provide those data, including the following, but far from exhaustive, list: the Dark Energy Survey (DES) \citep{DES_2005}, the Baryon Oscillation Spectroscopic Survey (BOSS) \citep{Dawson_2012}, the extended Baryon Oscillation Spectroscopic Survey (eBOSS) \citep{Alam_2021}, the Legacy Survey of Space and Time (LSST) \citep{LSST_2009}, Euclid \citep{Euclid_2011}, the DESI Survey \citep{DESI_1,Dey_2019}, and, of course, the Hobby-Eberly Telescope Dark Energy Experiment (HETDEX) \citep[][]{Ramsey1998,Gebhardt+2021,Hill+21}.

HETDEX is a multi-year untargeted spectroscopic survey designed to make new measurements of the Hubble Parameter, $H(z)$, and the Angular Diameter Distance, $D_{A}(z)$, at z$\sim$2.4 to better than 1\% accuracy in an effort to better characterize dark energy and look for possible evolution. HETDEX observations fall into two large, high galactic latitude fields. The $\sim$ 390 deg$^2$ "Spring" field is centered near (RA,Dec) 13h00m +53d00m and the $\sim$150 deg$^2$ "Fall" field is centered near 1h30m +0d00m \citep[][]{Gebhardt+2021}. Functionally, HETDEX seeks to map the 3D positions of some $10^6$ galaxies between $1.88 < z < 3.52$ and use their large scale clustering to derive $H(z)$ and $D_A(z)$. More specifically, the galaxies HETDEX is using for large-scale structure are identified by their bright, conveniently red-shifted into the optical, Lyman-$\alpha$ emission lines. These Lyman-$\alpha$ Emitters (LAEs) are generally small, blue, rapidly star-forming galaxies that, while uncommon in the local Universe, are present in large numbers in the HETDEX redshift search window \citep[][and many others]{Peebles_1968,Gawiser_2007, Nilsson_2007,Finkelstein_2010}.

The HETDEX Visible Integral-Field Replicable Unit Spectrographs \citep[VIRUS;][]{Hill+21} cover the wavelength range 3500-5500\ \AA\ with R$\sim$750--900, and are optimized to detect \lya\ flux down to $\sim4\times 10^{-17}$ \cgs\ (increasing to closer to $2 \times 10^{-16}$~\cgs\ at the extreme blue end of the range). This allows the detection of \lya\ luminosities down to about $10^{42.3}$ $\mathrm{erg~s^{-1}}$ for $z\sim2.4$. Since it is of utmost importance to know the redshift of the observed galaxies, the emission must be correctly identified. However, the relatively narrow wavelength range often limits our ability to capture multiple emission lines and the low spectral resolving power prohibits most doublet splitting, making classifications difficult. Around 95\% of HETDEX emission line detections\rv{\footnote{{HDR3 is limited to emission line detections with SNR $\geq$ 4.8, of which 95\% have only a single detected emission line. The fraction of detections with only a single line is partly a function of the SNR cut and other selection criteria used to define a sample. As in \citet{Cooper_2022}}, SNR $\geq$ 5.5 is commonly used as it is effectively free from noise detections (\S \ref{contamination_from_noise}). For SNR $\geq$ 5.5, 70\% of HETDEX spectra consist of only a single emission line and the entire sample is reduce by 60\%.}} are spectra containing only one, apparently single peaked (given the HETDEX spectral resolving power) emission line, and \lya\ is not the only emission line to fall into this observed wavelength range. Neutral hydrogen (and dust) in each source galaxy's Interstellar Medium (ISM) and in the Intergalactic Medium (IGM) along our line of sight effectively eliminate emission lines blueward of \lya\ at higher redshifts \citep{Haardt1995,Meiksin,Cowie1998,Overzier,Vanzella_2018}, leaving low-$z$ galaxies as the primary contaminate to be considered. 

In the relatively nearby universe, intrinsically small, line-emitting faint galaxies can be misidentified as their higher redshift cousins. In particular, at the low HETDEX spectral resolving power and with no strong lines in the wavelengths around it, the \OII\-3727\AA\ emission line can be confused with \lya\-1216\AA\, which similarly appears unique in its spectral neighborhood. In a common case, HETDEX observations detect only a single, fairly narrow, emission line and little or no continuum at the detection limits. Most likely the line is either \lya\ and originates from a high-$z$ galaxy, or \OII\ from a low-redshift interloper, and unfortunately, these two primary cases occur in roughly equivalent numbers \citep[][]{Adams2011a,Gebhardt+2021}. Since the HETDEX $H(z)$ and $D_A(z)$ measurements are sensitive to interloper clustering \citep{Leung2017,Gebhardt+2021,Farrow+2021}, contamination from \OII\ in the LAE sample needs to be $\lesssim$ 2\% \citep{Gebhardt+2021}. Historically, a 20\,\AA\ equivalent width cut (using the rest-frame of \lya) has been used to segregate \OII\ from \lya \citep[][]{Gronwall2007a,Adams2011a}, and indeed, this criterion is quite effective. However, used by itself, the discriminant can still lead to $>$4\% contamination and degrade the recovery of lower equivalent width \lya\ lines \citep{Acquaviva2014}. \cite{Leung2017} improves on the 20\,\AA\ cut by taking a Bayesian approach and including information on the luminosity functions and equivalent width distributions of \lya\ and \OII\null. From their modeled data, they report an expected contamination by \OII\ of between $\sim 0.5$\% and 3.0\% at a cost of $\sim$ 6.0\% to 2.4\% lost LAEs, depending on the methods used. This is a significant enhancement over the simpler 20\,\AA\ cut and, in this work, we are able to extend and improve on \citet{Leung2017} by (1) incorporating additional selection criteria, (2) considering other emission lines as contaminants, and (3) comparing directly against observational data.

The HETDEX Emission Line eXplorer (ELiXer) software incorporates and extends these classification works, integrates supplemental data and additional classification criteria, and expands the analysis to consider more than two dozen other emission lines. Its primary objective is to classify every HETDEX emission line detection by assigning the correct redshift to the observed emission lines. In addition to its primary function as an emission line classifier, ELiXer also provides diagnostic and data integrity checking to supplement that of the HETDEX pipeline \citep{Gebhardt+2021}, which is run prior to the ELiXer invocation and provides the detection coordinates, observation conditions, processed (calibrated, PSF weighted) spectra, emission line parameter measurements (flux, line width), and CCD information as ELiXer inputs. These features are useful for identifying and debugging some issues (e.g. errant sky subtraction, stuck/hot pixels, amplifier interference, etc) as well as in the manual inspection of individual detections. 

While ELiXer does classify all HETDEX detections regardless of magnitude, additional classification support is provided for continuum-bright sources via another software tool utilized by HETDEX called \textit{Diagnose}, developed for the Hobby Eberly Telescope VIRUS Parallel Survey (HETVIPS, \citet{Zeimann_2022}). For a further description of source classification and redshift assignment of HETDEX sources please see \citet{Cooper_2022}. Here, however, we focus only the bulk of the HETDEX detections, where ELiXer is the primary (or only) classifier. For this work, we reference ELiXer version 1.16 used in the generation of the most recent HETDEX detections catalog, HETDEX Data Release 3 (HDR3). This catalog contains more than 1.5 million entries and was released internally in April 2022 with a public version to be released in the future. We report a projected HETDEX LAE contamination rate from \OII\ of 1.2\% ($\pm$0.1\%) and an additional 0.8\% ($\pm$0.1\%) from all other sources, along with an LAE recovery rate of 99.1\% ($\pm$3.3\%) for the default classification configuration. ELiXer provides a tunable \lya\ classifier, allowing the balancing of contamination vs.\  completeness as needed for specific science goals (see \S \ref{results}). ELiXer is a work in progress and continues to evolve and improve as more data are collected, both from HETDEX and from other surveys, and as classification methods are added and refined.

The remainder of this paper is organized as follows: Section \ref{sec:catalogs} provides an overview of the various photometric catalogs currently included in ELiXer. Section \ref{sec:classification} describes the classification methodologies and supporting functions. Section \ref{testing} covers the selection of a Spectrocopic-z Assessment Sample (SzAS) providing spectroscopic redshifts from various imaging catalogs and the results of testing against that sample. Section \ref{discussion} presents a discussion of the results and the science implications. Section \ref{conclusions} summarizes the work and future enhancements. Example ELiXer detection reports are shown in Appendix-\ref{sec:elixer-report-APPENDIX} with descriptions provided for the major features.

Throughout the paper, the Planck 2018 cosmology \citep{planck2018} with $\Omega_{\mathrm{\Lambda}}$= 0.69, $\Omega_{\text{m}}$= 0.31 and H$_0$ = 67.7 $\mathrm{km~s^{-1}~Mpc^{-1}}$ is assumed. All magnitudes are in the AB system \citep[][]{OkeGunn1983} and coordinates are J2000.\\


\section{Imaging Catalogs} \label{sec:catalogs}

HETDEX is an untargeted spectroscopic survey, and the spectra alone provide most of the critical information for object classification. Coupled with the on-sky positions of the associated fibers, these data form the basis for the HETDEX cosmology measurements. For the brighter detections, a source's redshift and, to a lesser degree, its physical extent and morphology can be determined securely from the spectra. However, for the fainter emission line detections, additional information from archival photometric imaging, including an object's magnitude, color, angular/physical size, morphology, and even on-sky neighbors, can prove quite useful in ascertaining its identity. Even superimposing the HETDEX fiber positions on imaging data can provide diagnostic checks on the astrometry and the reduction pipeline. Given these substantial benefits, ELiXer attempts to match all HETDEX observations with multi-band archival photometry at the highest angular resolution and imaging depth available.\\

\subsection{Individual Catalog Summaries}

At the time of writing, ELiXer references 11 separate imaging catalogs, most with their own associated object catalog. These catalogs are of varying depth, resolution, band-coverage, and footprint. Additional catalogs can be added at any time and several new or expanded source lists are anticipated before the next HETDEX data release. With the exceptions of an $r$-band survey from the HyperSuprimeCam group (HSC-DEX) and a $g$-band survey from Kitt Peak National Observatory (KPNO;HETDEX-IM) that were specially designed and executed for HETDEX, all imaging and object catalogs are archival and publicly available. These catalogs are summarized in Table \ref{tab:Table of Imaging Catalogs} and in the list below.

\begin{deluxetable*}{ >{\centering}p{3.8cm} | >{\centering}m{1.8cm} | >{\centering}m{1cm} |  >{\centering}p{3.8cm} | >{\centering}m{1.8cm} | >{\centering}m{3.3cm} } [ht]
\tablecaption{Summary of the imaging surveys incorporated into ELiXer \label{tab:Table of Imaging Catalogs}}
\tablewidth{0pt}
\tablehead{
\colhead{Name}  & \colhead{HETDEX Field} & \colhead{\rv{Overlap\tablenotemark{\small{1}}}} & \colhead{Filters and Depth\tablenotemark{\small {2}}} & \colhead{PSF FWHM\tablenotemark{\small {3}}} & \colhead{Object Catalog\tablenotemark{\small {4}}} 
}
\startdata
Canada-France-Hawaii Telescope Legacy Survey (CFHTLS)  & Spring & 4\% & {Deep: $u$(26.3), $g$(26.0), $r$(25.6), $i$(25.4), $z$(25.0)\\ Wide: $u$(25.2), $g$(25.5), $r$(25.0), $i$(24.8), $z$(23.9)} & 0.6-1.0\arcs & phot-$z$ \tabularnewline
\hline
$HST$ Cosmic Assembly Near-infrared Deep Extragalactic Legacy Survey (CANDELS) in the Extended Groth Strip (EGS)  & Spring & $<$1\% & {ACS/WFC: F606W, F814W\\WFC3: F105W, F125W, F140W, F160W} & 0.08\arcs & spec-$z$, phot-$z$ \tabularnewline
\hline
$HST$ Cosmic Assembly Near-infrared Deep Extragalactic Legacy Survey (CANDELS) in the Great Observatories Origins Deep Survey, North (GOODS-N) & Spring & $<$1\% & {ACS/WFC: F435W, F606W, F775W, F814W\\WFC3: F105W, F125W, F160W} & 0.08\arcs & spec-$z$, phot-$z$ \tabularnewline
\hline
Hyper Suprime-Cam HETDEX Survey (HSC-DEX)  & Spring & 44\% & $r$(25.5) & 0.6-1.0\arcs & mag only \tabularnewline
\hline
Kitt Peak National Observatory HETDEX Imaging Survey (KPNO; HETDEX-IM) & Spring & 20\% & $g$(24.4) & 1.1-1.5\arcs  & mag only \tabularnewline
\hline
Cosmic Evolution Survey (COSMOS) with Dark Energy Camera (DECam) & Fall & 2\% & $g$(25.5), $r$(25.5) & 0.7-1.0\arcs & {(1) phot-$z$ (Laigle+2015)\\ (2) mag only}   \tabularnewline
\hline
Hyper Suprime-Cam Subaru Strategic Program (HSC-SSP)  & Fall & 29\% & {Deep $g$(27.5), $r$(27.1), $i$(26.8), $z$(26.3), $y$(25.3)\\ Wide $g$(26.5), $r$(26.1), $i$(25.9), $z$(25.1) ,$y$(24.4)} & 0.6-1.0\arcs & mag only \tabularnewline
\hline
Spitzer/HETDEX Exploratory Large-Area (SHELA) with Dark Energy Camera (DECam) & Fall & 25\% & $u$(25.4), $g$(25.1), $r$(24.7), $i$(24.0), $z$(23.7) & 0.7-1.0\arcs & mag only  \tabularnewline
\hline
Dark Energy Camera Legacy Survey (DECaLS) & Spring \& Fall & 17\% & $g$(24.0), $r$(23.4), $z$(22.5) & 1.2\arcs & {No} \tabularnewline
\hline
Panoramic Survey Telescope and Rapid Response System (Pan-STARRS) & Spring \& Fall & $<$1\% & $g$(23.3), $r$(23.2), $i$(23.1), $z$(22.3), $y$(21.3) & 1.0-1.3\arcs & No  \tabularnewline
\hline
Sloan Digital Sky Survey (SDSS) DR16 & Spring \& Fall & $<$1\% & $u$(22.0), $g$(23.1), $r$(22.7), $i$(22.2), $z$(20.7)  & 1.3\arcs & spec-$z$, phot-$z$ \tabularnewline
\enddata
\tablenotetext{1}{\small{\rv{Fraction of HETDEX Data Release 3 within each catalog footprint, except for DECaLS, Pan-STARRS, and SDSS which report only the fraction which does not also overlap with a previously listed catalog. Since multiple catalogs overlap, the column sums to $>$ 100\%.}}}
\tablenotetext{2}{\small{Approximate average AB depth over the whole catalog as reported, typically for point sources and 2\arcs apertures. For some $g$ and $r$ filters and some image tiles, ELiXer uses its own estimated depths at 1\arcs and 2\arcs apertures. Not all surveys use the same SDSS \textit{ugriz} filters, though for this purpose they are approximately similar. Only filters used by ELiXer are listed.}}
\tablenotetext{3}{\small{Typically in $r$-band}}
\tablenotetext{4}{\small{If not "No", also has an object catalog used by ELiXer with at least $g$ or $r$ magnitudes. Spec-$z$ and/or phot-$z$ redshifts are available where noted, but not necessarily for all object entries.}}
\end{deluxetable*}

\begin{itemize}
    \item \textit{Canada-France-Hawaii Telescope Legacy Survey (CFHTLS)}:  
    A multi-band ($ugriz$) imaging survey and joint venture of the National Research Council of Canada, the Institut National des Science de l'Univers of the Centre National de la Recherche Scientifique (CNRS) of France, and the  University of Hawaii, utilizing the MegaPrime/MegaCam on the 3.6m Canada-France-Hawaii Telescope (CFHT) on Mauna Kea. ELiXer uses the deep and wide fields, D3/W3 centered near RA 210$^{\circ}$, Dec +52$^{\circ}$.
    \citep[][]{Brimioulle_2008,CFHTLS}
    
    \item \textit{HST Cosmic Assembly Near-infrared Deep Extragalactic Legacy Survey (CANDELS) in the Extended Groth Strip (EGS)}: CANDELS is a deep  $HST$ survey (900+ orbits) with multiple filters in the optical (using the Advanced Camera for Surveys, ACS) and near-IR (using the Wide Field Camera 3, WFC3) studying on galaxy evolution with an emphasis on Cosmic Dawn and Cosmic High Noon. The EGS is one of the five fields of CANDELS and is centered near RA 215$^{\circ}$, Dec +53$^{\circ}$.  \citep[][]{CANDLES,Koekemoer_2011,candels_egs_Stefanon}. The photometric redshifts used in ELiXer are provided by Andrews, B., et al, ApJ submitted.

    \item \textit{HST Cosmic Assembly Near-infrared Deep Extragalactic Legacy Survey (CANDELS) in the Great Observatories Origins Deep Survey, North (GOODS-N)}: Another of the 5 CANDELS fields (see previous bullet), GOODS-N is centered near RA 189$^{\circ}$, Dec +62$^{\circ}$ \citep[][]{GOODSN,CANDLES,Koekemoer_2011,goodsn_barro} Again, the photometric redshifts used in ELiXer are provided by Andrews, B., et al, ApJ submitted.

    \item \textit{Hyper Suprime-Cam HETDEX Survey (HSC-DEX)}:  \label{sec:HSC-DEX}
    This survey consists of three nights of HSC $r$-band observations with the Subaru/HSC in 2015-2018 (PI: Andreas Schulze) and 2019-2020 (PI: Shiro Mukae) and covers the $\sim250$ \sqdeg\ area of the HETDEX Spring field. Data reduction and source detections were performed with version 6.7 of the HSC pipeline, hscPipe \citep{Bosch2018a}, and produced $r$-band images with a 10$\sigma$ limit of $r = 25.1$ mag in a $2\arcsec$ diameter circular aperture. 
    These HSC $r$-band images are complementary to 
    the existing imaging data of the Kitt Peak 4-m Mosaic camera and the CFHT Wide-Field Legacy survey.  
    
    \item \textit{Kitt Peak National Observatory HETDEX Imaging Survey (KPNO; HETDEX-IM)}:  A $g$-band survey with the Mosaic camera on the Mayall 4-m telescope at Kitt Peak National Observatory in 2011-2014 (PI: Robin Ciardullo).      
    \item \textit{Cosmic Evolution Survey (COSMOS) with Dark Energy Camera (DECam)}:
    The 3 deg$^2$ \textit{ugriz}-band COSMOS DECam catalog was generated with the same procedure used for the larger field of view SHELA DECam survey listed below \citep[][]{Wold2019}. This also overlaps with \cite{Laigle+2015}.

    \item \textit{Hyper Suprime-Cam Subaru Strategic Program (HSC-SSP)}: Multi-depth, multi-band, wide-field imaging survey using the Hyper Suprime-Cam on the 8.2m Subaru at the Mauna Kea Observatories. For HETDEX Data Release 3, ELiXer uses HSC-SSP Public Data Release 3 from August 2021. \citep[][]{HSC-SSP}
    
    \item \textit{Spitzer/HETDEX Exploratory Large-Area (SHELA) with Dark Energy Camera (DECam)}:
    This survey covers 17.5 deg$^2$ of the HETDEX Fall field within the Sloan Digital Sky Survey (SDSS) “Stripe 82” region. The ugriz-band DECam catalog is riz-band-selected and reaches a $5\sigma$ depth of $\sim24.5$ AB mag for point sources \citep[][]{Wold2019}. 

    \item \textit{Dark Energy Camera Legacy Survey (DECaLS)}: A multiband ($grz$) photometric survey, part of the Dark Energy Survey \citep[][]{des}, based at the  Cerro Tololo Inter-American Observatory using the Dark Energy Camera (DECam) on the 4m Blanco telescope. ELiXer uses Data Release 9 which also includes observations from the Beijing-Arizona Sky Survey (BASS) and the Mayall z-band Legacy Survey (MzLS).
    \citep[][]{DECaLS}
    
    \item \textit{Panoramic Survey Telescope and Rapid Response System (Pan-STARRS)}: Specifically, Pan-STARRS1, is a set of wide-field synoptic imaging surveys using the 1.8m PS1 optical telescope at the Haleakala Observatories. PS1 collected data from 2010 through 2014. \citep[][]{PanSTARRS}
    
    \item \textit{Sloan Digital Sky Survey (SDSS)}: Multiband ($ugriz$) wide-field survey in operation since 2000 using a 2.5m optical telescope at the Apache Point Observatory. ELiXer uses Data Release 16 from SDSS Phase-IV. \citep[][]{SDSS_DR16}\\
    
\end{itemize}

\subsection{ELiXer Aperture Photometry} \label{elixer_aperture_phot}

ELiXer directly uses the photometric imaging to gather aperture magnitudes for the HETDEX detections. While magnitudes are computed for each available filter, only $g$ and $r$ magnitudes are used in the classification process (\S \ref{sec:classification}). For each HETDEX detection, ELiXer identifies the catalogs with overlapping imaging and gathers postage-stamp ($9\arcs \times 9\arcs$ by default) imaging cutouts centered on the HETDEX detection's coordinates. Three sets of aperture magnitudes are then computed using the Python packages \textit{Astropy} \citep[][]{astropy}, \textit{Photutils} \citep[][]{photutils}, and \textit{Source Extraction and Photometry (SEP)} \citep[][]{sep_pkg}. The identified aperture(s) are used later to provide continuum estimates (\S \ref{continuum_estimates}) and size information (\S \ref{object_size_vote}).

First, ELiXer computes a magnitude for a dynamically sized circular aperture. We center the circular aperture on the HETDEX coordinates, compute the magnitude within the aperture, and allow the aperture to grow until the magnitude stabilizes \citep[\eg{}][]{Howell1989}. The initial size is set by a combination of the median seeing and pixel scale of the catalog+filter and is typically $\sim 1\arcs$ in diameter. The magnitude within the aperture is computed, with the background determined from an annulus $2\times$ to $3\times$ the defined maximum allowed object aperture (6$\arcs$ diameter by default, for an annulus of 12$\arcs$ to 18$\arcs$). The aperture is then grown in steps of $0\farcs 1$, with each measurement recorded, until the maximum diameter is reached. The smallest aperture size where the magnitude change to the next step up is less than 0.01 is assigned, and the corresponding magnitude is selected.

Next, ELiXer uses \textit{SEP} \citep[][]{sep_pkg}, which is based on the original \textit{Source Extractor} \citep[][]{source_extractor}, iterating over each cutout and records the magnitude, barycentric position, major and minor axes, and orientation of each identified object. ELiXer also computes and records the angular separation from each barycenter to the HETDEX coordinates and the separation to the nearest point on the bounding ellipse if the HETDEX position lies outside that ellipse. The object with the nearest barycenter to the HETDEX position whose bounding ellipse includes the HETDEX position is considered the best aperture match. If no object's ellipse includes the HETDEX position, then the object with the nearest ellipse point to the HETDEX position but no more than $0\farcs 5$ away is selected as the best match. If no object meets these criteria then no \textit{SEP} found object is selected and the best circular aperture (see previous paragraph) is used for the aperture photometry.

Lastly, at each \textit{SEP} identified barycenter, ELiXer computes and records the background subtracted magnitude in a fixed, $3\farcs 0$ diameter circular aperture.  These aperture magnitudes are intended for use in any fixed-aperture spectral energy distribution (SED)-fitting and color comparisons, but are not otherwise significantly used in the core ELiXer processing.\\

\subsection{Catalog Counterpart Matching} \label{catalog_match}

ELiXer also attempts to match each HETDEX detection to one or more objects in each imaging catalog with a particular focus on $g$ and $r$ magnitudes, which can provide additional measures for use in other ELiXer functions.  Object matching is based on a combination of barycenter position and agreement between the magnitudes reported by each catalog, the magnitudes computed within the ELiXer ellipses (\S \ref{elixer_aperture_phot}), and the HETDEX spectrum estimated $g$-band magnitude. 

The nearest catalog object to the HETDEX position that falls within the selected best aperture (\S \ref{elixer_aperture_phot}), or the nearest catalog object within $1\farcs 0$ of the HETDEX position if no object falls within the best aperture, is identified as the catalog match object. If the candidate object's reported magnitude is not compatible with the magnitude estimated from the HETDEX spectrum, then the next nearest object is evaluated until a match is found or the distance criteria are no longer satisfied. Compatibility with the HETDEX $g$ magnitude (\S \ref{hetdex_spectrum_continuum}) is defined as an absolute difference of 0.5 magnitudes; if the HETDEX $g$ magnitude is fainter than the HETDEX magnitude limit (about $25_{AB}$), then no faint-side restriction is imposed. On the other end, if both the counterpart and the HETDEX magnitudes are brighter than $22_{AB}$, they are considered compatible. For the purposes of this comparison, $g$ and $r$ are considered equivalent. There is at most one catalog match object per catalog+filter combination. This object is later used for additional information, including spec-$z$ and phot-$z$ assignments if available, in the classification process.\\

\section{Classification}\label{sec:classification}

Classifications in ELiXer are broadly interpreted as the identification of the redshifts of observed astrophysical objects. This properly requires the additional steps of correctly associating an observed spectrum with a single host object and furthermore identifying or bounding what constitutes that "single object". More fundamentally, given a spectrum and a specified emission line in that spectrum, what we hereafter call the "anchor line", ELiXer attempts to determine the identity, and thus the redshift, of that anchor line. Classification proceeds from the assumption that the anchor line is real and not spurious noise, an instrument or software artifact, or a misinterpretation of spectral data, such as the misidentification of continuum between two closely-separated absorption troughs. ELiXer initially assumes that the spectrum represents a single object (single redshift), though later analysis explores the possibility that a HETDEX spectrum is a blend of spectra from discrete but immediately adjacent or overlapping sources on sky (within a single, common detection aperture) at different redshifts. 

\rv{The focus of ELiXer's classification is placed on distinguishing \lya from \oii, by far the most common \lya contaminant in HETDEX data, and the bulk of the tests and conditions target that objective. Additional checks, described throughout this section, attempt to refine this bifurcated classification and identify the spectral line(s) as any one of those listen in Table \ref{tab:Table of Lines}. As will be discussed in \S \ref{discussion}, these "Other" lines are encountered much less frequently than \lya\ and \oii\ and, while they can be more challenging to identify, the HETDEX cosmology science is extremely robust against contamination from these misclassifications.}
 
The classification of HETDEX detections is organized to answer three increasingly general questions, with each answer incorporating the results of the previous question.
First, closely following the work of \citet[][]{Leung2017}, we evaluate the relative likelihood that the target emission line is \lya and rather than \oii\ \citep[][]{Adams2011a,Gebhardt+2021,Farrow+2021}. This is largely based on measurements of the emission line luminosity and equivalent width evaluated against luminosity and equivalent width distributions of \lya and \oii\ emitting galaxies from other publications interpolated at the redshift corresponding to the emission line wavelength (see \S \ref{plae_poii_sec}). Second, we determine the confidence of the initial classification by performing checks against more than two dozen other emission lines. Here a weighted voting scheme is used with many independent (or semi-independent) rules applied to measured and derived features of the spectrum and detection object. Third, we assign, with some rough measure of quality, the redshift and thus the specific identity of the emission line(s). This final step incorporates some additional rules and weights to combine all prior results. 

\rv{Broadly, ELiXer classifications build up evidence in a series of steps and then weighs the evidence to make a determination. The high level steps are fairly serial and often largely independent, with their results only combined toward the end of the process. These major steps are described in more detail, and in roughly the same order, in the subsections that follow.
\begin{enumerate}
    \item Find, fit, and score all emission and absorption lines and set the anchor line 
    \item Evaluate all combinations of found spectral lines for compatibility with redshifts, based on relative positions, strengths, etc
    \item Collect additional (aperture) photometric imaging information and any reported magnitude, spec-z, and phot-z measurements for the target object and its neighbors from non-HETDEX catalogs (Table \ref{tab:Table of Imaging Catalogs})
    \item Evaluate spectra shape, lines, and imaging for consistency with known astrophysical objects (star, White Dwarf, AGN, meteor, low-z galaxy)
    \item Examine HETDEX data for corruption, pipeline artifacts, and instrument issues.
    \item Test the compatibility of the anchor line with \lya
    \item Perform evaluations on the anchor line, including spectral and photometric information, to specifically distinguish \lya from \oii
    \item Perform separate evaluations on the anchor line, including spectral and photometric information, for consistency with lines other than \lya and \oii
    \item Combine all evaluations to determine and rank likely redshifts and line classifications
    \item Re-evaluate redshift classification based on clustering with ELiXer results from the other neighboring HETDEX detections
\end{enumerate}}

The figures in this section illustrating some of the voting criteria and thresholds pull their data from the Spectroscopic-$z$ Assessment Sample (SzAS) whose selection and composition is described in Section \ref{testing}. \\

\subsection{Line Finder} \label{ss_line_finder}

Emission (and absorption) line detection is implemented as both a layered, untargeted search and a targeted line fit assuming an "anchor" line. More details will follow in the next subsections, but briefly put, the untargeted search scans the full width of the spectrum from blue to red, marks the locations of possible emission line-centers, and attempts to fit a single Gaussian (in agreement with the measured instrumental resolution; Hill \etal 2021) to each position. The targeted search uses a single previously identified emission line (from the HETDEX input, user input, or the previous untargeted search) as an anchor and then assumes that anchor line is one of roughly two dozen potential emission lines (Table \ref{tab:Table of Lines}) and attempts to fit a Gaussian to the positions where other emission lines could be found, assuming that identify for the anchor line. The descriptions that follow are couched in terms of emission lines, as that is the primary use. A limited use of absorption lines is implemented and is described in \S \ref{linefinder_absorption}.\\

\begin{deluxetable}{>{\centering}p{1.4cm}  >{\centering}p{1cm}| >{\centering}p{1.4cm}  >{\centering}p{1cm} } [ht]
\tablecaption{Emission Line Candidates \label{tab:Table of Lines}}
\tablewidth{0pt}
\tablehead{
\colhead{Name}  & \colhead{rest-$\lambda$ [\AA]} & \colhead{Name} & \colhead{rest-$\lambda$ [\AA]} 
}
\startdata
\ion{O}{6} & 1035 & H{$\eta$} & 3835  \tabularnewline
\lya & 1216 & [\ion{Ne}{3}] & 3869  \tabularnewline
\ion{N}{5} & 1241 & H{$\zeta$} & 3889  \tabularnewline
\ion{Si}{2} & 1260 & (K) \ion{Ca}{2}\tablenotemark{\small{*}} & 3934  \tabularnewline
\ion{Si}{4} & 1400 & [\ion{Ne}{3}] & 3967  \tabularnewline
\ion{C}{4} & 1549 & (H) \ion{Ca}{2}\tablenotemark{\small{*}} & 3968  \tabularnewline
\ion{He}{2} & 1640 &  H{$\epsilon$} & 3970  \tabularnewline
\ion{C}{3}] & 1909 & H{$\delta$} & 4101 \tabularnewline
\ion{C}{2}] & 2326 & H{$\gamma$}  & 4340 \tabularnewline
\ion{Mg}{2} & 2799  & H{$\beta$} & 4861  \tabularnewline
\ [\ion{Ne}{5}] & 3346 & [\ion{O}{3}] & 4959 \tabularnewline
\ [\ion{Ne}{5}] & 3426 & \ion{Na}{1} & 4980  \tabularnewline
\ [\ion{O}{2}] & 3727 & [\ion{O}{3}] & 5007  \tabularnewline
           &      & \ion{Na}{1}\ & 5153
\enddata
\tablenotetext{*}{\small{Fit as an absorption line}}
\tablecomments{
Possible identifications for spectral lines found in the HETDEX spectra.
}
\end{deluxetable}

\subsubsection{Untargeted Search}\label{blind_search}
The untargeted search scans the entire 1D HETDEX spectrum to identify the positions and model the parameters of potential emission lines. It is used to (1) identify the strongest line as the reference or anchor line when no initial emission line is explicitly provided, (2) mark strong lines for consistency checks with redshift solutions and to help identify blended spectra, and (3) mark line positions for followup visual inspection, without respect to the selected solution.

Because Markov Chain Monte Carlo (MCMC) fits are relatively computationally expensive, and HETDEX spectra typically have only one or very few emission lines, we do not want to perform such fits at each pixel along the spectrum. Instead, we first conduct a quick examination to narrow the potential locations of emission lines. We do this using two independent algorithms and then combine the output positions into a single list for further examination. 

Two passes through the algorithms of this untargeted search are conducted. The first execution uses the native 2\,\AA\ binned HETDEX spectrum and focuses on identifying the common narrow spectral features. The second execution is performed after passing the original spectrum through a median filter (by default using a 5 pixel kernel), to smooth out some of the noise. This helps identify candidate emission lines that are wider than the $\sim 400$~km~s$^{-1}$ resolution of the VIRUS spectrographs and may have small noise peaks within their overall broad shape. 

The first algorithm searches for the basic shape of an emission feature, a general rise to a peak and then a decline. Due to the unavoidable noise in the data, the spectra are not smooth and the use of the first derivative to find zeros (and the second derivative to distinguish between an emission and absorption) results in more false detections than real spectral features. Instead, we look for the general shape of the lines (a rise and fall in the flux of minimum height over a minimum width), based on the spectral resolution, flux limits, and noise of HETDEX\null. Sets of contiguous pixels that are sufficiently wide in the spectral direction and have the expected rise-peak-fall pattern are recorded as possible emission lines, and their line centers are recorded.

The second algorithm counts contiguous pixels with flux values above some multiple of the corresponding noise (typically SNR $>$ 3, under the assumption that the flux uncertainty is distributed normally). Where the contiguous count of pixels above this noise is greater than some count (here, typically 3-5 pixels), the position of the highest flux value within that range is recorded as the possible emission line center. Essentially, this is just a SNR-cut over the spectrum. Unlike the first algorithm, the shape of the flux above the SNR-cut is irrelevant.

The line centers from each algorithm are then passed to fitting (\S \ref{gaussian_fitting}) and scoring routines (\S \ref{line_scoring}). When model fits to the flux at those positions are successful and the computed line score is sufficiently large, the feature is recorded to a list of potential spectral lines.

After both the standard and broad line searches are conducted, the list of potential emission and absorption lines are merged into a single list, and any neighboring lines with line centers within in 4\,\AA\ of each other are combined into single entries by keeping only the feature with the largest line score.

As a brief note: though this is not the normal operation of ELiXer under HETDEX, if no anchor line is specified for the spectrum to be classified, the line (emission or absorption) with the largest score (\S \ref{line_scoring}) found in this untargeted search is assumed as the anchor line. If the untargeted search fails to identify any spectral lines, the wavelength bin with the largest flux value is assigned as the anchor line position.\\

\subsubsection{Targeted Search}\label{targeted_search}

Unlike the untargeted search described above, the targeted search does not scan for potential emission or absorption lines, but instead attempts to fit for an emission or absorption feature at a specified position. Essentially, ELiXer attempts to fit spectral lines from a predefined list of common lines (Table \ref{tab:Table of Lines}) at their expected observed wavelength positions given an assumed identity or redshift for the anchor line. The redshift assumptions come from alternately interpreting the anchor line as each of the common lines and from any matching spectroscopic or photometric catalogs with a possible counterpart to the HETDEX detection. With each redshift assumption, all other lines in the subset that could occur within the HETDEX spectral window are fitted, allowing for some error in the systemic redshift (see \textit{Position Capture} under \S \ref{gaussian_fitting}). This is often redundant with the untargeted search in that, for higher signal-to-noise ratio (SNR) lines, the lines found in the targeted search are also found in the untargeted search. However lower SNR lines, \oiii\ $\lambda$4959 for example, can be missed in the initial sweep of the untargeted search. Fitting to a specific wavelength location helps avoids such misses.

Each successfully fitted line for each assumed identity of the anchor line is scored (\S \ref{line_scoring}) and associated with the redshift solution (\S \ref{redshift_solution_finder}) for that identification.\\

\subsubsection{Line Fitting}\label{gaussian_fitting}

ELiXer uses a simple, 4-parameter ($A$, $\mu$, $\sigma_{\rm Line}$, $y$) single Gaussian as the model to fit emission and absorption features:

\begin{equation} \label{eq:Gaussian}
    F(\lambda) = \frac{A}{\sigma_{\rm Line} \sqrt{2\pi}} \exp\left(- \frac{(\lambda-\mu)^{2}}{2\sigma_{\rm Line}^{2}}\right) + y,
\end{equation}

\noindent where $F(\lambda)$ is the flux per 2\,\AA\ wavelength bin, $A$ is the area under the curve or equivalently the integrated line-flux, $\mu$ is the line center, $\sigma_{\rm Line}$ is the measure of width, $y$ is the vertical offset, or flat continuum level, and $\lambda$ is the wavelength (at the midpoint of a 2\,\AA\ wide wavelength bin).

The flat continuum is a reasonable simplification, as \rv{no assumption is made as to the object type or its redshift}, most HETDEX detections have continua at or below the survey's continuum flux limit, and those objects with continua bright enough to have a shape typically have multiple emission lines or are too bright to support a \lya classification. \rv{This continuum estimate can be highly uncertain, especially for the noisier spectra, but as discussed later, multiple continuum estimates are combined to improve the uncertainty and for the non-detections, the resulting equivalent width estimates are lower limits that favor a low contamination \lya selection, at the cost of some completeness.}

Type~I AGN may have broad lines that are not well fitted by a single Gaussian \citep{liu_2022a}. Such detections are marked by ELiXer with warnings, but are not confused with the fainter, compact LAEs the software is designed to identify. We note, however, that it is possible that the simple emission line search can completely fail to find rare, extremely broad emission lines, as $\sim 3500$~\kms\ is the maximum FWHM that ELiXer attempts to fit.

More complex models, including the fitting of multiple emission and absorption lines within a single spectral feature, have either proven to be unreliable, too computationally costly, and/or of limited utility for the main goal of simply identifying redshifts when the vast majority of line detections are well fit by the simple, single Gaussian model. Fitting for an emission line doublet would be useful in the effort to distinguish between \lya\ and \oii\; however, given the low spectral resolving power of VIRUS, $\Delta \lambda / \lambda \sim 800$ \citep{Hill+21}, the \OII\ doublet (3726, 3729\ \AA) is unresolved as are most other doublets (\MgII (2796, 2803 \AA)  is sometimes marginally resolved). The increased run time of fitting these extra parameters is not justified. For smaller data sets, such as for the case of AGN exploration, more complex fitting is warranted \citep{liu_2022a}, but left to those specialized projects. For ELiXer's classification needs, a description of the spectral feature that is limited to its position (wavelength), equivalent width (approximate integrated line flux and local continuum), and line width are sufficient. Additional parameters, such as the model's skewness and kurtosis, and conditions combining those and other parameters have been explored but have not been found to improve the identification of real spectral features or aid in the classification, and are thus excluded from further discussion in this work.

With the exception of the anchor line on which an MCMC fit is always performed, if a least square (LSQ) model fit passes its quality checks, no MCMC fit is conducted. This is due to the increased runtime cost of MCMC fitting weighed against the relatively modest needs for classification. In all MCMC cases however, an LSQ fit is performed first and its results are used as initial conditions (with appropriate randomization) for the MCMC algorithm. ELiXer uses the Python \textit{scipy} package and its \textit{scipy.optimize.curve\_fit} \citep[][]{scipy} as the LSQ fitter; the MCMC fitter is from the Python \textit{emcee} package \citep[][]{Foreman_Mackey_2013}. Uncertainties in the LSQ fit are estimated using the square root of the diagonal of the covariance matrix. Uncertainties in the MCMC fit are estimated using the 68\% confidence interval in the parameter distribution.

A series of loose checks evaluates the quality of each fit as minimally good, marginal, or poor. Poor fits are rejected; good fits are scored (see \S\ref{line_scoring}) in preparation for building solutions. Marginal solutions from the LSQ fitter are passed to the MCMC algorithm for improved optimization and re-evaluated. If the subsequent MCMC fit is good, the fit is scored and made eligible for inclusion in redshift solutions. If the MCMC fit is not sufficiently improved over the LSQ fit, it is rejected. 

The quality checks include following conditions:

\begin{itemize}
\item \textit{Peak Capture}:  As a basic check, should the peak of the model fail to reproduce the most extreme measured data value near the line center within 50\%, the fit is rejected. If the model is within 25\% and 50\% of the most extreme value, it is flagged for an MCMC fit. Should that MCMC fit fail to be within 25\%, the fit is rejected and no line is assumed to be at that position.

\item \textit{Position Capture}:  If the fitted line center is greater than a configured maximum distance (in \AA) from the local data extremum, the fit is rejected. The maximum distance allowed can depend on the assumed line identification and its assumed position, with greater separations allowed for \lya which can be significantly offset from the systemic redshift \citep[][among others]{Shapley_2003,McLinden_2011,Verhamme_2018,Gurung_L_pez_2021}. During the untargeted search, no variations are allowed and a default of 8\,\AA\ ($\sim$ 500~\kms\ in the HETDEX spectral range) is used. 

\item \textit{Width Capture}:  If the fitted line width (here parameterized as $\sigma$) is less than 1.0\,\AA, i.e., significantly below the HETDEX spectral resolution of $\sim 2.0$\,\AA\ \citep{Hill+21}, or if the line width is greater than the configured maximum value of 17\,\AA\ ($\sim$ 2700 \kms\ FWHM) or 25\,\AA\ ($\sim$ 3500 \kms\ FWHM) for special, broad fit attempts, the fit is rejected.

\item \textit{Area Error}:  If the error on the line area (as estimated from the square root of the diagonal of the LSQ fit's covariance matrix or the 68\% confidence interval on the MCMC fit) is larger than the absolute value of the area (allowing for absorption or emission), the fit is rejected.

\item \textit{Local Uniqueness}:  This is used only in combination with other conditions. An emission or absorption line is considered unique if there is at most one other data extremum greater than 90\% of this line's peak between 1$\times$ FWHM and 1$\times$ FWHM + 10\,\AA\ to either side of the line center.
 
This is an alternate rough measure of local noise and is used primarily as a filter with low SNR lines. 

\item \textit{SNR and $\chi^{2}$}:  ELiXer uses the following definitions of SNR and $\chi^{2}$:

\begin{equation} \label{eq:SNR}
    \mathrm{SNR} = \frac{\sum{}{} \sqrt{(F(\lambda)-y)^2}}{\sqrt{\sum{}{} (\mathrm{error}^2)}},
\end{equation}

\begin{equation} \label{eq:chi2}
    \chi^{2} = \sum{}{} \left( \frac{\mathrm{data} - \mathrm{model}}{\mathrm{error}} \right) ^{2},
\end{equation}

\noindent where the summations are over the wavelength bins within $\pm 2\sigma$ of the fit line center. 
 $F(\lambda)$ and $y$ are from Eqn $\ref{eq:Gaussian}$. The model is the fitted flux evaluated at each corresponding wavelength bin for the data and the error is the uncertainty on the data.

The uncertainty on the SNR is computed via standard error propagation using the MCMC or LSQ uncertainties on each of the model's Gaussian parameters. 

If the LSQ fit is marginal given the previous conditions, it is rejected if (1) the SNR is less than 5.0 or (2) if the SNR is between 5.0 and 15.0 and the $\chi^{2}$ is greater than 2.0. These indicate poor fits to possibly noisy data and are generally not worth pursuing. Otherwise, the SNR and $\chi^{2}$ are recorded for use in line scoring.\\
\end{itemize}

\subsubsection{Line Scoring}\label{line_scoring}

Every successfully fitted emission and absorption line receives a score based only on its own properties, without consideration to the position or properties of any other fitted emission or absorption lines. If that score exceeds a minimum threshold, the line, with its score, is accepted into a list of potential line candidates for later use in redshift solution finder (\S\ref{redshift_solution_finder}). The minimum threshold is configurable and is set, by default, to an empirically determined value based on the manual examination of many tens of thousands of observed spectra and a simulation of spectra drawn from median HETDEX noise properties (\S\ref{spectra_simulation}). Redshift solutions that fit multiple lines to the spectrum receive a separate "solution score" (\S \ref{redshift_solution_finder}) that is based, in part, on these individual "line scores".

The line score attempts to capture and quantify features beyond just the signal-to-noise ratio, which is a less than ideal metric for broad emission lines fitted with a single Gaussian. The line score takes into account additional data including the magnitude of the integrated (fitted) line flux, the line position relative to expectations, and the uniqueness of the line within a local spectral region. The intent is to codify not just the presence of each potential emission line, but the consistency and significance of that line with respect to the spectrum at an assumed redshift.

The line score calculation is defined as:

\begin{equation}  \label{eq:line_score_eqn}
    S_L = \frac{\mathlarger{ S_{\mathrm{lim}} \cdot AN \cdot UN \cdot F_\lambda \cdot m_{\sigma} \cdot  m_{\mathrm{pix}}}}  {\mathlarger{1 + \mid\delta dx_0\mid}}
\end{equation}

\noindent where:
\begin{itemize}
    \item $S_L$ is the numerical line score. Noise peaks receive scores in the low single digits, typically less than 3.0. Weak emission lines (low SNR, low lineflux) typically receive scores in the 5.0 - 15.0 range. Extremely bright, high SNR lines can even exceed a score of 100.0, but are clipped to a maximum of 100.
    \item $S_{\mathrm{lim}}$ is the maximum allowed fitted SNR from a Gaussian fit, up to a configurable limit (20.0 by default). This helps scale the scoring by capping the maximum contribution of the SNR.
    \item $AN$ is the "Above Noise" factor, defined by the measured flux value of the emission peak divided by a noise estimate at that position and normalized by a configurable factor (by default, 5). The noise estimate used here is the standard deviation of the 3$\sigma$ clipped fluxes at the same wavelength over all (448) fibers on the detector. The value of $AN$ is clipped to the range [0,3].
    \item $UN$ is an estimate of how unique the line is relative to the nearby spectrum (i.e., the presence of several similarly narrow, low flux peaks in the same wavelength range likely indicate noise in the spectrum). This is an encoding of the \textit{Local Uniqueness} described in the previous subsection. If the candidate line is sufficiently broad, with a fit FWHM of greater than 6.5\,\AA\ or if fewer than 3 possible lines are found, the current candidate line is considered sufficiently unique and $UN$ takes on a value of 1, otherwise it takes on a value of 1/2. 
    \item $F_\lambda$ is the Gaussian fitted, continuum subtracted integrated line flux in units of $10^{-17}$ \cgs. There is no particular significance these units; they are simply used so that the value of the line score is generally in the range of 1-100.

    \item $m_{\sigma}$ encodes the minimum acceptable Gaussian fitted $\sigma$. Values of $\sigma$ greater than  1\,\AA\ result in $m_{\sigma}$ = 1, but values less than 1\,\AA\ receive a multiplicative penalty equal to the $\sigma$ value as they are unlikely to have been fit to a real emission line. This is equivalent to min$\left(\sigma,1 \right)$.

    \item $m_{\mathrm{pix}}$ encodes the minimum acceptable number of pixels ($N_{\mathrm{pix}}$) over which the SNR of the line is calculated. If the number of pixels is less than $N_{\mathrm{min}}$ (by default, 10 pixels to either side of the wavelength bin containing the line center), there is a multiplicative penalty imposed equal to $N_{\mathrm{pix}}$ / $N_{\mathrm{min}}$ . Low numbers of pixels in the SNR measurement may be due to masked or invalid pixels or a line location near the edge of the wavelength range. This is equivalent to min$\left( N_{\mathrm{pix}} / N_{\mathrm{min}},1 \right)$.
    \item $\delta d_{x0}$ is the offset, in \AA, of the fit line center from the expected location of the center line. 
    For features found by the untargeted search (\S\ref{blind_search}), this is the bin with the maximum (minimum, for absorption) flux within the spectrum slice being used to fit the line. For corroborating features as part of the "Targeted Search" (\S\ref{targeted_search}), it is the expected position of the assumed feature for the given redshift. 
\end{itemize}

An adjustment is made to the $S_L$ if the fit SNR is less than 8.0 and the $\chi^{2}$ is greater than 3.0. These are considered marginal fits that could have a large score due to the integrated line flux. In these cases, the score is reduced by a factor of ($\chi^{2} - 1$).

If the center of an emission line falls within a prominent sky line, specifically those centered at 3545\,\AA\ or 5462\,\AA, and if the FWHM does not extend past the sky line, the score is further reduced by a factor of 2, encoding the risk that the emission line is a relic of incomplete sky subtraction.

For very broad lines (fit FWHM > 20\,\AA), the scoring is modified by rejecting the line (setting the $S_L$ to 0) if the fitted SNR is less than a minimum threshold (by default, 19) and the $\chi^2$ of the Gaussian model is greater than a maximum (by default, 1.5). These fits tend to be poor, and caused either by artifacts in the data or the merging of multiple spectral features.

Since the focus is on faint galaxies with continuum below the HETDEX sensitivity, absorption features do not factor strongly in classification for most HETDEX catalog objects. As such, their base scoring value is scaled by a factor of 1/2 and optionally limited to a maximum value. \\

\subsubsection{Spectra Simulation and P(Noise)}\label{spectra_simulation}

As part of the scoring and in an effort to quantify the probability that a fitted line is simply the product of noise, we use the line finding code to analyze simulated spectra, treating all identified emission lines as false positives. The procedure is applied only to emission lines, not absorption lines, but the results are applicable to both.

As part of the configuration for ELiXer, we compute the PSF weighted spectral uncertainties versus wavelength from $10^4$ random, non-continuum detections from the entire HETDEX catalog, and generate the median uncertainty for each wavelength bin. We then simulate $10^4$ spectra, randomly drawing a flux for each wavelength bin (1036 random draws per spectrum over the range, 3470-5540\,\AA) according to the median uncertainty, and assuming a normal distribution about each uncertainty and no correlated noise between wavelength bins. Each simulated spectrum is passed through the line finding code and all identified emission lines are recorded with their line scores (\S\ref{line_scoring}). The line scores are binned in steps of 1.0 and normalized by the number of simulated spectra. This represents the simulated estimate of the probability that an emission line in a given scoring bin is the product of noise. This probability, $P$(Noise) monotonically decreases with increasing line score. Note that it is possible by this mechanism for a scoring bin to have a value of $P$(Noise) greater than 1.0, and that is the case for the lowest scoring bins. For such cases, the probability is cropped to 1.0 and any emission line with a score that fall in those bins is considered to be noise. Higher scoring bins are cropped once the $P$(Noise) falls below $5\times10^{-4}$, with that $P$(Noise) assumed for all emission lines with line scores above that value.

When applied to line detections in real data, any line score below the lowest score for the bin is assumed to be noise and is rejected, and any line detection with a score above the highest score receives the $P$(Noise) of the highest score for the bin. These $P$(Noise) estimates factor in the Solution Scoring (\S\ref{redshift_solution_finder}), described later.

Since the $P$(Noise) is based on the line scoring and on the uncertainties in the HETDEX PSF weighted spectra, any reformulation of the line scoring or any change to the HETDEX pipeline that results in a change in flux uncertainties necessitates a re-computation of this mapping.\\

\subsubsection{Absorption Lines}\label{linefinder_absorption}

As called out by its name, ELiXer is primarily designed to identify and act on emission lines. Continuum bright HETDEX detections ($g$ < 22) are also analyzed with an independent software package (\textit{Diagnose}, \citet{Zeimann_2022}). Nevertheless, ELiXer does currently include a limited use of absorption lines, triggered either explicitly at its invocation or automatically for detections with continuum greater than $2 \times 10^{-17}$ \cgsa. The same untargeted search (\S\ref{blind_search}) used for emission lines is executed for absorption lines, with the exception that the spectrum is first inverted by subtracting all the flux densities from the maximum flux density of the spectrum. This allows the fitter to treat the absorption lines as if they were emission lines, but only for purposes of line identification within the spectrum. The actual fitting (\S\ref{gaussian_fitting}) and initial scoring (\S\ref{line_scoring}) is performed on the original, non-inverted spectrum, with the appropriate sign changes to account for the different direction in the Gaussian model. And like the case for emission lines, the positions of absorption lines with scores above a configurable threshold are also marked in the 1D spectrum.

While there are 26 emission lines checked by ELiXer, only the \caii\ (H\&K) 3968,3934\,\AA\ absorption lines are explicitly fitted and used in spectral redshift identification. Additionally, these two lines are fit simultaneously and must appear together. If they occur at the edge of the spectral range, such that only one line could be found in the spectrum, the fit is not allowed. A simple assertion is made to the pair of lines, requiring them to be of similar flux and FWHM such that the difference in flux and FWHM must be with 50\% of the mean of their mean values. If the assertion fails, the fit is rejected. If the assertion passes, the lines are both accepted and contribute to the solution scoring (\S\ref{redshift_solution_finder}).\\

\subsection{Continuum Estimates} \label{continuum_estimates}

Much of the classification effort rests on an accurate measure of the emission line equivalent width, so a robust estimate of the continuum underlying the emission line is of major importance. There are several, independent and semi-independent estimates of the continuum which contribute to a single combined estimate. 

Since most of the independent estimates arise from photometric imaging, we calibrate our continuum derived classification properties (described later in this section) to the bandpass continuum estimates, all of which assume a flat spectrum over the bandpass with no emission or absorption line masking (see \S\ref{hetdex_spectrum_continuum}, \S\ref{aperture_phot_continuum}, \S\ref{catalog_counterpart_cont}, and \S\ref{combined_cont}). This means we are slightly biased to overestimate the continuum level. This is more pronounced for objects such as AGN with strong, broad emission, but given the objective of accurate classification, this is a non-issue with these objects being a rare subset of HETDEX data and unlikely to be confused with the typical, continuum faint LAE\null. In the general case that ELiXer is designed to address, our objects have faint or undetected continuum and a single, faint emission line so the bandpass overestimate is minimal and serves as an upper limit.

All continuum estimates from broadband photometry assume a flat spectrum point source over the bandpass and convert the magnitude to flux density at the emission line's observed wavelength rather than the filter's effective wavelength as: 

\begin{equation} \label{eq:mag_to_cgs}
    f_{\lambda} = c\ \lambda^{-2} \times (3631 \times 10^{-23}) \times 10^{-0.4 m}
\end{equation}

\noindent where $f_{\lambda}$ is the flux density at the observed wavelength (in ergs~cm$^{-2}$~s$^{-1}$~\AA$^{-1}$), $c$ is the speed of light in vacuum (\AA~s$^{-1}$), $\lambda$ is the fitted, observed wavelength center (\AA), and $m$ is the $g$ or $r$ magnitude. The literal constant is in units of ergs~cm$^{-2}$~s$^{-1}$~Hz$^{-1}$.  \rv{As most of the HETDEX emission line detections have either only $r$ coverage or are undetected in the imaging even when multiple bands are available, a color correction to the photometric continuum estimate is rarely possible. In limited testing where photometric detections are made in both $g$ and $r$ no improvement in the classification performance and no change in the classification rates is found, and so no color correction is included in this version of ELiXer.}\\

\subsubsection{HETDEX Spectrum}\label{hetdex_spectrum_continuum}

The HETDEX spectrum covers the entire $g$ bandpass and therefore can be used to estimate an object's $g$-band magnitude without the use of external data. Sky and background subtraction is very good and the continuum level is consistently measurable $\lesssim 10^{-18}$ \cgsa \citep{Gebhardt+2021}. We use two methods to derive the $g$ magnitude from the HETDEX 1D spectrum.  The first multiplies the HETDEX spectrum through the SDSS $g$ filter's throughput curve using the Python \textit{speclite} package \citep{speclite_pkg}. ELiXer runs 1000 realizations of the HETDEX spectrum, sampling over the flux errors, and assigns the biweight \citep{Beers_1990} of those realizations to define an estimated $g$-magnitude and its 68\% confidence interval. The second method sums the total flux in the HETDEX spectrum, again with propagated errors, and uses the mean flux density and an $f_{\lambda,\mathrm{eff}}$ of 4726\,\AA\ to set a continuum and the $g$-band magnitude.  In both cases, the object is assumed to be a point-source. The combined continuum mean is converted into a $g$ magnitude for ease of use and comparison to other catalog reported magnitudes.

While this estimate is reported as computed, it is used internally with an imposed flux density limit of $5.38\times 10^{-19}$ \cgsa\ ($g = 25$).  When our measured HETDEX continuum flux density is at least $1.2\times$ brighter than the limit, it receives the highest weight ($4\times$ standard) in the combined estimate (\S \ref{combined_cont}), as it is based on the same data that provides the line flux estimate. All other continuum estimates are from other data sources and matched by proximity. As the limit is approached, the weight rapidly drops to the standard vote weight and is considered a non-detection once the limit is reached. 

A second estimate of the continuum is obtained using the $y$ offset from the Gaussian  fit to the emission line (equation \ref{eq:Gaussian}). While this is the estimate nearest the emission line, it can also have a large uncertainty and the simple Gaussian model does not allow for asymmetric line flux or different continuum levels on either side of the line. When this estimate is brighter than the HETDEX limit, it receives a small, empirically set weight of $0.2\times$ the standard vote, otherwise it receives zero weight and is not included in the combined continuum estimate.

A third and final estimate is also recorded, but is not, by default, included in the combined continuum estimate. In this estimate, the continuum is still assumed to be flat in $f_{\nu}$, but all emission and absorption lines identified in the spectrum are masked at $\pm 2\sigma$ from the fitted line centers. The mean of the unmasked fluxes, with standard error propagation, is converted into a flux density and returned as the continuum estimate. With the exception of the continuum bright objects with multiple, broad spectral lines mentioned earlier, this estimate is not significantly different from the \textit{speclite} result and its inclusion in the combined estimate would be both redundant and somewhat inconsistent, given the other photometric estimates. It is, however, used internally in some diagnostic checks.\\

\subsubsection{Aperture Photometry} \label{aperture_phot_continuum}

The $g$ and $r$-bandpass continuum estimates come directly from run-time aperture photometry as described in section \ref{elixer_aperture_phot}. When an \textit{SEP} aperture matches that of the HETDEX detection, its magnitude is used. If no \textit{SEP} aperture is a match, then the smallest, stable ELiXer circular aperture provides the magnitude estimate. In either case, if the computed magnitude is fainter than the imaging limit, that limit is used and the continuum value is flagged as a non-detected upper limit. 

Since the HETDEX emission lines appear in the $g$-band, an optional correction is allowed for translating an $r$-band continuum estimate to $g$-band, however this is not used by default, as an examination of $g$ and $r$ continuum estimates where both are available from the same instrument for the same objects shows no consistent trend. Additionally, \cite{Leung2017} finds no advantage in using $g$ over $r$ and their simulated data actually suggest that LAE/\OII\ segregation is slightly improved with $r$, though this is not confirmed with the observed spectra in this work. 

If the measured aperture magnitude is brighter than the limiting magnitude of the image, it receives a full (1.0) weight in the final, combined estimate. If the measured aperture magnitude is fainter than the limit, it is treated as a non-detection and the limit is used in the combined estimate. When the \textit{limit} is used for the aperture magnitude, the weight in the combined estimate is scaled down linearly from 1.0 to 0.0 as the limit grows brighter from $26_{AB}$ to $24_{AB}$ and a non-detection in that increasingly bright limit provides less and less useful information (noting that the HETDEX spectra has a magnitude limit near $g=25$). The $26_{AB}$ and $24_{AB}$ boundaries selected to roughly cover the the magnitude range of maximal LAE and \OII\ galaxy $g$ magnitude overlap in HETDEX.\\

\subsubsection{Catalog Counterpart} \label{catalog_counterpart_cont}

Lastly, if a catalog counterpart can be matched to the HETDEX detection (\S \ref{catalog_match}), its reported bandpass magnitude (again, only $g$ or $r$) is added to the list of continuum estimates. A minimum 20\% flux uncertainty is assumed, even if no uncertainty is reported by the catalog. All catalog reported values are assumed to be a proper detection and receive a full (1.0) weight. \\

\subsubsection{Combined Continuum}\label{combined_cont}

The combined estimate is produced using the weighted mean of a subset of the individual continuum estimates, described in the immediately previous subsections, with less informative estimates and extreme outliers removed from consideration.

At most, a single upper limit estimate is allowed in the subset and is selected as the deepest (faintest) upper limit. This is typically the limit from the deepest photometric imaging where there is no detection or where the aperture magnitude is fainter than the image's limit. No upper limit is included if there exists a positive aperture detection. If there are three or more continuum estimates in the subset, a fairly aggressive clip is applied, which excludes the most extreme estimate(s) with values greater than 1.5$\times$ the weighted biweight scale \citep{Davis_2021} while retaining a minimum subset size of two. The final combined continuum estimate is then the weighted mean of the surviving continua in the subset:

\begin{equation} 
    \Bar{f_{\lambda}} = \frac {\sum_{i}^{} \left( f_{\lambda_i}\  w_i\  \sigma^{-2}_i \right)}{\sum_{i}^{}w_i\  \sigma^{-2}_i},
\end{equation}

\begin{equation} 
    \Delta\Bar{f_{\lambda}}= \sqrt{\frac {\sum_{i}^{} \left(w_i\  \sigma^2_i \right)}{\sum_{i}^{}w_i}},
\end{equation}

\noindent where $\Bar{f_{\lambda}}$ is the combined ("averaged") continuum estimate, $f_{\lambda_i}$ is an individual continuum estimate, $w_i$ is the associated weight, and $\sigma_i$ is the associated standard deviation. The error, $\Delta \Bar{f_{\lambda}}$, is the square root of the weighed average of the variances.

This defines the distribution over which the continuum is sampled for the P(LAE)/P(OII) classifier in the next subsection.\\

\subsection{Redshift Solutions} \label{redshift_solution_finder}

Distilled to its most basic functions, ELiXer's raison d'\^etre is to assign the correct redshift to every detection as the operative analog to the classification of the target emission line. The core approach to this objective is the testing and ranking (or scoring) of many possible redshift solutions. Clearly the most secure, and consequently the highest scoring, solutions are those with multiple identified spectral lines consistent with known rest-frame features at an assumed redshift. ELiXer's initial set of redshift solutions is generated by iterating over the lines in Table \ref{tab:Table of Lines} and assuming, in turn, that each one represents the target emission line identification (note that the H\&K absorption lines are handled differently per \S \ref{linefinder_absorption}). With each assumed redshift, ELiXer attempts to fit all in the list, and accumulates a total solution score based on the number and quality of the successes (\S \ref{line_scoring}). At this stage, only the relative line positions are considered, with flux ratios, required lines, and other criteria considered in later steps. The more lines that are found, the more robust the solution. Unfortunately, only about 5\% of ELiXer classifications are established with more than one identifiable emission line, so additional methods must be applied to confidently identify the target emission lines and assign the corresponding redshift.\\

\subsubsection{Catalog Redshift Match}\label{ss_catalog_redshift_match}

When ELiXer matches a HETDEX detection to one (or more) catalog objects (\S \ref{catalog_match}) that have associated spectroscopic and/or photometric redshift assignments, that information is evaluated in the context of the emission and absorption lines identified in the HETDEX spectrum. The catalog supplied redshift, with its error, is applied to the target emission line and all other ELiXer identified lines and the resulting rest-frame wavelengths are checked for consistency with those in Table \ref{tab:Table of Lines}. If the catalog redshift results in rest-frame wavelength matches, it boosts any previously assigned ELiXer score (\S \ref{redshift_solution_finder}) for that redshift, with a larger weight given to spec-$z$ (+100 to the redshift solution raw score, \S \ref{redshift_solution_scoring}) than to phot-$z$ (+5 to the redshift solution raw score). If an ELiXer redshift solution for that catalog redshift does not exist, one is created and scored in the same way. Approximately 0.1\% of the HDR3 detections have a catalog matched spec-$z$ counterpart and 1.5\% have a phot-$z$ counterpart.\\

\subsubsection{Large Galaxy Mask} \label{galaxy_mask}

In addition to matching redshift catalogs, ELiXer also compares the sky position and wavelength of each detection against an internal HETDEX catalog of large galaxies. We define this large galaxy catalog by searching the most recent versions of the RC3 \citep[][]{rc3}\footnote{available at: http://haroldcorwin.net/rc3/} and the UGC \citep[][]{ugc}\footnote{https://heasarc.gsfc.nasa.gov/W3Browse/galaxy-catalog/ugc.html} galaxy catalogs for objects larger than 1 arcminute in diameter within our survey area. In total, we find 644 large galaxies in the Spring field, and 447 in the Fall field.  For each system, we adopt the catalog's basic parameters for position, position angle, ellipticity, and D25 semi-major axis (i.e., the size of the galaxy defined by its $B$-band isophote at 25.0 mag~arcsec$^{-2}$). Prior to inclusion in the large galaxy mask, each galaxy is manually inspected to confirm that these values are reasonable.  Where values of these parameters are uncertain, they are corrected to values listed in the NASA/IPAC Extragalactic Database\footnote{http://ned.ipac.caltech.edu} or through visual inspection of the galaxy in SDSS $g$-band images.  Any HETDEX detection falling within $3 \times$ the D25 isophotal radius of a large galaxy is tested against the spectral features expected for the system's redshift.  This matching is performed in exactly the same way as for the catalog matching in the previous section, except that the scoring is scaled inversely by the distance in multiples of D25. The overall area of this large galaxy mask is dominated by a handful of nearby galaxies (NGC 5457 and NGC 4258 in the Spring field, and IC~1613 and NGC~474 in the Fall Field). \\

\subsubsection{Special Handling for [O III]} \label{ss_special_handling_oiii}

The \OIII 5007\,\AA\ line can be problematic \rv{to identify by equivalent width based methods} when other oxygen or Balmer lines are not detected as it can have a large equivalent width and appear similar to \lya. Low-$z$ compact star forming galaxies, planetary nebulae (PNe), extragalactic \HII regions, and the outer star forming regions of resolved galaxies could sometimes have detectable \OIII\ 5007\,\AA, but with \OIII 4959\,\AA, \OII 3727\,\AA, and \Hb\ that do not reach the threshold for \rv{a standard HETDEX} detection.  Such objects could be classified as \lya\ by the base algorithms. To protect against such misclassifications, additional tests are needed.

For observed emission lines redward of 5007\,\AA, but without any other nominally detected emission feature, a lower threshold for emission line detection is allowed at the expected positions of \OIII 4959\AA, \OII 3727\AA, and \Hb. If one or more of those lines are detected at this reduced stringency, a redshift solution is created with a score of at least the minimum acceptable threshold, and a flag is set for followup manual inspection.

If one or more of the above lines are found and there is no identified imaging counterpart, a flag is also set to indicate that this could be a planetary nebula, either in the Galaxy or in intergalactic space. Given the HETDEX lines of sight are out of the plane of the Galaxy, the likelihood of encountering Galactic planetary nebulae is reduced but is certainly not zero and several known Galactic planetaries are located in the HETDEX footprint.  Given their physical proximity, most of these objects will have sizes of several arc-minutes, and we test for this by looking for large spatial clusterings of emission at 5007\,\AA\null.  When found, these regions are masked from use in HETDEX cosmology.  A potentially more pernicious issue is planetary nebulae in the halos of nearby galaxies and intergalactic PNe within galaxies groups and clusters. These could be misinterpreted as background LAEs, though this risk is ameliorated via the check against the large galaxy mask (\S \ref{galaxy_mask}) and neighbor clustering (\S \ref{clustering}). Conversely, this comes at a (small) cost of the loss of some background LAEs with observed \lya\ redshifted to match the \oiii\ 5007 \AA\ line of on-sky adjacent foreground galaxies. 

\rv{We note that \OIII 5007\,\AA\ makes up only 1\% of the SzAS detections and none are misidentified by ELiXer.}\\

\subsubsection{Object Classifications Labels} \label{classification_labels}

Based on combinations of spectral features (with examples given later in this subsection), some HETDEX detections are assigned classification labels.  These labels indicate only that a detection is consistent with the class of object indicated by the label within the parameters defined for that class.  Classifications are not mutually exclusive and are applied simply if the corresponding conditions are met. If none of the specific classification conditions are met, then no extra classification label is applied to the detection. The classification is not Boolean, but is scored, with the strength of the classification based on the number and quality of the conditions that are met. A negative classification can also be made if the failure to meet conditions is sufficiently extreme such that a classification is excluded (i.e., if the detection's properties are grossly inconsistent with the given classification).

Strongly consistent object classifications can be used to increase the score of the corresponding redshift solution, while strongly inconsistent classifications decrease the score of the corresponding solution. In this way, the object classification $can$ modify the P(Ly$\alpha$) result (\S \ref{p_lya}) by altering the score of a multi-line solution available to the P(Ly$\alpha$) routines. However, the conditions are relatively strict and the overall impact of labeling is small, with only $\sim$4\% of detections actually meet the conditions to receive an object classification label.

Additionally, a few generic labels are applied for ELiXer detections that are associated with unique object in a photometric catalog (\S \ref{sec:catalogs}).  These labels are only provided as suggestions and do not impact the scoring of the multi-line solutions. 

The ELiXer assigned labels are:
\begin{itemize}
    \item \textit{AGN} ("agn") The "agn" label is set if a HETDEX spectrum contains (possibly broadened) emission lines consistent with those seen in AGN\null. These reference emission lines are: 
    \ion{O}{6} (1035\,\AA), 
    Ly$\alpha$ (1216\,\angstrom), 
    \ion{N}{5} (1241\,\AA), 
    \ion{Si}{2} (1260\,\AA), 
    \ion{Si}{4} (1400\,\AA), 
    \ion{C}{4} (1549\,\AA), 
    \ion{He}{2} (1640\,\AA), 
    \ion{C}{3}] (1909\,\AA), 
    \ion{C}{2} (2326\,\AA), 
    \ion{Mg}{2} (2799\,\AA), and
    [\ion{O}{2}] (3727\,\AA)\null. For some pairs of lines, bounds on relative line fluxes must be met and certain lines must be present to support the identification of other lines. For example, if a line assumed to be \ion{C}{4} is observed at 5000\,\AA, then a line for Ly$\alpha$ must also be found at 3295\,\AA\ and it should be at least as strong and have a similar FWHM as \ion{C}{4}\null.  If no line is observed at 3295\,\AA\ or if the feature is much weaker than the assumed  \ion{C}{4} line, then the identification is inconsistent with that of an AGN and the \ion{C}{4} solution receives a reduced score.   
    
    \item \textit{Low-$z$ Galaxy} ("lzg")  The "lzg" logic is largely the same as the "agn" but with a different set of reference lines:
    [\ion{O}{2}] (3727\,\AA), 
    H${\eta}$ (3835\,\AA), 
    H${\zeta}$ (3889\,\AA), 
    H${\epsilon}$/ion{C}{2} (3970\,\AA), 
    H${\delta}$ (4101\,\AA), 
    H${\gamma}$ (4340\,\AA), 
    H${\beta}$ (4861\,\AA), 
    [\ion{O}{3}] (4959\,\AA), and \ion{O}{3}] (5007\,\AA)\null. As with AGN, some bounds on line strengths must be met.  For example, if a line assumed to be [\ion{O}{3}] 5007 \AA\ is observed at 5300\,\AA, then another line at 5249\,\AA\ must be observed at one-third the strength.   Similarly, for HETDEX detections with strong continuum, if an absorption line is assumed to be calcium H at 3968\,\AA, calcium K at 3934\,\AA\ must also be present with at a similar equivalent width.  If these criteria are satisfied, then the detection will be labeled "lzg".  Moreover, an additional label of "o32" will be assigned to objects with an [\ion{O}{3}] 5007 \AA\ to \ion{O}{2} 3727 \AA\ flux ratio  greater than 5:1.

    \item \textit{Meteor} ("meteor")  \label{meteor_classification}
    
With any wide-field, long-term survey,  meteor intrusions on the extra-galactic observations are inevitable, and if not identified, they can be a significant nuisance source of emission (and sometimes of continuum) detections. A combination of methods are used to identify meteors in the detection catalog \citep[][]{Cooper_2022}.

Since ELiXer processes only single detections in isolation, its meteor identification methodology focuses on the transient nature of the phenomenon and their fairly distinctive emission line signatures. To identify a meteor emission, we divide a spectrum into 9 non-overlapping, non-contiguous regions by wavelength (in \AA) where meteor emission lines are common: [3570,3590], [3715,3745], [3824,3844], [3852,3864], [3926,3942], [3960,3976], [4210,4250], [4400,4450], and [5160,5220]. For the visually confirmed meteors in HETDEX, these regions often include bright features from Mg (3832, 3838, 5172, and 5183\,\AA) as well as typically fainter emission from Al, Ca, and Fe. Spectra that contain multiple emission lines that are within these ranges and are detected in only one of the three dithered exposures used for an observation are labeled as meteors.

\item \textit{White Dwarf} ("wd")  \label{wd_classification} The white dwarf label logic is very basic and simply looks for the Hydrogen series absorption lines for DA and DAB types, the Helium series for DB types, and Carbon and Oxygen for DQ types. Additionally, to be classified as a white dwarf, the spectrum must have a blue spectral slope. Since the shape and width of the absorption features are not taken into account, nor are the presence of other features (such as pronounced H and K (\caii) lines), it is possible to mislabel a main sequence star, particularly an A-type, as a white dwarf.  However, given the high Galactic latitude of the HETDEX survey, we do not expect the set of HETDEX detections to contain many early-type stars.

    \item \textit{Catalog Labels} ("gal", "star", "agn") These are recorded as suggestions when matched to an external photometric catalog, but they do not influence any of the ELiXer logic. For example, an "agn" label from a photometric catalog matched to a HETDEX detection is considered separately from the ELiXer "agn" label logic described above and will appear in the classification labels even if the ELiXer spectral features analysis does not result in an "agn" label.\\
\end{itemize}

\subsubsection{Redshift Solution Scoring} \label{redshift_solution_scoring}

Each redshift solution receives three scores, a raw score, a (normalized) fractional score, and a scale score, so that the solutions can be rank ordered and assessed in terms of their viability. The raw score is the unweighted sum of the individual line scores (\S \ref{line_scoring}) of the spectral lines included in the solution, excluding the anchor line (which is common to all solutions), and including a multiplier based on the number of identified spectral lines and any multipliers from classification labels (\S \ref{classification_labels}), where they are strongly consistent or inconsistent. It is defined as:

\begin{equation} \label{eq:ml_raw_score}
    rs = \left( \sum_{i}^{n} ls_{i} \right) \times \textbf{min}\left(1,\frac{1}{2} \left(n^2 - n\right)\right) \times b,
\end{equation}

\noindent where $rs$ is the solution raw score, $ls$ is a line score of an included spectral line, $n$ is the total number of spectral lines included in the solution not counting the anchor line, and $b$ is any multiplier from the object classification label logic (typically 0.25 to 2.0).

The raw score is normalized to produce the fractional score by dividing it by the sum of the raw scores of all redshift solutions.

Lastly, a scale score is produced from the weighted sum of the probability that the solution is comprised of noise, the raw score,  and the fractional score as:

\begin{equation}  \label{eq:ml_scale_score}
\begin{aligned}\quad
ss = & \left( 1- \prod_{i}^{n} P(\mathrm{noise})_i \right) \times w_{\mathrm{noise}} \\
 + ~ & \textbf{min}\left(1.0, rs / F \right) \times w_{\mathrm{raw}} \\
 + ~ & fs \times w_{\mathrm{frac}},
\end{aligned}
\end{equation}

\noindent where $ss$ is the scale score, $P(\mathrm{noise})_{i}$ is the probability that the included spectral line is noise (\S \ref{spectra_simulation}), $w_{\mathrm{noise}}$ is the weight for this first term (by default, 0.40), $rs$ is the raw score from Eqn (\ref{eq:ml_raw_score}), $F$ is the configured raw score scale factor (by default, 50.0), $w_{\mathrm{raw}}$ is the weight for this second term (by default, 0.50), $fs$ is the fractional score, and $w_{\mathrm{frac}}$ is the weight of this third term (by default, 0.10).\\

\subsection{P(LAE)/P(OII)} \label{plae_poii_sec}

P(LAE)/P(OII) (sometimes as PLAE/POII in other documentation) represents the ratio of the relative probability that given a set of measured characteristics, an emission line is \lya\ (representing an LAE) rather than \oii. These probabilities are based on the number of galaxies expected at the volume sampled by the redshift slices assuming the emission line is either \lya\ or \oii\ given the measured line flux and equivalent width. The expected number of galaxies derives from the equivalent width distributions of \lya\ and \oii\ conditioned on the luminosity functions found in \cite{Gronwall_2014} and \cite{Ciardullo2013a} respectively, interpolated or extrapolated as needed (see also \citet[][Figure 2]{Leung2017}).

This is an improvement on the commonly used 20\,\AA\ equivalent width cut \citep[][]{Gronwall2007a,Adams2011a} and is based largely on the analysis of \cite{Leung2017}, and using the specific translation and implementation described in \citep[][primarily in Section 2]{Farrow+2021}. ELiXer slightly updates \cite{Farrow+2021} by (1) using multiple independent or semi-independent estimates of the continuum (\S \ref{continuum_estimates}), (2) combining those estimates into a single, best-fit continuum value, and (3) sampling over the uncertainties in the measured line flux and continuum estimates to generate a (68\%) confidence interval around each P(LAE)/P(OII) measurement. Partly for convenience and partly as a representation of the practical limits of this method, the ratio is cropped to values between $0.001 \leq \mathrm{P(LAE)/P(OII)} \leq 1000$.

The interpretation of the P(LAE)/P(OII) value is not quite straightforward. While LAE evolution between $ 2 < z < 4$ appears somewhat muted \citep[][]{Blanc_2011,Santos_2021}, there is more redshift evolution of the \oii\ systems \citep[][]{Gallego_2002,Comparat_2016,Saito_2020,Park_2015,Gao_2021} for $z<0.5$.  This evolution may be underrepresented in the base P(LAE)/P(OII) code and lead to a deviation from the expectation that a ratio near 1 should be interpreted as the likelihood of the emission line being \lya\ or \oii\ is approximately equal.  
Building on the suggestion in \cite{Leung2017} of using different thresholds for the P(LAE)/P(OII) ratio at different observed wavelengths, ELiXer adopts an empirical threshold relation (\S \ref{plae_poii_vote}).

The overall combined P(LAE)/P(OII) value and its confidence interval factor significantly in the final automated classification of the emission line. It can frequently be the most influential (and sometimes the only) metric that is used in that classification (\S \ref{p_lya}).\\

\subsection{P(LyA)} \label{p_lya}

Using some of the features/measurements already described, along with a set of additional features described below, ELiXer synthesizes an aggregate confidence in the classification of the anchor emission line as \lya\ or \notlya. For familiarity, this is couched in terms of a probability, labeled as P(\lya) with values between 0 (definitely not \lya) and 1 (definitely \lya), but is not a true probability in the formal sense. P(\lya) is the result of a weighted voting system where each of the features described in this section provides a vote (typically 0 or 1, but can be in between) and that vote is given a weight based on the robustness or confidence of the measurement. With specifically noted exceptions, features that do not produce a clear preference are given zero or very little weight. The final P(\lya) value is then simply the sum over all votes multiplied by their respective weights:
\begin{equation} \label{eq:weighted_voting}
    P(\mathrm{Ly}\alpha) = \frac {\displaystyle\sum_{i}^{} \left( \mathrm{vote}_i \times \mathrm{weight}_i \right)}{\displaystyle\sum_{i}^{}\mathrm{weight}_i},
\end{equation}
Note that the sum of the weights alone is not normalized and can exceed 1. In the relatively rare cases where the sum of all weights is less than 1, a special "uncertainty" vote is added with a value of 0.5 and a weight equal to $1-\sum{\mathrm{weights}}$, so that the weights do sum to 1. This helps capture the uncertainty in the classification and prevents one or two votes with very low weights from being dominant.

The selection of voting criteria and the weights applied to the votes is the result of empirical analysis and trial-and-error testing and is discussed in Section \ref{testing}. \rv{This is a little bit of the Central Limit Theorem and the Wisdom of the Crowd, even though the votes are not entirely independent as several incorporate similar elements and some are designed to handle edge cases not well covered by the others. No single vote is universally dominant, though each can be decisive under the right circumstances, such as the high weight of \S \ref{ss_ml_z_solution_votes} when multiple emission lines are present or even a low weight vote from \S \ref{simplified_ew_vote} for some moderate equivalent widths when the rest of the vote tally is near 0.5.}

As a word on the notation in this section; often \oii\ is used in place of "\notlya" as \oii\ is the most common contaminant. Votes "for \oii" are really votes for "\notlya". Further, the figures in this subsection all show only those assessment sample detection emission lines that are \lya\ or \oii, so \oii\ is equivalent to "\notlya".\\

\subsubsection{Object Size Vote} \label{object_size_vote}
In cases where a counterpart is identified and resolved in the $g$- or $r$-band imaging, the angular and physical extent of the counterpart contributes a vote. For this purpose, an object is considered resolved if the angular major diameter is greater than 1.1$\times$ the seeing FWHM\null. This includes artificially enlarged footprints in the imaging due to the "blooming" of bright sources that have saturated the detector. The proper physical diameter is computed assuming the redshift of \oii, as larger objects tend to be more evolved and at lower redshift. The emission line FWHM is used to help break the size degeneracy between larger, lower-$z$ objects and saturated, higher-redshift sources, via the assumption that the latter are AGN with a large emission line FWHM.

\begin{figure}[ht]   
    \centering
    \advance\leftskip-0.75cm
    \includegraphics[width=0.5\textwidth]{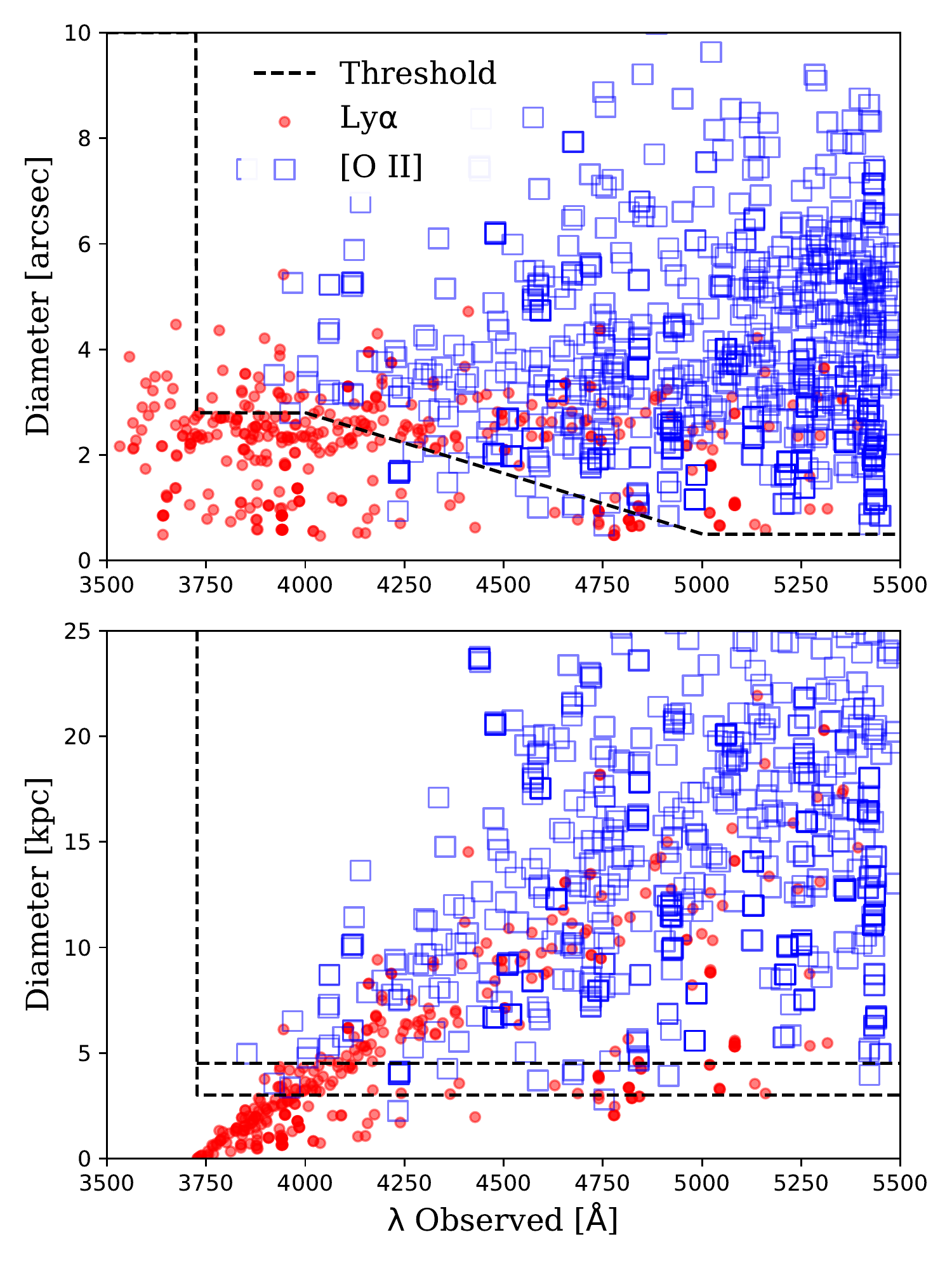}
    \caption{The separation of \lya\ from \oii\ in the assessment sample (SzAS, \S \ref{testing}) based on the angular (upper panel) and physical (lower panel) diameters. Errors are $\sim 0\farcs2$. The dashed line corresponds to the thresholds defined in Table\ref{tab:angular_size_votes}. There are no points blue-ward of 3727\,\AA\ in the lower figure since the physical diameter is computed based on the assumption that the emission line is \oii. The lower panel is cropped to a maximum of 25 kpc for readability and shows two horizontal thresholds at 3.0 and 4.5 kpc, corresponding to the first two conditions in Table \ref{tab:angular_size_votes}. 
    This generates a vote for 70\% of the SzAS with a 4\% contamination of \lya\ in those votes.
    }
    \label{fig:angular_diam_vs_wave}
\end{figure}

The parameter thresholds are set from a manual partitioning of classifications in scatter plots of angular and physical diameter versus the observed wavelength of the anchor emission line, as shown in Figure \ref{fig:angular_diam_vs_wave}. The conditions and their associated votes are summarized in Table \ref{tab:angular_size_votes}. \rv{The specific limiting values of the FWHM help distinguish possible AGN with a broadened emission line, from lower redshift galaxies. It is reasonable for an AGN to receive a vote for \lya, but an angularly large object with a more narrow emission line is more likely an \oii\  emitter. The gap between the conditions avoids a vote where it is unclear.} The angular diameter threshold (in arcseconds), $\theta_{\lambda}$, is a piece-wise linear function:

\begin{equation} \label{eq:size_threshold_vote}
    \theta_{\lambda} = \left\{
                        \begin{aligned}\quad
                      & 2.8, & 3727 \angstrom < \lambda \leq 4000\angstrom  \\
                      & -0.0018\lambda + 10.0, & 4000\angstrom < \lambda \leq 5000\angstrom \\
                      & 1.0, & \lambda > 5000\angstrom  
                      \end{aligned}
                      \right.
\end{equation}

{\renewcommand{\arraystretch}{1.2}}
\begin{deluxetable}{ c |c  |c } [ht]
\tablecaption{Angular and Physical Diameter Votes\label{tab:angular_size_votes}}
\tablewidth{0pt}
\tablehead{
\colhead{Condition}  & \colhead{Vote} & \colhead{Weight}
}
\startdata
$d_p$ $<$ 3.0 kpc or $\theta$ $<$ $\theta_{\lambda}$  & 1.0 & 0.25 \\
$d_p$ $<$ 4.5 kpc  & 1.0 & 0.10 \\
$\theta$ $<$ 2\farcs5 and FWHM $>$ 1000 \kms & 1.0 & 0.25 \\
$\theta$ $>$ 2\farcs5 and FWHM \rv{$<$} 800\kms & 0.0 & 0.25 \\
else no vote & NA & 0.00
\enddata
\tablecomments{
Summary of angular and physical size votes. The conditions are ordered such that the logical evaluation results in at most one unique vote. If no conditions are met, there is no vote.\\
$d_p$ is the proper diameter in kpc.\\
$\theta$ is the angular diameter in arcsec.\\
$\theta_{\lambda}$ is the minimum expected angular size for an \oii\ galaxy for the observed anchor emission line wavelength.\\
$FWHM$ refers to the emission line.
}
\end{deluxetable}

The object size criteria results in a cast vote for 70\% of the SzAS (down-selected to only contain \lya\ and \oii), where the separation of \oii\ from \lya\ is effective, with a \lya\ contamination rate of 4\% in those votes. \\

\subsubsection{Multi-line Redshift Solutions Votes} \label{ss_ml_z_solution_votes}

This criterion can generate multiple votes, one for each potential redshift solution (\S \ref{redshift_solution_finder}) based on the positions and fluxes of the fitted spectral lines.\rv{There must be two or more found spectral lines, with the scores based largely on the number of lines and their strengths (see \S \ref{redshift_solution_scoring}). However, as is shown later in this subsection, solutions incorporating three or more lines receive an increased voting weight.} At most, there will be a single \lya (1.0) vote if there is a solution that supports the classification of the anchor line as \lya. All other redshift solutions necessarily require the anchor line to be something other than \lya, and therefore cast a not-\lya (0.0) vote. The weight each vote receives depends on the scaled solution score assigned multiplied through a sigmoid:
\begin{equation} \label{multiline_weight_sigmoid}
    w_0 = ss / (1 + \exp(0.75m - rs)) 
\end{equation}
where $w_0$ is the initial voting weight, $ss$ is the redshift solution scale score (Eqn \ref{eq:ml_scale_score}), $m$ is the minimum acceptable score (25, by default), and $rs$ is the redshift solution raw score (Eqn \ref{eq:ml_raw_score}).

An additional multiplier is applied for exceptionally strong redshift solutions with 3 or more contributing spectral lines:
\begin{equation}
    w = w_0 \times \textbf{min} (\rv{rs} / m, 10)
\end{equation}
where $w$ is the modified voting weight, $w_0$ is the original weight (Eqn \ref{multiline_weight_sigmoid}), \rv{$rs$} is the raw solution score, and $m$ is the minimum acceptable score. This multiplier is always greater than 1 since, by definition, a qualifying redshift solution must have a raw solution score greater than minimum acceptable value. The maximum value of $w$ is limited to 10$\times$ the original number, but that allows this vote to dominate with a high confidence redshift solution comprised of multiple, strong spectral lines.

\rv{This criteria does not often trigger a vote, casting one for only 7\% of the SzAS, down-selected to only contain \lya\ and \oii, and 12\% for the entire SzAS, but has no contamination of \lya\ for those votes.} Due to the bright skew in SzAS (see \S \ref{testing} and \S \ref{bias_corrections}), this voting rate is exaggerated and is only cast for 2\% of the $g > $22 detections in HETDEX.\\

\subsubsection{P(LAE)/P(OII) Vote} \label{plae_poii_vote}

As most HETDEX detections are faint, single emission lines, the above criteria rarely produce strong redshift solutions, and the P(LAE)/P(OII) computation is often the most significant vote. The value (0 or 1) of the vote depends on which side of a wavelength dependent midpoint the P(LAE)/P(OII) ratio falls, and the weight of the vote increases with the distance of the ratio from that midpoint. The midpoint value, which separates the \oii\ (0) and \lya (1) vote, is a modification of the binary condition suggested in \cite{Leung2017},

\begin{equation}
    \mu = \left\{
            \begin{aligned}\quad
                      & 1.38, & \lambda \leq 4255\angstrom  \\
                      & 10.3, & \lambda > 4255\angstrom  
            \end{aligned}
        \right.
\end{equation}

and is defined as

\begin{equation} \label{eqn:elixer_plaepoii_thresh}
    \mu = \left\{
            \begin{aligned}\quad
                      & 1.0, & \lambda \leq 4000\angstrom  \\
                      & 0.018\lambda - 71, & 4000\angstrom < \lambda \leq 4500\angstrom \\
                      & 10.0, & \lambda > 4500\angstrom  
            \end{aligned}
        \right.
\end{equation}

\noindent where $\mu$ is the midpoint or vote threshold and $\lambda$ is the wavelength of the anchor emission line. Ratios nearer the midpoint suggest an increasingly equal likelihood that the source emission line is \oii\ or \lya and, as such, add little evidence for a classification. This is reflected in a low voting weight ($w$) built from a Gaussian,

\begin{equation}
    w = 1 - \exp\left( -\left(\frac{P - \mu}{\sqrt{2}\ \sigma}\right)^2\right) \times (1 - i),
\end{equation}

where

\begin{equation}
    P = \left\{
            \begin{aligned}\quad
                      & \mathrm{P(LAE)/P(OII)}, &  \mathrm{for\ } \mathrm{P(LAE)/P(OII)} \geq 1 \\
                      & \mathrm{P(OII)/P(LAE)}, &\mathrm{for\ } \mathrm{P(LAE)/P(OII)} < 1  
            \end{aligned}
        \right. 
\end{equation}

\noindent and $\mu$ is the midpoint and $\sigma$ is the usual Gaussian width (here set to 5.0, which is tuned by hand to give balanced voting weights).  The parameter $i$ is an ersatz standard deviation from the scaled 68\% confidence interval around the P(LAE)/P(OII) (\S \ref{plae_poii_sec}) and is defined as:

\begin{equation} \label{eqn:elixer_plaepoii_ersatz_sd}
i = \frac{1}{2} \times \left(\frac{U}{U+1} - \frac{L}{L+1}\right) 
\end{equation}

\noindent where $U$ is the upper bound of the confidence interval and $L$ is the lower bound. As the P(LAE)/P(OII) ratio moves farther from the midpoint in either direction, the weight of the vote increases and rapidly asymptotes to 1.

Alone, the P(LAE)/P(OII) vote is effective, with a 4\% \lya\ contamination rate (by \oii) in the SzAS (down-selected to only contain \lya\ and \oii), voting 90\% of the time. As with the other equivalent width based votes, though, it struggles to identify \lya\ emission lines when originating from non-LAE (i.e. low-EW \lya\ emitting galaxies) (see also \S \ref{missing_agn_lbg}). \rv{As the P(LAE)/P(OII) computation includes the volumes sampled by the two assumed redshifts, it can become a less effective discriminator as the observed wavelengths approach the rest wavelength of \oii\ and that volume shrinks (\S \ref{plae_poii_sec} and \citet{Leung2017,Farrow+2021}). The other votes, including two more based partly on the emission line equivalent width, \S \ref{simplified_ew_vote} in particular, help compensate.}
\\

\begin{figure}[ht]   
    \centering
    \advance\leftskip-0.75cm
    \includegraphics[width=0.5\textwidth]{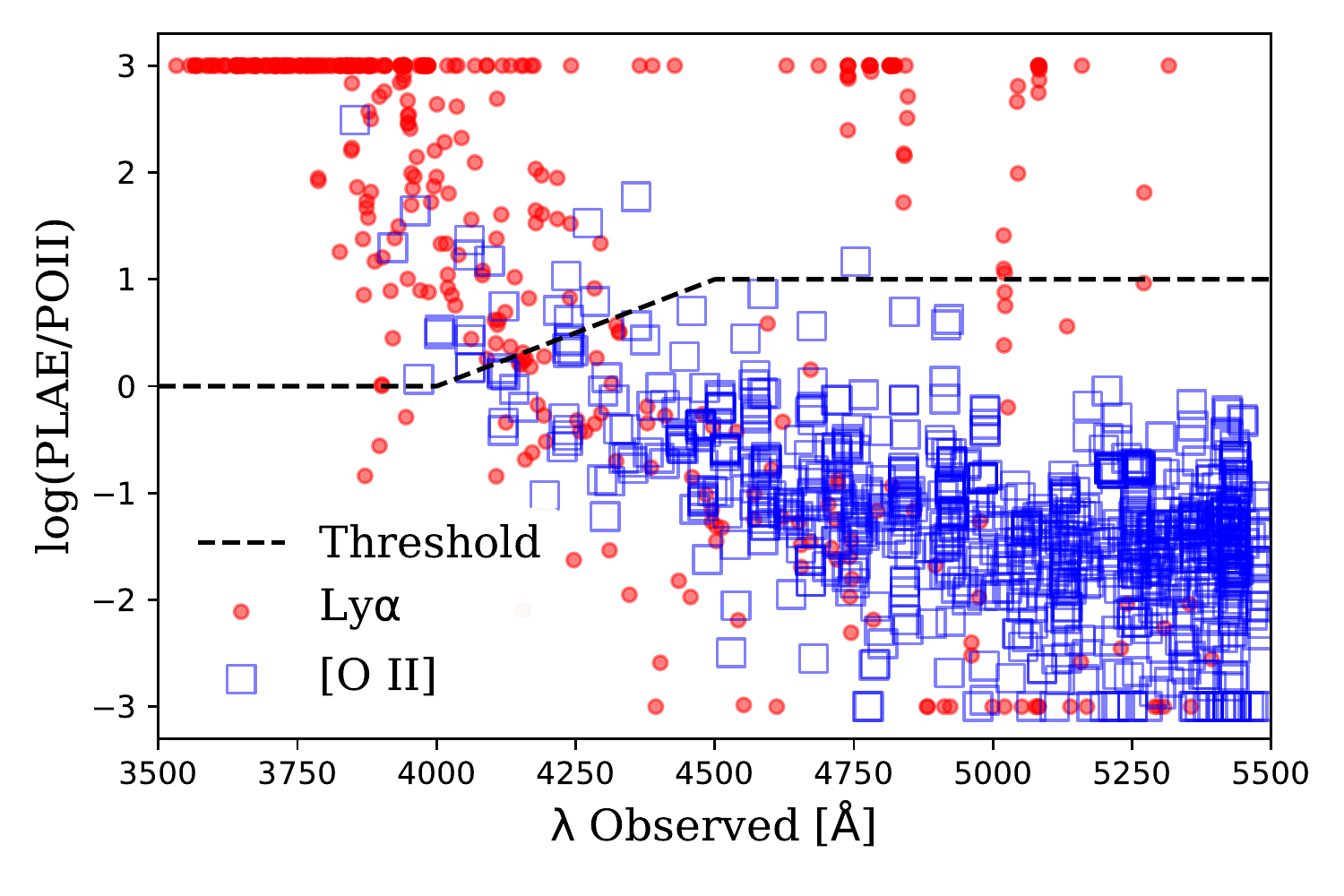}
    \caption{P(LAE)/P(OII) distribution (clipped to $10^{\pm3}$) in the assessment sample (SzAS, \S \ref{testing}) shown without the 68\% confidence intervals (\S \ref{plae_poii_vote}). The dashed line is the midpoint of the segregation threshold (Eqns \ref{eqn:elixer_plaepoii_thresh} -  \ref{eqn:elixer_plaepoii_ersatz_sd}) with points above the line receiving a vote for \lya\ and those below for \notlya\ with weights based on the distance from the threshold. This vote has a 4\% contamination rate of \lya\ by \oii\ in the SzAS.
     }
    \label{fig:plae_ploii_vote}
\end{figure}

\subsubsection{Line FWHM Vote} \label{line_fwhm_vote}
\begin{figure}[ht]   
    \centering
    \advance\leftskip-0.75cm
    \includegraphics[width=0.5\textwidth]{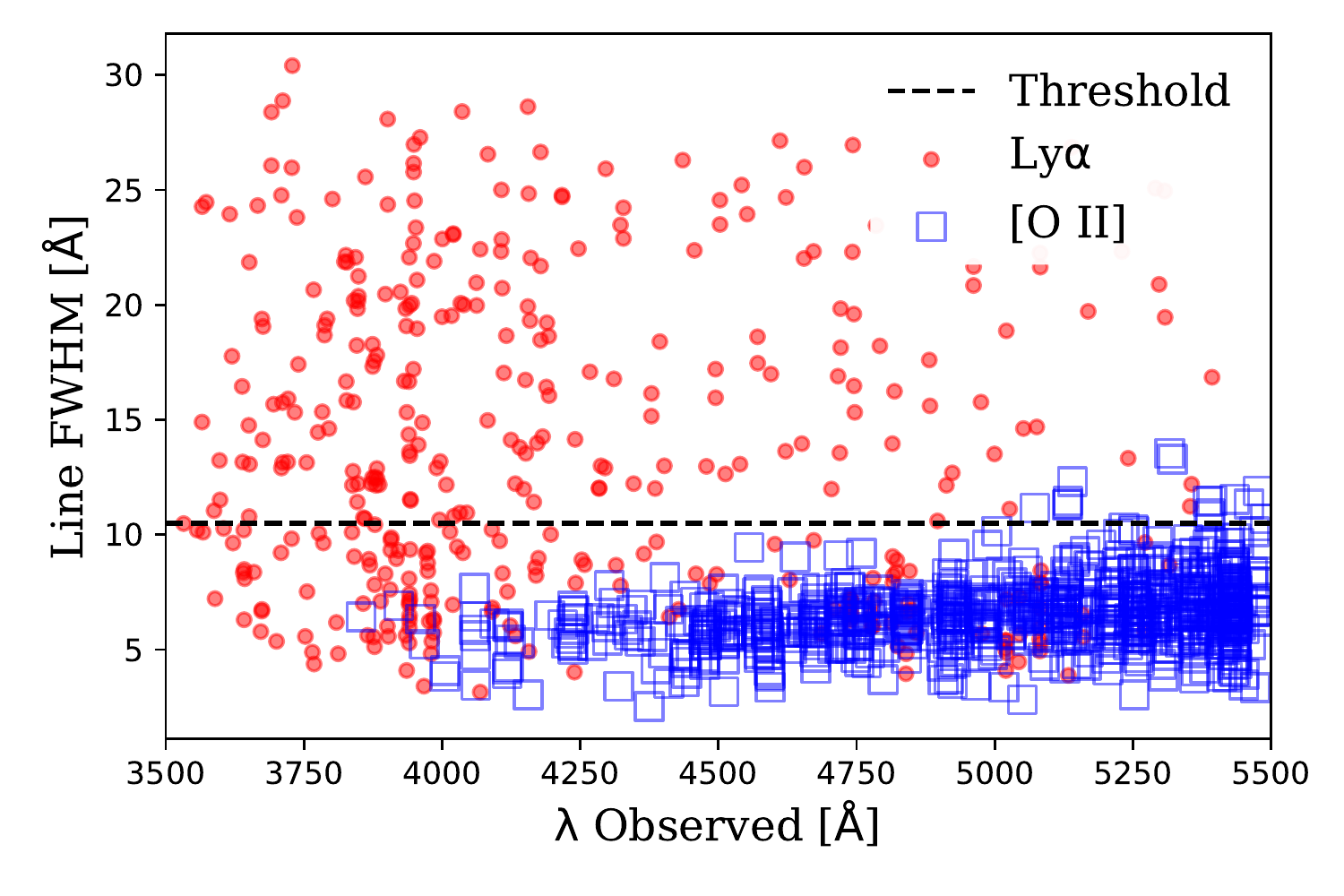}
    \caption{\lya\ and \oii\ separation in the assessment sample (SzAS, \S \ref{testing}) based on the emission line FWHM\null.  The data points are shown without their uncertainties ($\sim$ 14\% ). The horizontal dashed line represent the minimum threshold to receive a vote for \lya\ as described by Eqn \ref{eq:fwhm_vote_threshold}. }
    \label{fig:line_fwhm_vote}
\end{figure}

This is logically one of the simplest votes. 
If the emission line FWHM is larger than 10.5\,\angstrom, as seen in Figure \ref{fig:line_fwhm_vote}, the line receives a \lya vote (1) with a weight as high as 1.0 using

\begin{equation} \label{eq:fwhm_vote_threshold}
 w = \textbf{min} (\mathrm{FWHM} / 10.5 - 1.0, 1.0),
\end{equation}

\noindent where $w$ is the assigned weight of the line and FWHM is line's  fitted full-width at half-maximum. As the contamination rate decreases with larger FWHM thresholds, the voting weight increases. If the lower uncertainty bound of the fitted FWHM, here defined as the fitted FWHM minus the uncertainty derived from standard error propagation, exceeds a configurable minimum (15.3 \angstrom\ by default), the vote weight is set to the 1.0 maximum value, as \oii\ emission lines are rarely that broad. Also, as a consequence of the increasing FWHM threshold, these higher weighted votes tend to favor AGN selection and thereby helps reduce the confusion caused by lower AGN emission line equivalent widths. \rv{In short, it helps improve the recovery of \lya\ (and decrease the misclassification as \oii) from AGN that can fail the other voting criteria based on equivalent width (\S \ref{plae_poii_vote}, \S \ref{simplified_ew_vote}), bandpass magnitude (\S \ref {mag_ew_vote}), and angular size (\S \ref{object_size_vote}).}

This criteria casts a vote for 23\% of the down-sampled SzAS (containing only \lya\ and \oii) with a total \lya\ contamination of 11\%. This drops to 3\% when considering votes with weights above 0.3 (received by 18\% of the down-selected SzAS) and is contamination free for votes with weights above 0.7 (received by 12\% of the down-selected SzAS).

We note that while this particular vote is a good discriminator against \oii, it can confuse \lya with other broad AGN lines, such as \ciii\ or \civ. We largely address this issue using multi-line redshift solutions (\S \ref{redshift_solution_finder}) and clustering (\S \ref{clustering}).\\

\subsubsection{Simplified Equivalent Width Vote} \label{simplified_ew_vote}

This vote is somewhat redundant with the full P(LAE)/P(OII) vote (\S \ref{plae_poii_vote}), but does not consider the redshift based population distributions or observed wavelength variations. It slightly moderates the P(LAE)/P(OII) vote and can help push away from an uncertain classification where the P(LAE)/P(OII) vote has a low weight. It can also push toward an uncertain classification if the P(LAE)/P(OII) vote and this vote have similar weights, but different votes, allowing other voting criteria to have more influence. These two votes agree 95\% of the time and this simplified equivalent width vote is only important in these boundary cases.

This simplified vote uses EW$_{Ly\alpha}$, which is defined by the Gaussian fitted line flux (\S \ref{gaussian_fitting}) and the combined continuum estimate (\S \ref{continuum_estimates}). For EW$_{Ly\alpha}$ much greater or much less than 20\,\angstrom, this reinforces the P(LAE)/P(OII) vote and helps nudge the solution away from the P(LAE)/P(OII) midpoint. If the EW$_{Ly\alpha}$ is greater than 30\,\angstrom, then the vote is for \lya (1); if the  EW$_{Ly\alpha}$ is less than 20\,\angstrom, the vote is for \oii\ (0). All other EW$_{Ly\alpha}$ values do not generate a vote.

The assigned voting weights are based on the EW$_{Ly\alpha}$ lower (EW$^{-}_{Ly\alpha}$) and upper (EW$^{+}_{Ly\alpha}$) bounds and increase with conditions where the contamination is reduced. The maximum weight is limited to 0.5 so that the P(LAE)/P(OII) vote is dominant when both votes approach their maximum weights. 

In the pro-\lya case, the weight is either 0 or between 0.1 and 0.5 as:
\begin{equation}
    w = \left\{
            \begin{aligned}\quad
                      & 0.0, & r^- \leq 0.0  \\
                      & 0.1, & 0.0 < r^- < 1.0 \\
                      & \textbf{max}(0.1,\textbf{min}(0.5,r^--1.0)), & r^- \geq 1.0  
            \end{aligned}
        \right.
\end{equation}
where $w$ is the assigned weight and $r^-$ = $\frac{1}{25} \times EW^{-}_{Ly\alpha}$.\\

In the pro-\oii\ case, the weight is between 0.1 and 0.5 as:
\begin{equation}
    w = \left\{
            \begin{aligned}\quad
                      & 0.1, &   r^+ < 1.0 \\
                      & \textbf{min}(0.5,\textbf{max}(0.1,f)) , & r^+ \geq 1.0  
            \end{aligned}
        \right.
\end{equation}
\begin{equation}
f = -0.04\times EW^{+}_{Ly\alpha} + 0.9
\end{equation}
where $w$ is the assigned weight and $r^+$ = 20 / $EW^{+}_{Ly\alpha}$.\\

\begin{figure}[ht]   
    \centering
    \advance\leftskip-0.75cm
    \includegraphics[width=0.5\textwidth]{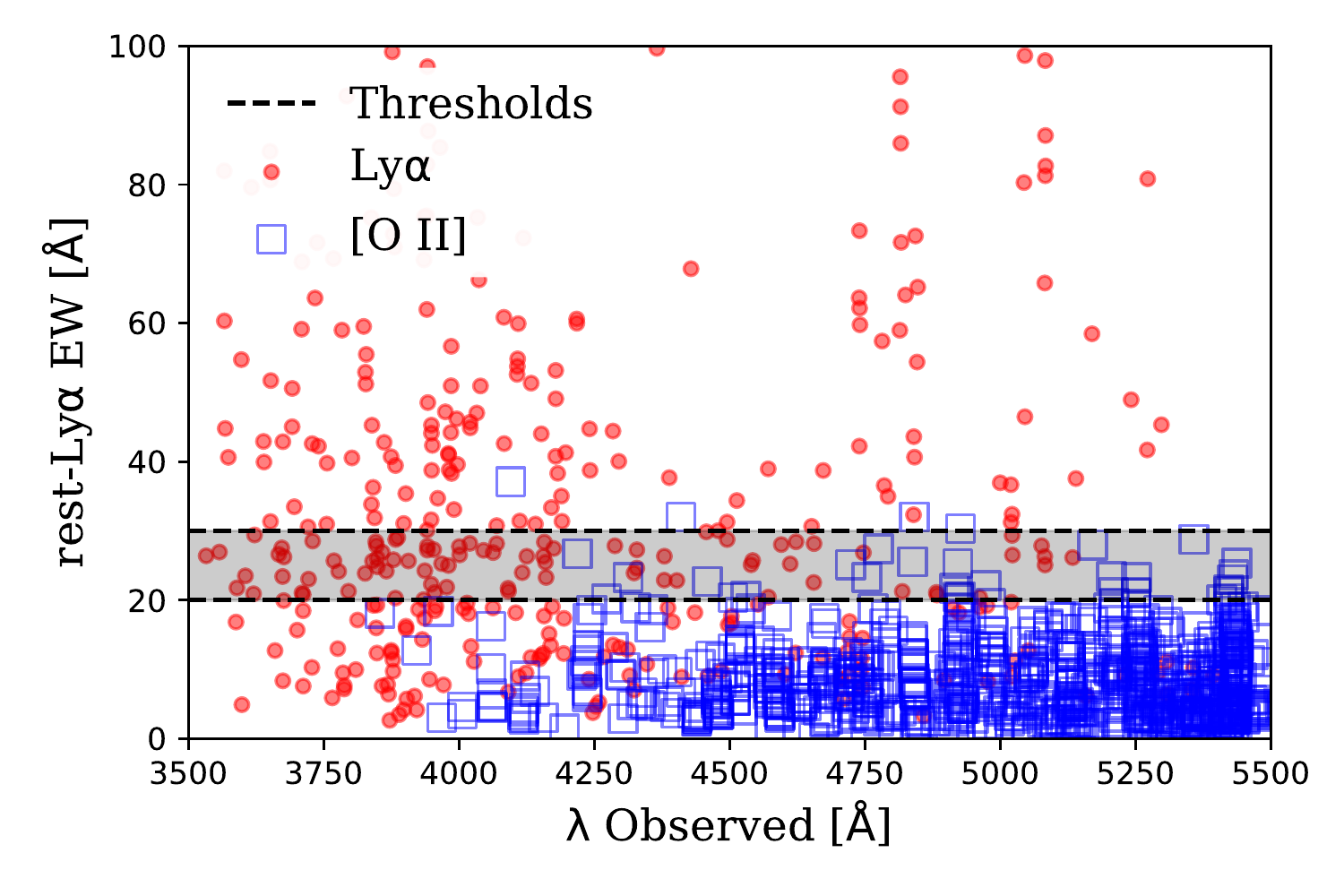}
    \caption{Simplified rest-\lya\ equivalent width vote applied to the assessment sample (SzAS, \S \ref{testing}). The figure is cropped to a maximum EW of 100\,\AA\ for readability and plotted without the $\sim$ 16\% errors. The SzAS contains no spectroscopically confirmed \OII\ emission lines with rest-\lya\ EW > 80\,\AA\null. Detections with EWs falling in the gray shaded region between 20 and 30\,\AA\ receive no vote while those above receive a \lya\ vote and those below an \OII\ vote with the weight of the vote modulated by the distance to the nearest threshold (\S \ref{simplified_ew_vote}).}
    \label{fig:simplified_ew_vote}
\end{figure}

Figure \ref{fig:simplified_ew_vote} shows the \lya\ and \OII\ SzAS detections with rest-\lya\ EW less than 100\,\AA\ (this includes all SzAS \OII\ emission lines) with the voting thresholds marked.  

This criteria votes in 80\% of the down-selected SzAS (containing only \lya\ and \oii) with a \lya\ contamination rate of 2\%. Superficially, this is superior \lya/\oii\ segregation compared to the P(LAE)/P(OII) vote (\S \ref{plae_poii_vote}, but by design, avoids voting in the difficult EW transition region (shaded region in Figure \ref{fig:simplified_ew_vote}).
\\

\subsubsection{Catalog Photometric Redshift Vote} \label{ss_cat_photz_vote}

The photometric redshifts fits from the various included catalogs (\S \ref{sec:catalogs}) are often too broad to confidently pin down a tight redshift constraint. However, they can be sufficient to distinguish between low-$z$ ($z \lesssim$ 0.7) and high-$z$ (1.7 $\lesssim z \lesssim$ 3.7) objects and thus help separate \oii\ from \lya. If there are any photometric redshifts for a HETDEX detection, this vote simply takes the arithmetic mean of all phot-$z$ measurements of the matched catalog counterpart from all contributing catalogs and compares it to the low-$z$ and high-$z$ ranges quoted above. If the mean falls within either range and is within a redshift distance of 0.5 of \oii\ or \lya respectively, then the corresponding vote is cast with a weight of 0.5. If the mean falls outside of both ranges or if the redshift separation between the mean and an assumption of \oii\ or \lya is greater than 0.5, then no vote is cast.

Only $\sim$ 1.5\% of HDR3 sources have at least one phot-$z$ catalog counterpart match, so this vote rarely contributes to the \plya\ logic. For the SzAS testing, since contributions from catalog phot-$z$ and spec-$z$ are necessarily turned off, this vote is never cast.
\\

\subsubsection{Apparent Magnitude and Equivalent Width Vote} \label{mag_ew_vote}

This vote is largely predicated on the observation that the HETDEX LAEs tend to be fainter than \oii\ galaxies. However, there certainly exist bright LAEs (including AGN) and faint \oii\ galaxies, so the EW$_{Ly\alpha}$ is also incorporated into the decision \rv{to moderate it}.

\begin{figure}[ht]
    \centering
    \advance\leftskip-0.75cm
    \includegraphics[width=0.5\textwidth]{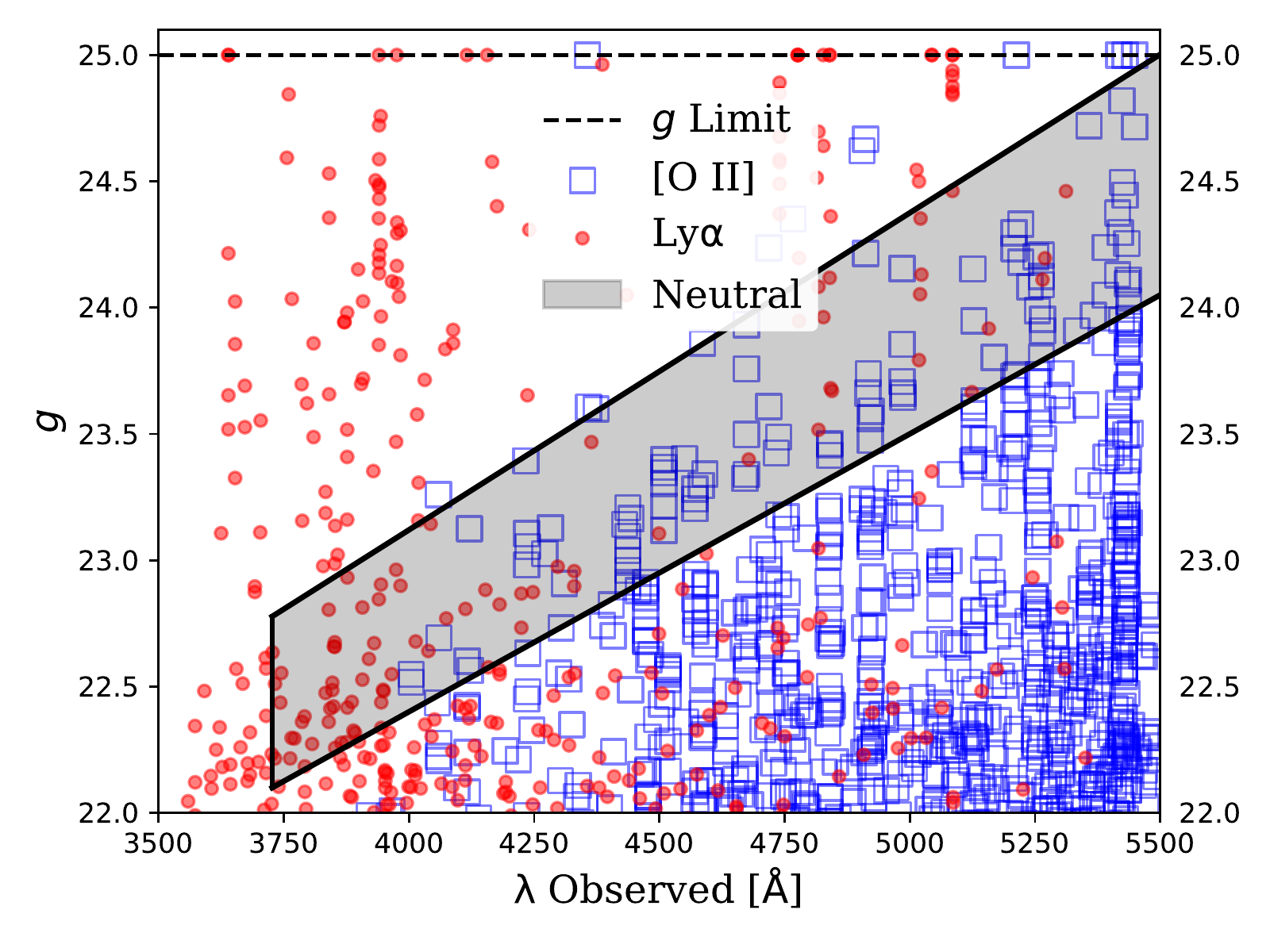}
    \caption{ The apparent magnitude (error $\sim 0.1$) and equivalent width vote, by itself, is highly effective at segregating \lya\ from \oii\ against the assessment sample here (SzAS, \S \ref{testing}). The \textit{Neutral} region is defined by the lines of Eqns (\ref{eq: ew_bright}) and (\ref{eq: ew_faint}) as the lower and upper bounds respectively, and extends from 3727\AA\ to the red edge of the HETDEX spectral window. \lya\ emitters represent 97\% of the down-selected SzAS above the \textit{Neutral} region, 44\% inside the \textit{Neutral} region, and 14\% below the \textit{Neutral} region.  
    }
    \label{fig:plya_vote_gmag_ew} 
\end{figure}

The apparent magnitude used in this vote is the $g$-band magnitude derived from the HETDEX spectrum (\S \ref{continuum_estimates}), which has a limiting magnitude of $\sim$25$_{AB}$. The magnitude threshold between votes for \oii\ and for \lya\ are defined by a pair of lines whose parameters are set to optimize the segregation of those two samples. Objects with $g$ magnitudes fainter than the upper line of Figure~\ref{fig:plya_vote_gmag_ew} are more likely to be \lya, while those brighter than the lower line of Figure~\ref{fig:plya_vote_gmag_ew} are more likely to be \oii.  The classification of objects, defined by their SzAS spectroscopic redshifts, lying between these two regimes is uncertain. The optimization over the slope and intercept parameters of these lines was performed using a simple grid search that maximizes the \lya\ accuracy in one case and the \oii\ accuracy in the other. While an MCMC fit could be more precise, given the uncertainties in the data features and the desire to avoid over-fitting to the specific test set, the grid search is preferred. The accuracy is defined as:

\begin{equation}
    \mathrm{accuracy} = 1 - \frac{\xi + \epsilon}{\Omega}
\end{equation}

\noindent where $\xi$ is the number of "true" \lya\ (\oii) detections (here as the spec-z counterparts in the SzAS test sample), that are not identified by the selection, $\epsilon$ is the number of incorrectly classified \lya\ (\oii) detections, and $\Omega$ is the total number of \lya\ (\oii) classified detections. Here, "true" is assumed as the catalog based spectroscopic redshifts (\S \ref{testing}).

The two lines are defined as:
\begin{equation} \label{eq: ew_bright}
    g^{+}_T = 1.10\times 10^{-3} \lambda + 18.0, \rv{(lower line, Fig. \ref{fig:plya_vote_gmag_ew})}
\end{equation}
\begin{equation} \label{eq: ew_faint}
    g^{-}_T = 1.26\times 10^{-3} \lambda + 18.1, \rv{(upper line, Fig. \ref{fig:plya_vote_gmag_ew})}
\end{equation}
where $g^{+}_T$ is the faint magnitude threshold, $g^{-}_T$ is the bright magnitude threshold, and $\lambda$ is the wavelength (\angstrom) of the anchor emission line.

We also define upper (faint) and lower (bright) bounds for the measured $g$ magnitude of each detection ($g^{+}$ and $g^{-}$, respectively) based on the propagated errors of the HETDEX spectroscopically-measured $g$-band magnitude.

The votes and their weights for this criterion are summarized in Table \ref{tab:gmag_and_ew_votes} with Figure \ref{fig:plya_vote_gmag_ew} showing the segregation of \lya\ and \OII\ with Eqns \ref{eq: ew_bright} and \ref{eq: ew_faint}. As the $g$ magnitude becomes brighter, the voting weights for \lya\ decrease and those for \oii\ increase. Large anchor line EW$_{Ly\alpha}$ favor \lya\ and small EW$_{Ly\alpha}$ favor \oii. With the exception of spectra associated with objects having faint $g$ magnitudes, those spectra with anchor line EW$_{Ly\alpha}$ (with error) between 15\,\angstrom\ and 30\,\angstrom\ receive no vote either way. Contamination of \lya\ by \OII\ for the down-selected SzAS is low with \lya\ comprising 97\% of the detections above the \textit{Neutral} region in Figure \ref{fig:plya_vote_gmag_ew}). Conversely, \lya\ represents only 44\% within the \textit{Neutral} region, where no vote is cast, and 14\% below it, where the vote is cast for \OII. \\

{\renewcommand{\arraystretch}{1.2}}
\begin{deluxetable}{ c |c  |c } [ht]
\tablecaption{Apparent Magnitude and EW Votes\label{tab:gmag_and_ew_votes}}
\tablewidth{0pt}
\tablehead{
\colhead{Condition}  & \colhead{Vote} & \colhead{Weight}
}
\startdata
$g^{-} > g^{+}_T$ & 1.0 & 0.50\tabularnewline
\hline
\rv{$g^{-}_T < g^{-} < g^{+}_T < g^{+}$ ~~{and}~~ $EW^{-} > 80$} & 1.0 & 0.50 \tabularnewline
\rv{$g^{-}_T < g^{-} < g^{+}_T < g^{+}$ ~~{and}~~ $EW^- > 30$} & 1.0 & 0.30 \tabularnewline
\rv{$g^{-}_T < g^{-} < g^{+}_T < g^{+}$ ~~{and}~~ $EW^+ \leq 15$} & 0.0 & 0.25 \tabularnewline
\hline
\rv{$g^{-} < g^{-}_T < g^{+} < g^{+}_T$ ~~{and}~~ $EW^- > 80$ }& 1.0 & 0.30 \tabularnewline
\rv{$g^{-} < g^{-}_T < g^{+} < g^{+}_T$ ~~{and}~~ $EW^- > 30$ }& 1.0 & 0.15 \tabularnewline
\rv{$g^{-} < g^{-}_T < g^{+} < g^{+}_T$~~{and}~~ $EW^+ \leq 15$ }& 0.0 & 0.40 \tabularnewline
\hline
\rv{$g^{+} < g^{-}_T$ ~~{and}~~ $EW^{-} > 80$ }& 1.0 & 0.25\tabularnewline
\rv{$g^{+} < g^{-}_T$ ~~{and}~~ $EW^{-} > 30$ }& 1.0 & 0.10\tabularnewline
\rv{$g^{+} < g^{-}_T$ ~~{and}~~ $EW^{+} \leq 15$ }& 0.0 & 0.50 \tabularnewline
\hline
else no vote & NA & 0.00\tabularnewline
\enddata
\tablecomments{
Summary of apparent magnitude and equivalent width votes. The conditions are ordered such that the logical evaluation results in at most one unique vote. If no conditions are met, there is no vote. The apparent $g$ magnitude becomes brighter moving down the table. \\
$g^{+}_T$ is the \rv{upper} (faint) $g$ threshold as a function of $\lambda$.\\
$g^{-}_T$ is the \rv{lower} (bright) $g$ threshold as a function of $\lambda$.\\
$g^{+}$ is the upper bound (faint) $g$ for the detection.\\
$g^{-}$ is the lower bound (bright) $g$ for the detection.\\
$EW^{+}$ is the upper bound restframe EW in \angstrom, assuming \lya.\\
$EW^{-}$ is the lower bound restframe EW in \angstrom, assuming \lya.\\
}
\end{deluxetable}

\subsubsection{Disqualifications} \label{disqualifications}

Disqualification conditions are a set of special classifications and data integrity issues that can either contribute additional weighted votes against a \lya\ classification or, in extreme cases, completely override the \plya\ results.

\begin{itemize}
    \item \textit{Meteor}: If the detections has a possible classification as a meteor (\S \ref{meteor_classification}, a vote against \lya\ is added with a weight equal to the strength of the meteor classification (0.0 - 5.0). Given its potentially large weight, this vote can be dominant. Regardless of the final result of the vote, the "meteor" label is attached to the detection.
    \item \textit{Bad Pixel Flat}: If a bad pixel flat is indicated by pixel-to-pixel variations or pixel flux values outside the acceptable range for an emission-line on that part of the CCD, then the emission line may be entirely due to, or at least enhanced by, this artifact. The detection will thus receive a vote against \lya\ with a weight equal to 1.0 plus the sum of the relative weights of those fibers contributing to the spectrum that have a bad pixel flat. The total weight for this vote is between 1.0 and 2.0. However, if the sum of the fiber weights exceeds a threshold, 0.50 by default, the entire \plya\ vote is disqualified. Independent of the vote, the bad pixel flat flag is associated with the detection and shown on the ELiXer report. 
    \item \textit{Duplicate Fibers}: If duplicated fibers (identified by repeated fiber identifiers or identical flux and error data vectors) appear in the detection spectra, the \plya\ vote is disqualified. This is an indication of a data reduction problem.
    \item \textit{Grossly Negative Spectrum}: If less than 10\% of the wavelength bins contain non-negative integrated flux values, the spectrum is considered "grossly negative" and suggests some issue in the reduction.  In this case, the detection and the \plya\ vote is disqualified.
    \item \textit{Poor Observation}: If the seeing FWHM is worse than a threshold (3\arcs\ by default) or the throughput response, as defined by \citep{Gebhardt+2021}, is less than a threshold (0.08, by default), the input observation is considered too poor to make a meaningful classification attempt and the vote is disqualified.
    \item \textit{Bad Dither Norm}: If the dither-to-dither normalization \citep{Gebhardt+2021} for the detection is above a threshold (3.0$\times$ by default), a potentially severe observation or reduction issue is indicated and the vote is disqualified.
\end{itemize}

\subsection{Best-z and Q(z)} \label{best_z}

Unless there is a serious error or a disqualification (\S \ref{disqualifications}), ELiXer assigns a single, best guess redshift, "Best-$z$", along with a quality score, "$Q(z)$", as an indication of the confidence in that redshift. The assignment of the Best-$z$ incorporates all prior information and analysis including the \plya, catalog spec-$z$ and phot-$z$, and any multi-line redshift solutions (\S \ref{redshift_solution_finder}). The $Q(z)$ value takes on a continuous value between 0 and 1, with 1 meaning "\rv{full confidence}" and 0 meaning "no confidence" (i.e., the redshift is effectively a guess). Where the \plya\ analysis is limited only to a determination as to whether the emission line is \lya\, the Best-$z$ logic attempts to fully specify the redshift. In the ideal scenario, there are multiple high-SNR emission lines within the HETDEX spectrum, each corresponding to a known line at a consistent redshift. In such a case, the Best-$z$ is clear and the corresponding $Q(z)$ is 1.0.  Such objects are rather rare, but they do define the starting benchmark.

The Best-$z$ is set as (1) the redshift from a qualified multi-line spec-$z$ solution, (2) the \lya\ redshift when there is no spec-$z$ solution but \plya\ favors \lya, or typically, (3) the \OII\ redshift. In the last case, the redshift can be set to \ion{C}{3}] or \ion{Mg}{2} when the line is broad and occurs at a wavelength within the HETDEX spectral window where no other strong feature is expected to be found.

The $Q(z)$ confidence value is set based on the Best-$z$ selection condition. It is primarily a function of the \plya\ value and the multi-line solution score. $Q(z)$ is maximized by \plya\ when \plya\ is near 0 or 1 and minimized when \plya\ is near 0.5. The effect of the multi-line solution score, on the other hand, is a monotonic increase with the multi-line solution score.  $Q(z)$ may also have penalties and caps imposed on it based on specific circumstances and flags, such as the detection being near a spatially extended, bright object or if the various continuum estimates (\S \ref{continuum_estimates}) disagree. If the multi-line solution and \plya\ agree, the $Q(z)$ score increases; if the two measures disagree, the $Q(z)$ score is decreased based on the relative difference between the multi-line solution and \plya\ strengths. The selection logic and $Q(z)$ assignment is summarized in Table \ref{tab:best_z_q_z}.

Since the majority of HETDEX objects are faint, with a single detected emission line, most ($\sim$ 80\%) receive a $Q(z)$ score less than 0.5 with $\sim$ 35\% in the lowest $Q(z)$ bin (0-0.1). These are still usually correctly classified as is shown in Sections \ref{results} and \ref{discussion}, but rely on less evidence and thus have a low $Q(z)$ value.\\

{\renewcommand{\arraystretch}{1.2}}
\begin{deluxetable}{ >{\centering}p{4cm} | >{\centering}p{1.5cm}  | >{\centering}p{2.5cm} } [ht]
\tablecaption{Best-$z$, $Q(z)$ Summary\label{tab:best_z_q_z}}
\tablewidth{0pt}
\tablehead{
\colhead{Condition}  & \colhead{Best-$z$} & \colhead{$Q(z)$}
}
\startdata
Strong, multi-line spec-$z$ solution consistent with \plya\ & multi-line spec-$z$ & \ 4-5$\star$ \tabularnewline
\hline
Strong, multi-line spec-$z$ solution not consistent with \plya\ & multi-line spec-$z$ & \ 0-3$\star$ \tabularnewline
\hline
Weak, multi-line spec-$z$ solution consistent with \plya\ & multi-line spec-$z$ & \ 2-4$\star$ \tabularnewline
\hline
Weak, multi-line spec-$z$ solution not consistent with \plya\ & multi-line spec-$z$ & \ 1-3$\star$ \tabularnewline
\hline
\plya\ only, $\gtrsim$ 0.7 & \lya\ & \ 3-4$\star$ \tabularnewline
\hline
\plya\ only, $\gtrsim$ 0.5 & \lya\ & \ 0-2$\star$ \tabularnewline
\hline
\plya\ $\lesssim$ 0.5 with single, broad emission line & \oii\, \ion{Mg}{2}, \ion{C}{3}] & \ 0-1$\star$ \tabularnewline
\hline
\plya\ only, $\lesssim$ 0.5 & \oii\ & \ 0-1$\star$ \tabularnewline
\hline
\plya\ only, $\lesssim$ 0.3 & \oii\ & \ 0-2$\star$ \tabularnewline
\enddata
\tablecomments{
Summary of the Best-$z$ and $Q(z)$ logic. Specific values (0.0-1.0) of the $Q(z)$ are not shown as they depend on details omitted, but are expressed as these qualitative descriptors: 5$\star$ ($\sim$1.0), 4$\star$ ($\sim$0.80), 3$\star$ ($\sim$0.50), 2$\star$ ($\sim$0.35), 1$\star$ ($\sim$0.25), 0$\star$ ($\sim$0).
}
\end{deluxetable}

\subsection{Clustering/Neighbor Redshift Matching} \label{clustering}

In the low-surface brightness outer regions of spatially resolved galaxies, HETDEX detections with low, PSF-weighted line fluxes (commonly arising from faint \ion{H}{2} regions and planetary nebulae) may be incorrectly classified by ELiXer as \lya.  To address this issue, ELiXer can optionally compare a detection against other nearby HETDEX detections and look for consistencies. When invoked, ELiXer examines all \rv{HETDEX emission line} detections within 15\arcs\ (by default) of the \rv{current detection} under consideration, and tests for $g$-band magnitudes brighter than 23$_{AB}$ with matching observed emission line(s) of higher line score (\S \ref{line_scoring}). The presumption, which is borne out in testing, is that the brighter, higher-scoring detections are (1) better centered on the object and (2) more likely to receive the correct classification. The requirement to match the observed emission line wavelength(s) in addition to the on-sky proximity helps preserve the classification of background objects with lines of sight passing near the brighter, foreground source. When more than one match is found, the highest scoring redshift solution is selected and if the selected object is brighter and higher scoring than the current detection's solution, that neighbor's classification is used as a replacement. \rv{In other words, faint, low scoring detections can be assigned the more secure redshift of an immediately adjacent, brighter, higher scoring "neighbor" detection when they share matching observed-frame emission lines and are assumed to represent different detections of the same object.} When this update is made, the altered detection is marked with a flag and the detection ID number of the matching neighbor detection.

This clustering has a relatively small effect, modifying less than 0.5\% of all HETDEX emission line detections. The algorithm does not link nor otherwise combine the individual detections; all detections remain uniquely reported.\\

\section{Testing and Results} \label{testing}

All the effort made toward classification is effectively meaningless without appropriate testing and a selection of a reasonable spec-$z$ assessment sample (SzAS) against which to test. As HETDEX is a large and unique survey with no pre-selection of targets, it is impossible to collect an overlapping observational dataset of known redshifts of even remotely similar size (in terms of numbers of unique astrophysical objects) and continuum depth. Beyond polling experts for classifications based on visual inspection, and comparing ELiXer results against those of simulated objects, the best we can do is match HETDEX sources against spectroscopic redshift catalogs produced by other surveys.

The assessment sample for this work is a composite of matched HETDEX detections from the public, archival catalogs described in Section \ref{sec:catalogs} and in \cite{Cooper_2022}. In all cases, these are spectroscopic redshifts only; no photo-$z$ estimates are used in this assessment sample. For the catalog provided redshifts, source matching to HETDEX is based on sky position and apparent magnitude. The catalog source position must be inside or within 0\farcs5 of the edge of the \textit{SEP} aperture associated with the HETDEX detection if an aperture match is made (\S \ref{elixer_aperture_phot}), or within 0\farcs75 of the HETDEX position if the object is fainter than $g=24.5$ and no \textit{SEP} aperture is matched. The catalog matched spectroscopic redshifts are accepted as true.

The assessment sample is down-selected to only those detections fainter than $g=22$ with redshifts that match any of the emission lines in Table \ref{tab:Table of Lines} to within $\pm 4$\,\AA\null.  Though the magnitude distribution still significantly skews to brighter objects, this filtering helps refine the selection to better align with the more common, fainter HETDEX detections. The result is a dataset consisting of 834 \OII\ emission lines, 384 \lya\ lines, and 402 "Other" lines, including \CIV, \CIII, \MgII, and \Hb as reported in the SzAS. 
Each redshift corresponds to a unique HETDEX detection, however, these are not necessarily unique galaxies. For brighter, extended galaxies there can be more than one overlapping HETDEX emission line detection, and where there are multiple observations covering the same position, the same galaxy may be detected more than once. Since ELiXer operates on each HETDEX detection individually, this is as intended.\\

\subsection{Definitions} \label{definitions}
For the remainder of this work, we make the following definitions:
\begin{itemize}
    \item Accuracy:  The number of agreements between the ELiXer assigned classification and the SzAS classification divided by the number of ELiXer detections of that classification.  A match is counted if the rest-frame wavelengths from the HETDEX observed wavelength and the SzAS and ELiXer assigned redshifts agree within $\pm 4$\,\AA.
    \item Recovery:  A fraction roughly equivalent to completeness, but with no correction made for survey biases. Here we refer to the number of detections of a particular emission line identified by the ELiXer software that are matched 1:1 to that of the SzAS divided by the number of those emission lines in the SzAS.
    \item Contamination:  The fraction of detections within some defined range that are incorrectly classified. This may be further refined to the fraction of misclassifications by a particular emission line. For example, we will discuss the contamination in the \lya\ sample by \OII\ as a function of $g$-magnitude.
\end{itemize}
Accuracy can be slightly under reported for broad, noisy lines where the fitted line center can be offset from the true center and where winds and radiative transfer effects can create a significant velocity offset from the systemic redshift. The $\pm 4$\,\AA\ allowance covers all but the most extreme cases so the impact is minimal. Accuracy and contamination are direct inverses and, for any given emission line, they necessarily sum to unity. Accuracy and recovery are similar, but differ by the base divisor. For the recovery of detections, any contamination of one emission line comes at the direct cost to the recovery of another emission line. Conversely, the recovery counts of an emission line is also one minus the sum of the contaminations of all other emission line types. Notice that the relationship does not directly hold for recovery and contamination \textit{rates}, as each of those rates have different divisors. \\

\subsection{Calibration} \label{calibration}

Testing and calibration are combined in a highly iterative process. ELiXer is executed on the detections of the test dataset, but with catalog matching spec-$z$ and phot-$z$ turned off. That is, for the test runs, ELiXer does not include or consider the catalog reported spectroscopic redshifts that would, in a standard run, factor into the classification. The ELiXer output, specifically the P(Ly$\alpha$) values and the redshifts, are then compared to the test sample and checked for contamination, recovery, and accuracy. Disagreements between the ELiXer results and the assessment sample are examined, and manual adjustments to the individual votes and voting weights (\S \ref{sec:classification}, and \S \ref{p_lya} in particular), are made as warranted. Considerations against over-tuning and potentially incorrect test sample redshifts are addressed with deliberately loose fitting, low-order segmentation thresholds and by varying the composition of the test sample by creating random and targeted (in apparent magnitude, line FWHM, observation field, etc.) subsets. The process is repeated until there is good agreement (generally, matching 90-95\% or better) between the ELiXer assigned redshifts and the test sample redshifts. With the focus on P(Ly$\alpha$) as the primary classification metric and with its flexible threshold selection, what constitutes "good" agreement is somewhat subjective but is also highly adaptable to the specific scientific needs. For example, the stacking of spectra to measure Lyman Continuum in \citet{Davis_2021} is very sensitive to contamination but does not specifically require a highly complete sample and so utilizes a \plya\ selection of 0.8 and greater. On the other hand, the $H(z)$ and $D_A(z)$ precision goals for the primary HETDEX science is less sensitive to contamination but needs to be largely complete \citep[][]{Gebhardt+2021,Farrow+2021} and a \plya\ threshold of 0.5, or even lower, is more appropriate.\\

\subsection{Additional Testing} \label{Parallel_Efforts}

To supplement the catalog spec-$z$ testing, several other testing and feedback efforts are actively used. Though the mechanics vary, all provide checks on the ELiXer classifications with targeted detection subsets. As with the SzAS, the detections where these alternate methods and ELiXer disagree are manually inspected and adjustments to the ELiXer classification algorithm(s) are made as warranted.

These supplementary efforts fall into two categories. The first are automated machine learning classifiers, both supervised and (sometimes) unsupervised. These are all in early development and explore various classification frameworks, with both T-distributed Stochastic Neighborhood Embedding (tSNE) \citep{van_der_maaten_2008} and Autoencoder Neural Network \citep{Wang_2014} techniques showing good promise.  

The second category relies on manual, visual vetting. The first efforts focused on HETDEX collaboration experts and university students (after receiving training). A more recent science outreach effort has opened classification and general exploration to the public in a citizen science project on \textit{Zooniverse} (\url{https://www.zooniverse.org/}). One workflow of the \textit{Dark Energy Explorers} (\url{https://www.zooniverse.org/projects/erinmc/dark-energy-explorers}) project \citep[][]{House_2022} tasks its citizen scientists to classify HETDEX detections as either being at low-$z$ ("Nearby Galaxy or Star") or possibly high-$z$ ("Distant Galaxy or nothing") using a reduced ELiXer report that contains only sections of 2D fiber cutouts, single band ($g$ or $r$) photometric imaging, and a Gaussian fit to the emission line. Each detection receives 15 responses with the aggregate classification reported as the mean of those responses. Even with this reduced information, these broad categories match with the ELiXer classification more than 92\% of the time with House, et al, estimating 7.7\% contamination and 90.7\% recovery of high-$z$ galaxies. As with the other methods, select disagreements between ELiXer and \textit{Zooniverse} are reviewed for potential classification failures by ELiXer.

\subsection{Results Summary} \label{results}

\begin{figure}[ht]
    \centering
    \advance\leftskip-0.75cm
    \includegraphics[width=0.5\textwidth]{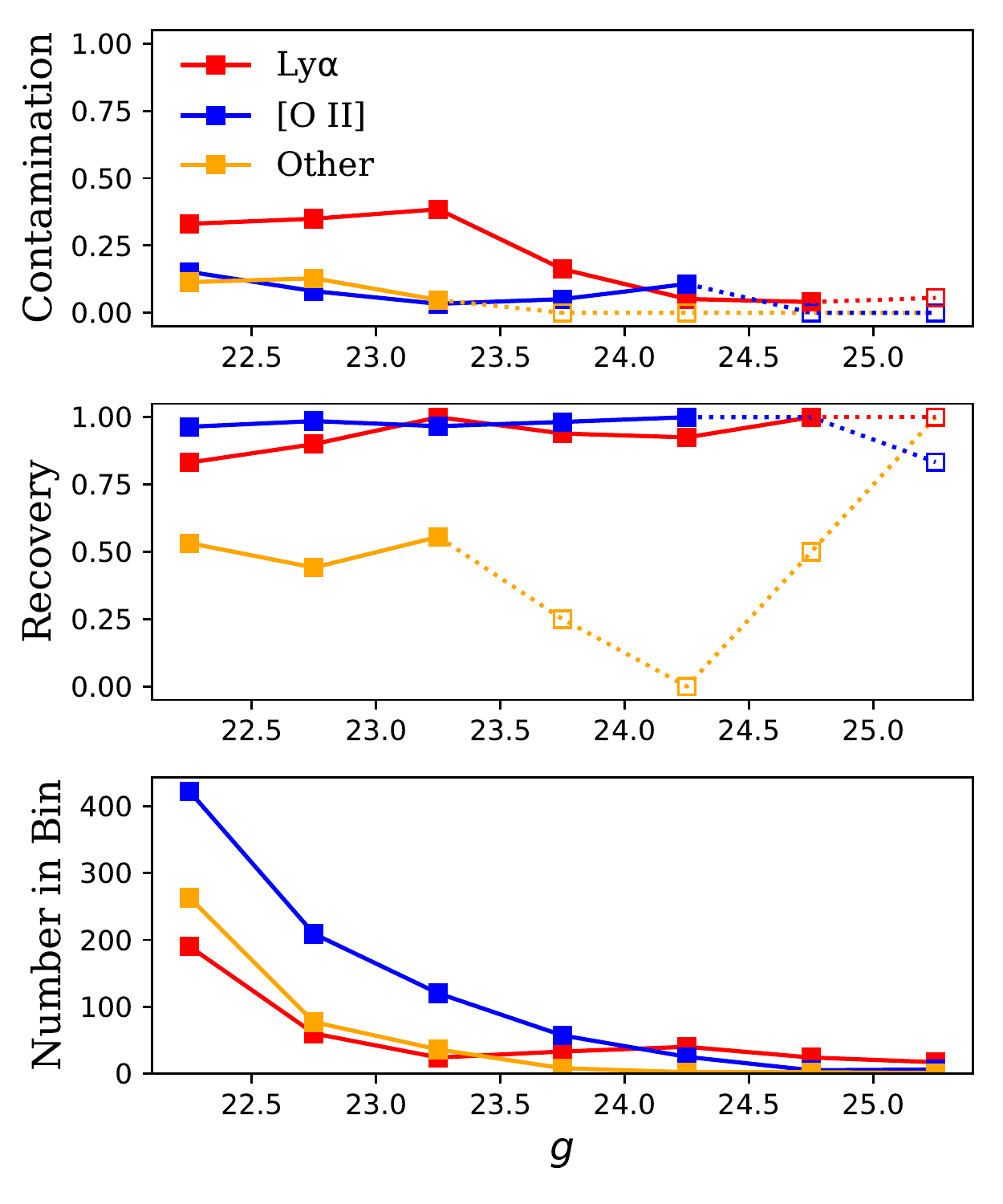}
    \caption{Performance summary of ELiXer classification and redshift assignment vs.\ the SzAS in $g$-magnitude bins. ELiXer does very well with \lya\ and \OII, as intended, but struggles with the "Other" lines, such as \ion{C}{4} $\lambda 1550$, \ion{C}{3}] $\lambda 1909$, and \ion{Mg}{2} $\lambda 2800$ Note that the results for the faintest bin for \lya, the faintest 2 bins for \OII\ and the faintest 4 bins for Other lines, denoted with open markers and dotted lines, have too few SzAS counts to be meaningful. The high contamination rate in \lya\ at brighter magnitudes is a result of the biases in the SzAS and is discussed in section \ref{discussion}. }
    \label{fig:accuracy_truth_sample}
\end{figure}

A comparison of the ELiXer classification/redshift assignments with those of the SzAS are summarized in Figure \ref{fig:accuracy_truth_sample} and in Table \ref{tab:truth_performance}. Figure \ref{fig:accuracy_truth_sample} breaks out the contamination and recovery rates by $g$-magnitude, with the counts of each type shown as a reference in the bottom panel. When there are very few classifications of a given type, such as faint \OII\ and "Other" lines, the accuracy and recovery rates are not meaningful. Against the SzAS, ELiXer performs very well on \lya\ and \OII\ classifications, but is challenged by the "Other" emission lines. As will be discussed later, the elevated contamination in the \lya\ detections at bright magnitudes is a function of the biases in the SzAS as compared to the HETDEX survey.  

Table \ref{tab:truth_performance} summarizes the cumulative performance of several different \lya/\OII\ segregation methods against the SzAS identifications of the \lya\ or \OII\ line. This down-selection is made so that the comparisons of the ELiXer P(Ly$\alpha$) method (\S \ref{p_lya}) at several selection thresholds is equitable, as 20\,\AA\ equivalent width cut and the P(LAE)/P(OII) method do not classify lines other than \lya\ and \OII\null.  It is clear that each method is an effective classifier.   Except at the extreme thresholds, the P(Ly$\alpha$) methods produce the lowest contamination and highest recovery rates, with P(Ly$\alpha$) $> 0.5$ yielding a good balance of contamination and recovery fraction.  This is the default input for the ELiXer Best-$z$ assignment (\S \ref{best_z}).  Given the biases in the SzAS for bright objects and AGN, though, these results cannot be directly applied to the whole of HETDEX. However, a correction for these biases is made and discussed later in \S \ref{bias_corrections}. We also caution that the detections in the SzAS factor significantly in the calibration of the votes and weights of the \plya\ metric. Although efforts are made to avoid over-fitting, these results could still be less reflective of HDR3 in general. 

The contamination rate of \lya\ by \OII\ is effectively flat as a function of the observed wavelength of the emission line. However, the recovery rate of \lya\ sources trends lower as the observed wavelength moves redward. At the blue end of the HETDEX spectral range, $ \lambda_{\mathrm{obs}} \lesssim$ 4200\AA, the recovery rate is $\sim$97\%; in the middle range, 4200 $\lesssim \lambda_{\mathrm{obs}} \lesssim$ 4800, the rate is $\sim$91\%; and at the red end, $4800 \lesssim \lambda_{\mathrm{obs}}$ the rate is $\sim$81\%. This is an effect of larger numbers of faint \OII\ emitting galaxies and fewer numbers of LAEs in their respective higher redshift regions. These \OII\ galaxies are more similar in appearance to LAEs based on several of the metrics used in ELiXer, $g$ and $r$ magnitudes, angular size, and even EW and line width to a lesser extent (see Sections \ref{object_size_vote}, \ref{line_fwhm_vote}, \ref{simplified_ew_vote}, and \ref{mag_ew_vote} and their figures). The observed emission line wavelength factors in the related votes help keep the \lya\ contamination rate flat and low, but at the cost of the loss of some LAEs to \OII\ classifications. As shown in Table \ref{tab:truth_performance}, this can be tuned to improve the \lya\ recovery rate at the expense of a higher contamination rate as dictated by particular science needs. \\

\begin{deluxetable}{l | c c} [ht]
\tablecaption{\lya\ vs \OII\ Segregation on Assessment Sample \label{tab:truth_performance}}
\tablewidth{0pt}
\tablehead{
\colhead{Method}  & \colhead{\lya\ Contamination} & \colhead{\lya\ Recovery} 
}
\startdata
\lya\ rest EW $>$ 20\AA\ & 0.084 & 0.708 \tabularnewline
P(LAE)/P(OII) default \tablenotemark{\footnotesize 1} &  0.090 & 0.763\tabularnewline
P(LAE)/P(OII) optimized \tablenotemark{\footnotesize 1} &  0.056 & 0.724\tabularnewline
P(LAE)/P(OII) ELiXer \tablenotemark{\footnotesize 2} &  0.042 & 0.705\tabularnewline
P(\lya) $>$ 0.7 & 0.005 & 0.752\tabularnewline
P(\lya) $>$ 0.6 &  0.007 & 0.797 \tabularnewline
P(\lya) $>$ 0.5 \tablenotemark{\footnotesize 3} & 0.010 & 0.903\tabularnewline
P(\lya) $>$ 0.4 & 0.027 & 0.926\tabularnewline
P(\lya) $>$ 0.3 & 0.056 & 0.940\tabularnewline
\enddata
\tablecomments{The cumulative performance of various methods against the SzAS down-selected to only include \OII\ (834 detections) and \lya (384 detections). This allows a fairer comparison of P(Ly$\alpha$) (\S \ref{p_lya}) to the first three methods, which do not consider other lines. The SzAS is biased to bright objects, with an over representation of AGN, so these results do not directly translate to the larger population of HETDEX detections. An adjustment for these biases are made and discussed later in \S \ref{bias_corrections}. Additionally, though efforts are made to avoid over-fitting to the SzAS, its detections significantly contribute to the determination of the votes and weights of the \plya\ metric, so these results may not be as representative when considering all HETDEX detections.
\tablenotetext{1}{\citet{Leung2017}}
\tablenotetext{2}{Modified P(LAE)/P(OII) optimized used in ELiXer (\S \ref{plae_poii_vote})}
\tablenotetext{3}{Default input to Best-$z$ logic (\S \ref{best_z})}
}
\end{deluxetable}

\section{Discussion} \label{discussion}

\begin{figure*}[ht]
    \centering
    \includegraphics[width=1.0 \textwidth]{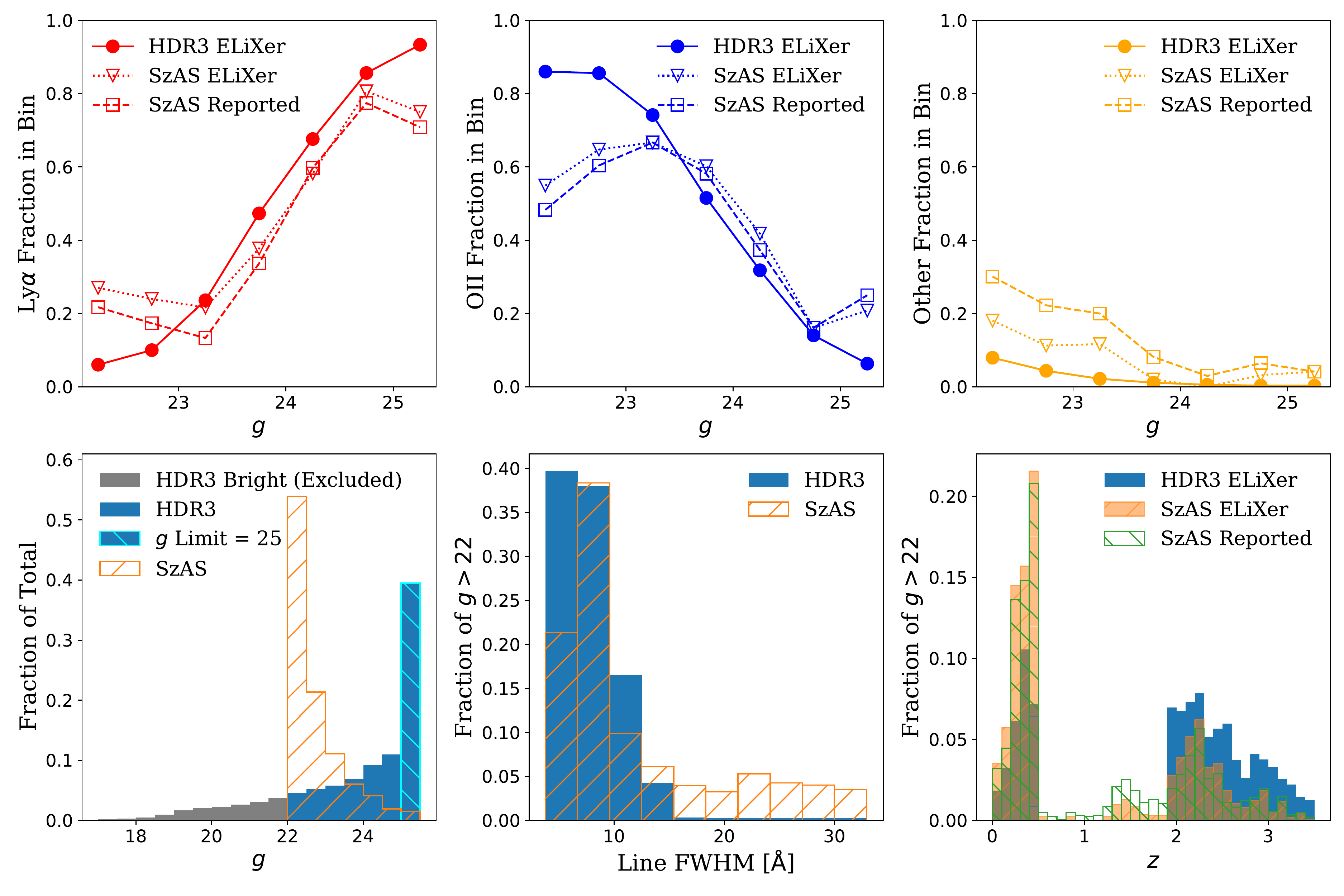}
    \caption{
    Summary of the $\sim1600$ emission line detections in the Spec-$z$ Assessment Sample (SzAS) compared to the $\sim1.5\times10^6$ detections in the HETDEX Data Release 3 (HDR3). The top panels show the relative fraction of \lya, \OII, and Other emission line detections as a function of  $g$-magnitude, as classified by ELiXer and as reported by archival spec-$z$ measurements in the SzAS.   The ELiXer reported classifications represent more of an "apples to apples" comparison, as it is clear that the SzAS is skewed towards brighter magnitudes and is significantly overabundant in Other emission line detections. The \lya\ and \OII\ distributions are very similar fainter than about 23.5$_{AB}$, but diverge at the brighter end. The lower left panel illustrates the bright bias. The lower-center panel shows an excess in the SzAS for broad emission lines; though not explicitly shown here, these broad lines are predominantly \lya, \CIV 1549\,\AA, and \CIII 1909\,\AA\ and originate from brighter, probably AGN, objects. The lower right panel echoes the over abundance of the Other emission lines, showing an increase in the fraction of $1.0 \lesssim z \lesssim 2.0$ detections, likely AGN, compared to HDR3. }
    \label{fig:truth_3x2}
\end{figure*}

As can be seen from Figure \ref{fig:truth_3x2}, the sample we use for spectroscopic assessment, SzAS, is highly biased to brighter detections, somewhat biased to broader lines, and contains an over representation of emission lines other than \lya\ and \OII, as compared to HETDEX as a whole.  At its bright end, the sample is under-abundant in \OII\ and over-abundant in \lya with the reverse at the faint end. Since these spectroscopic redshifts come from existing archival surveys (\S \ref{testing}) and spectroscopy is historically expensive, it stands to reason that the available spectra would favor brighter, rarer objects. An expansion of the SzAS is underway in collaboration with DESI (\citep[][]{desi1,desi2}) which will provide higher spectral resolving power (R$\sim$2000-5000) and a redder wavelength coverage (3600-9800\AA) to selected HETDEX detections. This will increase the number of faint ($g>25$) spectra in future assessment samples and bring their distributions more in line with HETDEX.

While not completely devoid of faint objects, the SzAS contains a smaller fraction of its detections in the faintest bins compared to the full HETDEX sample. This is not unexpected and is not a significant issue. Given the methodology of the classification, ELiXer is likely to classify anything fainter than $g\sim$25 as an LAE in the $1.9 < z < 3.5$ redshift range. While there are certainly \OII\ emission-line galaxies with $z < 0.5$ and $g > 25$, if we assume that this emission has a rest-frame equivalent width of less than 20\,\AA, then \OII\ can be expected to be, at most, $\sim 3\times 10^{-17}$ \cgs. This maximum value is $\sim$2$\times$ fainter than the 50\% flux limits for HETDEX \citep[][]{Gebhardt+2021}, making it unlikely that HETDEX would even detect an \OII\ emission line from such a galaxy.  Thus the reduced fraction of $g\gtrsim25$ objects in the SzAS, compared to HDR3, is largely moot.

Nevertheless, the other biases cannot be ignored. While an uncorrected assessment sample can serve as a development test set and provide reasonable limits on the expected contamination, recovery, and accuracy rates for ELiXer classifications, a correction is needed to extrapolate to the entire HETDEX emission line sample.\\

\subsection{Bias Correction to the Full HETDEX Catalog} \label {bias_corrections}

Given the clearly biased distribution of the assessment sample as compared to the full HETDEX catalog, it is prudent to apply some measure of correction before extending the results from the SzAS to the full catalog. The correction chosen is relatively simplistic and, as will be shown a little later, has effectively no impact on the overall sample results.

As seen earlier, the SzAS dataset is subdivided into \lya, \OII, and Other emission line detections, and each subset is binned by $g$ magnitude from $22_{AB}$ to $25_{AB}$ in steps of 0.5, with the last bin containing all detections fainter than the 25$_{AB}$ flux limit. The contamination (by type) for each of the three classifications is computed against the SzAS in each $g$ bin as defined in \S \ref{definitions}.

To correct for the population biases in the SzAS compared to the full HDR3 sample, we consider the contamination rates in the SzAS to be functions of the per bin fractions of the contaminant, and the target type as classified by ELiXer. This allows us to use the same ELiXer classification rates in the full HDR3 sample as a correction to the SzAS rates. The applied correction to the SzAS values then is: 

\begin{equation} \label{contamination_correction}
    C'_{i,j} = \frac{\left( \displaystyle\sum_{k} C_{i,j,k} \times \frac{E_{H,j,k}}{E_{S,j,k}} \times N_{H,i,k} \right)} {\displaystyle\sum_{k} N_{H,i,k}} 
\end{equation}

\noindent where: 
\begin{itemize}
    \item $C'_{i,j}$ is the corrected contamination rate of the target type $i$ (\lya, \OII, or Other) by contaminant $j$, such that $i \ne j$.
    \item $C_{i,j,k}$ is the directly computed contamination rate in the SzAS per $g$-magnitude bin, $k$ (matching the bins in Figure \ref{fig:accuracy_truth_sample}).
    \item $E_{H,i,k}$ is the ELiXer classification fraction of the target type in HDR3 per $g$-magnitude bin.  
    \item $E_{S,j,k}$ is the ELiXer classification fraction of the contamination type per $g$-magnitude bin in the SzAS. 
    \item $N_{H,i,k}$ is the number of target ELiXer classifications in HDR3 per $g$-magnitude bin.
\end{itemize}

An additional simple correction is also applied to help account for false positive (FPN) detections caused by noise interpreted as an emission line by the HETDEX line-finding algorithm \citep{Gebhardt+2021}. \rv{These are random fluctuations in the PSF weighted spectrum from thermal electrons in the CCDs, stray photons, read noise, etc, that happen to scatter up and pass the various filtering thresholds in the line-finding code and masquerade as low SNR emission lines. They do not represent real astrophysical sources but when interpreted as such, they map to random locations in (RA, Dec, z)-space. As the candidate emission line SNR increases toward 5.5, the incidence of these FPN rapidly approaches zero. As discussed later in \S \ref{contamination_from_noise}, this has only a minimal impact on the HETDEX cosmological measurements. As an approximate correction,} the ELiXer classification ratios in Eqn (\ref{contamination_correction}) for HDR3 are modified by assuming 30\% of all detections with SNR $<$ 5.0 and 15\% of all detections with 5.0 $\leq$ SNR $<$ 5.5 are false positives and simply removing those from all summed counts. Early indications are that the true FPN rates may be significantly less than this, 

\citep[][]{Cooper_2022}, so we believe the assumed FPN rates are overestimates. \\

\begin{figure}[ht]
    \centering
    \advance\leftskip-0.75cm
    \includegraphics[width=0.5\textwidth]{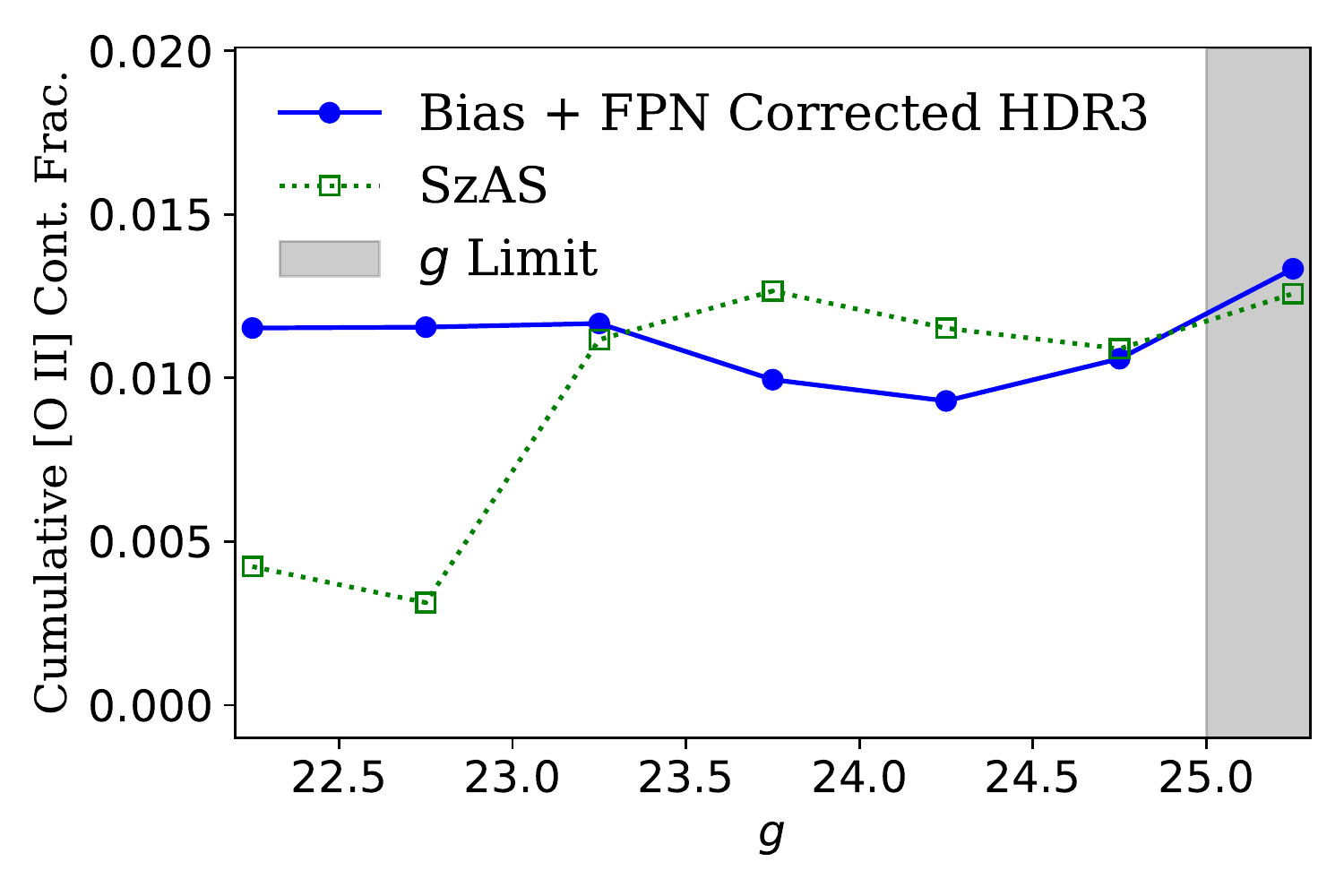}
    \caption{Cumulative (bright to faint) contamination of \lya\ by \OII\ as a function of $g$ magnitude using the default ELiXer configuration. The Bias + FPN Corrected HDR3 curve attempts to compensate for the biases in the SzAS (compared to all of HDR3) and account for false positives in the low-SNR regime (\S \ref{bias_corrections}).} 
    \label{fig:truth_cumulative_oii_contamination_vs_gmag}
\end{figure}

\begin{figure}[ht]
    \centering
    \advance\leftskip-0.75cm
    \includegraphics[width=0.5\textwidth]{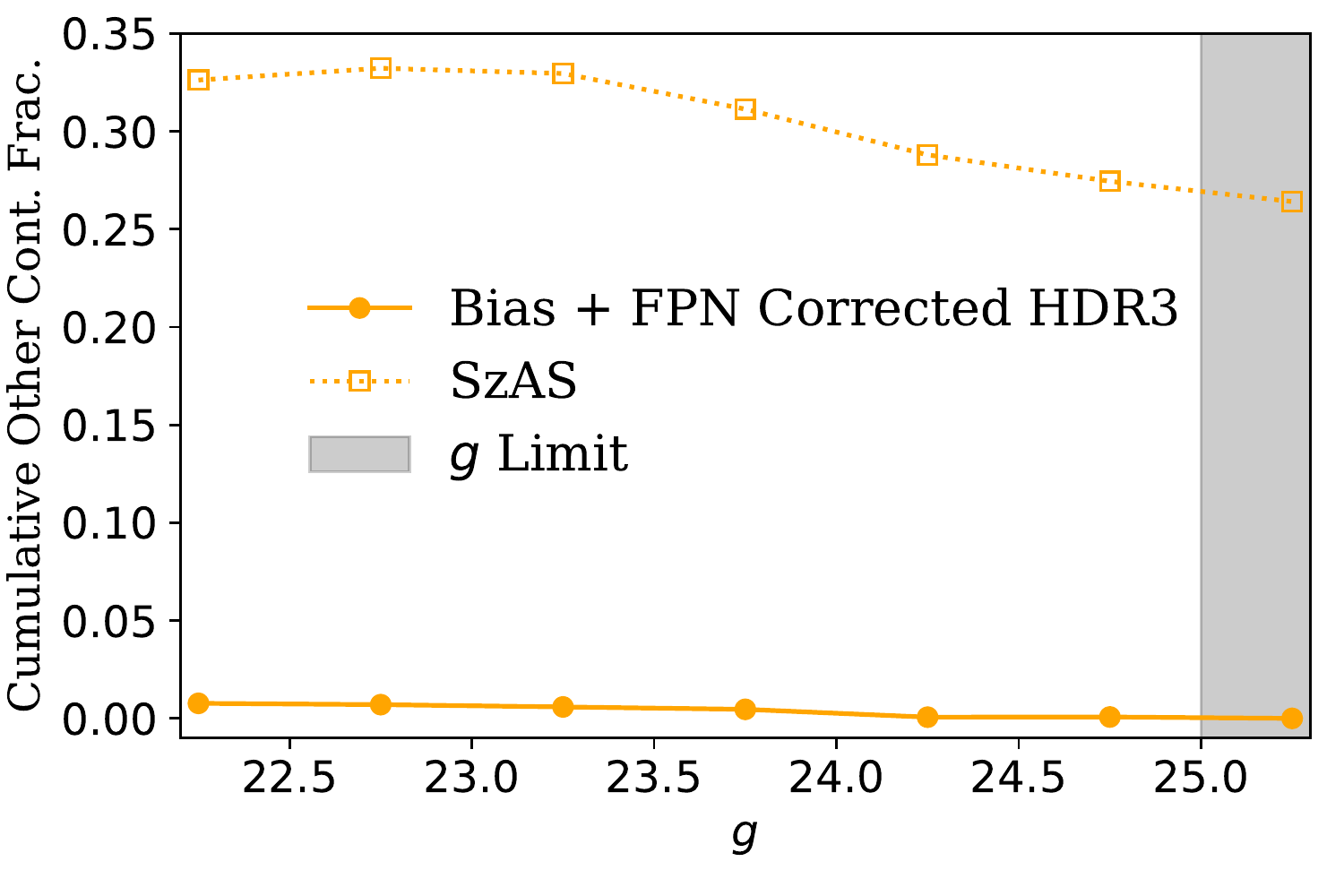}
    \caption{Cumulative (bright to faint) contamination of \lya\ by emission lines other than \OII\ for $g>22$ using the default ELiXer configuration. The FPN + Bias Corrected HDR3 curve attempts to compensate for the biases in the SzAS and account for false positives due to random noise in the low-SNR regime (\S \ref{bias_corrections}). The much larger contamination rate in the SzAS is largely driven by confusion of \lya\ vs.\ \CIII\ and \CIV, where the AGN population is significantly over represented (see Figure \ref{fig:truth_3x2}, \S \ref{performance} and \S \ref{missing_agn_lbg}).} 
    \label{fig:truth_cumulative_oth_contamination_vs_gmag}
\end{figure}

\subsection{Performance} \label{performance}
 
Figures \ref{fig:truth_cumulative_oii_contamination_vs_gmag} and  \ref{fig:truth_cumulative_oth_contamination_vs_gmag} show the cumulative (bright to faint) contamination fraction of \lya\ by \OII\ and all "Other" lines respectively, both for the SzAS and for the $g > 22$ HDR3 dataset. Table \ref{tab:overall_performance} reports the cumulative contamination rates from those two figures (highlighted by \textbf{bold} type face), provides summary information on the contamination in \OII\ and the "Other" lines, and gives the accuracy and recovery rates for all discussed line types. Note that the values for the SzAS corresponding to Table \ref{tab:truth_performance} are slightly different, since the detections for that table are down selected to only include \lya\ and \OII\null.  Overall, ELiXer performs extremely well in mitigating the contamination in the \lya\ classification, and excels at the faint end against the primary contaminant, \OII. This is what ELiXer is tuned to do. At brighter magnitudes, non-\OII\ contaminants are more problematic, though they represent only a small fraction of the total HETDEX dataset (Figure \ref{fig:truth_3x2}).  For the HETDEX data releases, the final classification of these objects is assisted by the supplemental program, \textit{Diagnose} \citep{Zeimann_2022} (see also \S \ref{missing_agn_lbg}).

The cumulative fractional contamination from \OII\ has a peak between $g\sim23.0$ and $g\sim24.0$, where the numbers of \OII\ and \lya\ emitters are most similar. 
The total contamination rate sits at only 1.3\% for the SzAS even with the \oii\ emitters outnumbering LAEs in that sample by more than 2:1. For HDR3, when corrected for the SzAS distribution bias and predicted false positives from noise, the predicted contamination rate is 1.2\%. While this already meets the HETDEX requirements, planned ELiXer enhancements, including updated \lya\ and \OII\ luminosity functions 
for the P(LAE)/P(OII) analysis (\S \ref{plae_poii_sec}) and run-time phot-$z$ fitting, should further decrease the contamination rate and improve overall accuracy.

The cumulative fractional contamination of \lya\ from all other lines in the SzAS is substantial at 26.4\%. This, however, is significantly inflated due to the over representation of AGN and \CIII and \CIV emission lines in the SzAS (Figure \ref{fig:accuracy_truth_sample}, upper right panel). 
When projected onto the HDR3 distribution and corrected for the SzAS distribution bias and noise driven false positives, this cumulative contamination fraction falls to a predicted 0.8\% for the full HDR3 dataset. This is even better than the \oii\ contamination. \rv{However, given the large correction from the SzAS results (Figure \ref{fig:truth_cumulative_oth_contamination_vs_gmag}), it is prudent to estimate a worst case contamination by these other lines by alternate means. These misclassifications in the SzAS are dominated by \ciii\ and \civ\ and are characterized by bright magnitudes and large line widths -- median $g$ = 22.5 $\pm$ 0.5 and median emission line FWHM = 22 $\pm$ 8 \angstrom. Using these properties as a guide, we select the fraction of HDR3 detections with emission line FWHM $>$ 14 \angstrom\ and $g < 23$, yielding 5.8\% of HDR3 detections, of which we assume 1/3 are misclassified as \lya. 
With 47\% of detections classified as \lya, we then estimate the worst case contamination rate by Other lines at 4\% (e.g.: $ \frac{1}{3} \cdot 0.058 ~/~ 0.47 = 0.04$).}

\rv{While this is $5\times$ the \textit{Bias + FPN Corrected} contamination rate of 0.8\%, this is still relatively small and the impact is far less than that of \oii\ contamination. The small scale clustering of \oii\ emitters projects to large scale clustering when misinterpreted as higher-z \lya. This is greatly diminished with \ciii\ and \civ\ as the contamination sources shift to higher redshift and scales proportionally to the square of the ratio of the co-moving angular diameter distances \citep{Grasshorn_Gebhardt_2019, Farrow+2021}. This means the HETDEX cosmology is some 6.5$\times$ less sensitive to \ciii\ contamination than \oii\ contamination and can tolerate $\sim$13\% (or $\sim$16\% for \civ) at the desired uncertainty.} \rv{So, even the worst case contamination is well within the required tolerances.} Additionally, the focus for ELiXer has been on the largest contaminant, \OII, as the contamination rate of other lines is expected to decrease with future improvements targeting their identification.

Overall, the ELiXer accuracy is good in the HDR3 dataset, while that in SzAS is poorer. The weaker performance in the SzAS set is due to the bright-magnitude and broad-emission line biases in the SzAS;  this is where ELiXer does not perform as well. The stronger (estimated) accuracy in the full HETDEX population is bolstered by the large numbers of faint end detections that are highly biased towards being \lya.

The results for ELiXer recovery rates are similarly mixed. The numbers are good for \lya\ and \OII, which are, by far, the most common emission lines found by HETDEX\null. The recovery of all other emission lines is rather poor, and is largely an issue of the default behavior of the classification algorithms. When there is only a single line in a HETDEX spectrum, ELiXer heavily weights the various \lya\ / \OII\ segregation methods which, as stated above, assume no contamination other than \OII\null.  In this case, ELiXer delivers a binary result, \lya\ vs \notlya, at the expense of all other emission lines.  Moreover, when analyzing particularly broad lines, ELiXer favors \lya\ (often suggestive of an AGN) over \OII; this also leads to the enhanced contamination of \lya\ by such "Other" lines. Additional identification metrics such as limited run-time phot-$z$, spectral slope, and multi-Gaussian fits, could help improve these rates and will be explored in future versions. 

A preliminary evaluation of an assessment sample expanded with $\sim$ 1000 DESI provided spectroscopic redshifts, 3/4 of which are for $g>24$ objects, is consistent with the HETDEX classification results of this work. The resulting assessment sample more closely matches the HETDEX magnitude and emission line distributions. After the observations are complete, the full, detailed results will be presented in Landriau et al, in preparation.
\\

\begin{deluxetable} {l| l  l} [ht] 
\tablecaption{Cumulative Classification Performance for HDR3 \label{tab:overall_performance}}
\tablewidth{0pt}
\tablehead{
\colhead{Metric} & \colhead{SzAS} & \colhead{ Bias + FPN Corrected}}
\startdata
\lya\ Accuracy               & 0.723   & ~~~~~~ 0.981 $\pm 0.034$ \tabularnewline
\lya\ Recovery               & 0.892   & ~~~~~~ 0.991 $\pm 0.033$ \tabularnewline
\textbf{\lya\ Contamination by \OII}   & \textbf{0.013}   & ~~~~~~ \textbf{0.012 $\pm 0.001$}\tabularnewline
\textbf{\lya\ Contamination by Other}  & \textbf{0.264}   & ~~~~~~ \textbf{0.008 $\pm 0.001$}\tablenotemark{\small{*}}\tabularnewline
\hline 
\OII\ Accuracy               & 0.890   & ~~~~~~ 0.965 $\pm 0.034$ \tabularnewline
\OII\ Recovery               & 0.972   & ~~~~~~ 0.970  $\pm 0.034$ \tabularnewline
\OII\ Contamination by \lya  & 0.039   & ~~~~~~ 0.021  $\pm 0.001$ \tabularnewline
\OII\ Contamination by Other & 0.071   & ~~~~~~ 0.014  $\pm 0.001$ \tabularnewline
\hline
Other Accuracy               & 0.892   & ~~~~~~ 0.916 $\pm 0.032$ \tabularnewline
Other Recovery               & 0.509   & ~~~~~~ 0.294 $\pm 0.010$ \tabularnewline
Other Contamination by \lya  & 0.027   & ~~~~~~ 0.006 $\pm 0.001$ \tabularnewline
Other Contamination by \OII  & 0.081   & ~~~~~~ 0.078 $\pm 0.003$ \tabularnewline
\enddata
\tablecomments{The cumulative performance of the ELiXer classifications on the SzAS and predictions for the full HDR3 dataset for detections with $g > 22$ and using the default ELiXer configuration. The Bias + FNP Corrected column corrects for the sample biases in the SzAS dataset and for false positives in the full HDR3 dataset, assuming 30\% false positive rate below emission line SNR of 5.0 and 15\% rate between $5.0 < \mathrm{SNR} < 5.5$. The values in the first column are slightly different than those in Table \ref{tab:truth_performance} since that table is down selected to only consider \lya\ and \OII\ detections. The \textbf{bold} type face rows correspond to the cumulative data points in the right-most (faintest) bins in Figure \ref{fig:truth_cumulative_oii_contamination_vs_gmag} and Figure \ref{fig:truth_cumulative_oth_contamination_vs_gmag}. \tablenotetext{*}{\small{\rv{0.04 worst case estimate. See \S\ref{performance} for a discussion.}}}
}
\end{deluxetable}

\subsection{Missing AGN and LBGs} \label{missing_agn_lbg}

Since ELiXer largely relies on equivalent width to classify most single-line spectra, the program currently does not perform well with \lya emitting objects that are not classical LAEs, i.e., broad-line AGN and Lyman-break galaxies (LBGs) which may have small \lya equivalent widths \citep[e.g.,][]{Shapley_2003}.  
Moreover, ELiXer can also fail to find some of the broad emission lines associated with the AGN, which can result in misclassifications that would otherwise be correctly assigned by the multi-line redshift solutions (\S \ref{redshift_solution_finder}). This is particularly noticeable in the bright end of the SzAS (Figure \ref{fig:truth_cumulative_oth_contamination_vs_gmag}), which has a disproportionately large number of AGN\null.   Moreover, in AGN, ELiXer can confuse \lya\ with \CIII\ when \CIII\ is the only significant emission line in the HETDEX spectral window ($0.96 \lesssim z \lesssim 1.25$) or with \CIV\ when the line fit to \CIII\ fails. Other approaches are taken to identify and recover AGN missed or misclassified by ELiXer \citep{liu_2022a} and future updates to ELiXer should improve upon its classification performance with these emission lines.

ELiXer also struggles to classify low \lya\-EW LBGs. On the whole, given their name-defining detection methodology \citep[][]{Guhathakurta_1990,Madau_1996,Steidel_1996}, LBGs tend to be more massive and more evolved than the typical LAE \citep[][]{Stark_2010,Kornei_2010,Jose_2013,Vargas_2014,Steidel_2018} and, consequently, may contain more dust to inhibit the escape of \lya. While some LBGs also meet the definition of an LAE and are likely to be detected and correctly identified as such by ELiXer, the more massive objects may often be confused with low-$z$ \OII\ emitters or even overlooked completely if they exhibit weak \lya\ emission or \lya\ absorption. While relatively few in number compared to LAEs, the more massive LBGs do represent a highly biased mass tracer and are of value to HETDEX, so it is desirable to recover and correctly identify as many of them as possible. This means using methods that do not use equivalent width as their primary discriminant.  To that end, several machine learning approaches (both supervised and unsupervised) are being explored, as are direct enhancements to ELiXer that incorporate additional classification methods, such as run-time photo-$z$ estimation. \\

\subsection{Contamination from Noise} \label{contamination_from_noise}

As stated earlier, ELiXer assumes an emission line detection is real, and not the result of noise or an artifact of the data reduction. As the SNR of an emission line detection decreases, it does become more likely that the feature is the result of noise. However, unlike real, incorrectly classified emission lines, false positives from noise are not expected to cluster \rv{(they occur in random spectra at random wavelengths and thus map to random sky positions at random redshifts)} and should only increase the uncertainty in the HETDEX cosmological measurements and not introduce a bias. As such, it is of lesser concern than misclassifications. Nevertheless, as described earlier, a (likely overly) aggressive false positives correction (\S \ref{bias_corrections}) is used for Figures \ref{fig:truth_cumulative_oii_contamination_vs_gmag} and  \ref{fig:truth_cumulative_oth_contamination_vs_gmag} and for Table \ref{tab:overall_performance} to better estimate the classification performance of ELiXer against the full HETDEX dataset.

Separate efforts to identify the noise driven false positive rate include repeat observations of low SNR sources (based on the premise that random noise will not cause a repeat detection at the same position and wavelength; \citet{Cooper_2022}) and various machine learning techniques.  Their goal is to allow a more accurate model of contamination from noise.\\

\subsection{Uncertainties}

The performance of ELiXer presented in the prior sections are shown without statistical uncertainties, though some uncertainty is implicit in its predictions for the whole of HETDEX Data Release 3. 

For the SzAS results in this work, the ELiXer classifications have been taken as absolute, as the quality of the classifications has not yet been calibrated to a proper probability. (This is a planned enhancement.) Since classifications are based on votes and weights, some of which have an MCMC element with a weak dependency on the initial random seed vectors, individual executions can occasionally result in a different classification due to conditions falling just to either side of a threshold (though the quality score ($Q(z)$) is generally unaffected; see \S \ref{best_z}). Similarly the catalog reported spec-$z$ values are taken as truth, and matching against the reported values is done as described in \S \ref{testing}, with a $\pm4\,\angstrom$ allowance, independent of the uncertainties in the spec-$z$ or the fitted emission line center (\S \ref{gaussian_fitting}). Nevertheless, many realizations of ELiXer classification runs compared against the SzAS have shown the results to be highly stable and repeatable.

In projecting the SzAS results onto the full HDR3 dataset, a few additional sources of uncertainty arise, such as the assumed false positive rate, which is binned only as a function of SNR. However, as with the SzAS, we assume the ELiXer classifications to be strictly categorical and the reported fractions subject only to rounding error. Anticipated expansion and improvements to the SzAS, including better matching to the HETDEX magnitude and emission line width distributions, will help address the systematics between the SzAS and the full HETDEX sample beyond the simplified corrections of \S \ref{bias_corrections}. 

As rough estimate on the uncertainties in the accuracy, recovery, and contamination rates reported for HDR3, we use the fraction of detections that are most susceptible to classification changes as described in this subsection. This is effectively captured by the largest factor in the classifications, P(\lya), where P(\lya) is least certain and least stable against change due to randomness in sampling (i.e., near 0.5). As 7\% of HDR3 detections have $0.4 <$ P(\lya) $< 0.6$, we assume a $\pm 3.5\%$ uncertainty on those rates. \\

\section{Summary} \label{conclusions}

As the primary emission line classifier for HETDEX, ELiXer must produce quality redshift identifications that are highly accurate, complete, and with minimal contamination. With a resolving power ranging from 750--950, HETDEX cannot split the \OII\ doublet, so object classification must rely heavily on continuum information combined with equivalent width distributions. By incorporating improvements to established \lya/\OII\ separation mechanics, from the 20\,\AA\ equivalent width cut \citep[][]{Gronwall2007a,Adams2011a} to the P(LAE)/P(OII) ratio \citep[][]{Leung2017}, and by combining additional partitioning techniques, ELiXer produces classifications that outperform the HETDEX science requirements for \lya\ contamination by its principle low-$z$ interloper, \OII 3727\,\AA, while providing a good recovery rate (Table \ref{tab:overall_performance}). The lower than required 1.2\% contamination of \lya\ by \OII\ affords the option to loosen the project's strict classification thresholds in exchange for gains in the \lya\ recovery fraction or completeness. 

Though they occupy a small fraction of HETDEX emission line detections, lines other than \OII 3727\AA, such \CIII 1909\,\AA, and \CIV 1549\,\AA\ represent a larger source of \lya\ contamination \rv{in the biased SzAS}. \rv{However, as described in \S \ref{performance},} these lines are not expected to produce a significant clustering signal or bias in the $z=2.4$ measures of $H(z)$ and $D_A(z)$. Regardless, planned enhancements to ELiXer and a larger spectroscopic redshift test sample (more aligned with the HETDEX distribution) will improve these classifications and further reduce \lya\ contamination.

The HETDEX project is continuing to work towards  reducing the rate of false positive detections as a function of the emission line signal-to-noise ratio \citep[][]{Cooper_2022}. Early indications suggest the contamination from noise is small above the 4.8-5.0 SNR acceptance threshold for detections. Regardless, these noise driven false positives should only add white noise to the LAE cluster signal.  Although this increases the uncertainty in the HETDEX measurements, it should not introduce specific features in the galaxy power spectrum.

ELiXer continues to evolve. Future enhancements and revised voting criteria will be tested against expanded assessment samples drawn from forthcoming data releases. 
This will improve the current classification capabilities, enabling new and higher precision science. Although ELiXer is designed for and calibrated to HETDEX, the methodology developed in this work can be adapted to other low-resolution, narrow wavelength range spectroscopic surveys. \\


\acknowledgments

The authors thank the anonymous reviewer for the helpful feedback which assisted in improving this manuscript.

HETDEX is led by the University of Texas at Austin McDonald Observatory and Department of Astronomy with participation from the Ludwig-Maximilians-Universit\"at M\"unchen, Max-Planck-Institut f\"ur Extraterrestrische Physik (MPE), Leibniz-Institut f\"ur Astrophysik Potsdam (AIP), Texas A\&M University, The Pennsylvania State University, Institut f\"ur Astrophysik G\"ottingen, The University of Oxford, Max-Planck-Institut f\"ur Astrophysik (MPA), The University of Tokyo, and Missouri University of Science and Technology. In addition to Institutional support, HETDEX is funded by the National Science Foundation (grant AST-0926815), the State of Texas, the US Air Force (AFRL FA9451-04-2-0355), and generous support from private individuals and foundations.

Observations were obtained with the Hobby-Eberly Telescope (HET), which is a joint project of the University of Texas at Austin, the Pennsylvania State University, Ludwig-Maximilians-Universit\"at M\"unchen, and Georg-August-Universit\"at G\"ottingen. The HET is named in honor of its principal benefactors, William P. Hobby and Robert E. Eberly.

VIRUS is a joint project of the University of Texas at Austin, Leibniz-Institut f\"ur Astrophysik Potsdam (AIP), Texas A\&M University (TAMU), Max-Planck-Institut f\"ur Extraterrestrische Physik (MPE), Ludwig-Maximilians-Universit\"at Muenchen, Pennsylvania State University, Institut fur Astrophysik G\"ottingen, University of Oxford, and the Max-Planck-Institut f\"ur Astrophysik (MPA). In addition to Institutional support, VIRUS was partially funded by the National Science Foundation, the State of Texas, and generous support from private individuals and foundations.

The authors acknowledge the Texas Advanced Computing Center (TACC) at The University of Texas at Austin for providing high performance computing, visualization, and storage resources that have contributed to the research results reported within this paper. URL:http://www.tacc.utexas.edu

The Institute for Gravitation and the Cosmos is supported by the Eberly College of Science and the Office of the Senior Vice President for Research at the Pennsylvania State University.

KG acknowledges support from NSF-2008793.

This research benefits from the open-source projects Python \citep{pythonref}, astropy \citep{astropy:2018}, numpy \citep{harris2020array}, photutils \citep[][]{photutils}, and others in the open-source community.\\

\bibliography{elixer.bib}
\clearpage

\appendix
 
\section{Example ELiXer Detection Reports} \label{sec:elixer-report-APPENDIX}

\begin{figure*}[ht]
\centering
\fbox{\includegraphics[width=1.0\textwidth]{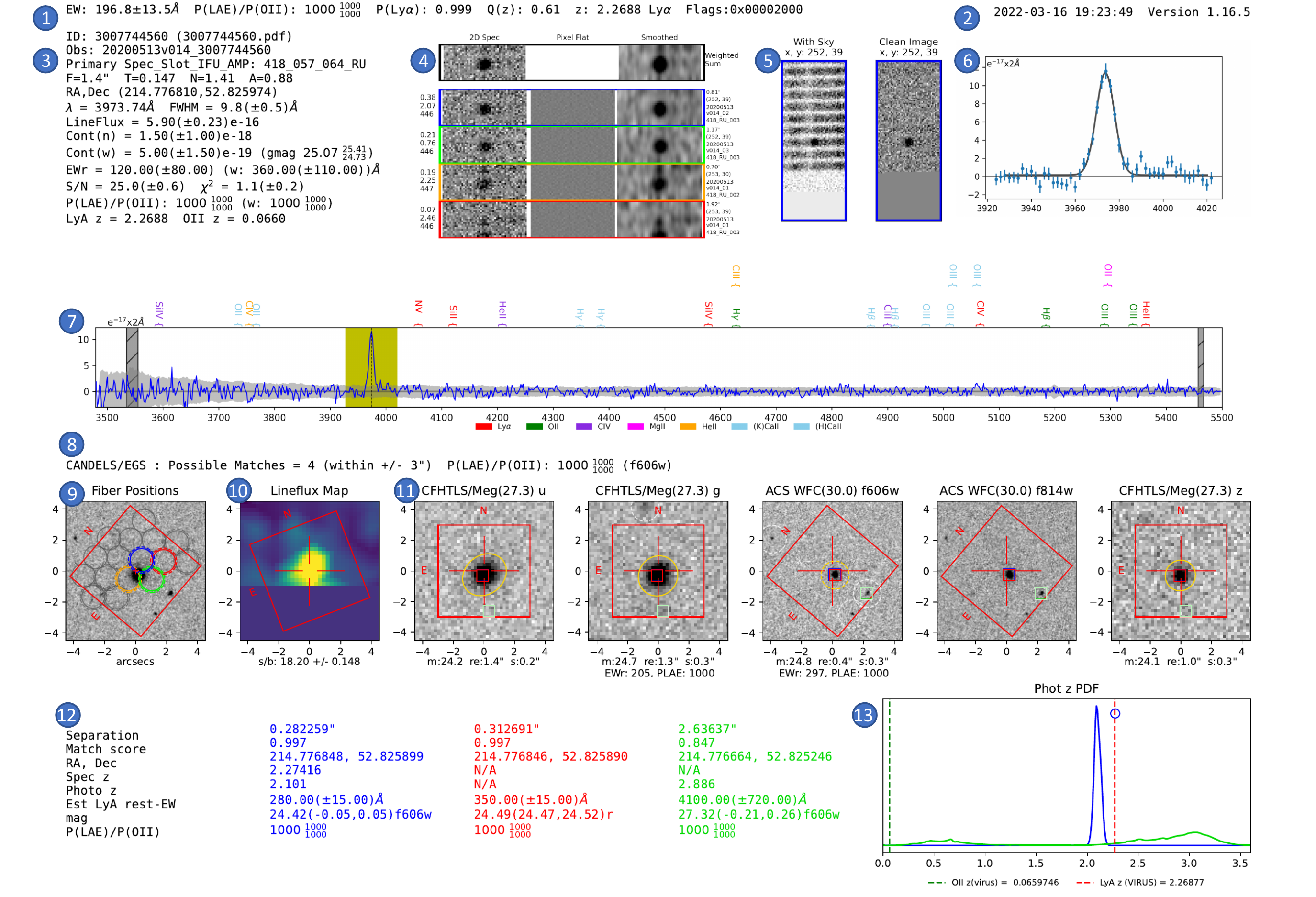}}
\caption{Example ELiXer detection report. This is a somewhat uncommon example selected to illustrate elements that are not always present for an individual detection, such as the classification label, warning flags, multiple catalog references, and photometric redshift PDFs. Descriptions of the bulleted features are provided below.
}
\label{fig:example_elixer1}
\end{figure*}

\begin{figure*}[ht]
\centering
\fbox{\includegraphics[width=1.0\textwidth]{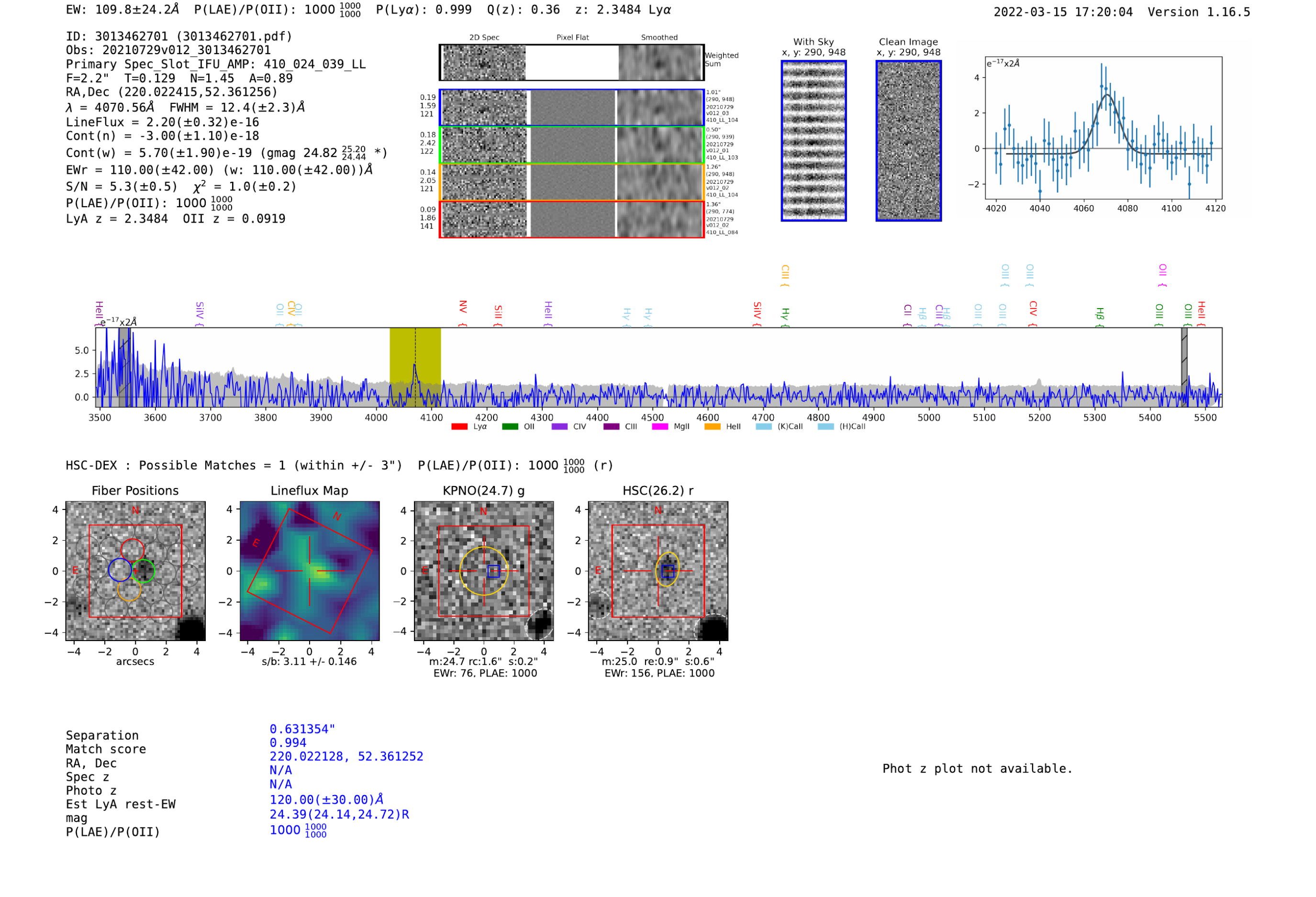}}
\caption{The ELiXer report of a typical HETDEX LAE\null. Note that this region of sky has fewer and shallower imaging data, and more limited catalog data compared to Figure \ref{fig:example_elixer1}. It is included here as a counter to the more illustrative, but less common example of Figure \ref{fig:example_elixer1}.
}
\label{fig:example_elixer2}
\end{figure*}

In this Appendix, we include two ELiXer detection reports as examples of those used for visual inspection and diagnostics. The first, Figure \ref{fig:example_elixer1}, is a somewhat unusual HETDEX LAE: it has a very high emission line SNR, it is matched to a source contained in multiple catalogs (\S \ref{sec:catalogs}), and has several photometric and spectroscopic redshift determinations, and it lies in an area of sky with deep \textit{HST} imaging. It is presented to illustrate the various sections within an ELiXer report. The second, Figure \ref{fig:example_elixer2}, is more representative of the typical HETDEX LAE and is shown here to that end. 

\begin{enumerate}
    \item \textit{Summary} - From left to right: (1) computed Equivalent Width of the emission line in the rest-frame of \lya, the combined continuum estimate (\S \ref{combined_cont}), (2) P(LAE)/P(OII) (\S \ref{plae_poii_sec}) and 68\% confidence interval using the combined continuum estimate, (3) P(\lya) score (\S \ref{p_lya}), (4) Quality score for the Best-$z$ redshift (\S \ref{best_z}), (5) Best-$z$ redshift, (6) Classification labels (\S \ref{classification_labels}) if any,  (7) Error/Warning Flags\footnote{Flags are not explicitly described in this work but are part of data release documentation} if any; in this example, there is a warning flag indicating a small disagreement in the $g$-magnitudes calculated from the spectrum. 
    \item \textit{Timestamp + Version} - Displays the date and time of the creation of this report and the ELiXer version number.
    \item \textit{Detection Details} - A block of information about the HETDEX observation and the emission line detection. From top to bottom: (1) Detection ID number and file name, (2) Observation ID, (3) IFU+Amp address of the fiber nearest the detection center, (4) 'F' = seeing FWHM in arcsecs, 'T' = effective throughput at 4540\,\angstrom, 'N' = dither to dither normalization, 'A' = aperture correction (divisor), (5) J2000 equatorial coordinates of the PSF weighted detection center in decimal degrees, (6) emission line wavelength center and FWHM, (7) integrated emission line flux, (8) continuum estimate (\S \ref{continuum_estimates}) from the spectrum within $\pm 40$\,\angstrom\ of the line center, (9) continuum estimate and $g$-magnitude from the full width of the spectrum, (10) equivalent width in \lya\ rest-frame with the continuum estimates from (8) and (9) respectively, (10) signal-to-noise ratio and $\chi^{2}$ of the emission line fit, (11) P(LAE)/P(OII) using the continuum estimates from (8) and (9) respectively, (12) redshifts assuming \lya\ and \oii, (13) multi-line emission line identification (\S \ref{redshift_solution_finder}), if one is selected, with its quality score, name, rest-wavelength, redshift, and equivalent width in its own restframe using the continuum estimate in (9).
    \item \textit{2D Fiber Cutouts} - $5\times 3$ grid of cutouts within $\pm 40$\,\AA\ of the detection line center in the spectral direction and $\pm$1 fiber in the CCD direction\footnote{Fibers adjacent on the CCD are not necessarily adjacent on sky}. The left most column is the pre-smoothing cutout with all rectifications and sky subtraction. The center column is the pixel flat, with any significant deviations marked in red (none in this example). The right most column is the same as the left most column but smoothed with a $2\times 2$ Gaussian filter. The top row (highlighed in black) is the weighted sum of all contributing fibers. The rows below (blue, green, orange, red) are the highest four fibers as weighted by PSF modeled flux. The values (in very small print) to the left of the grid represent (from top to bottom): the normalized fiber weight in the PSF, the $\chi^{2}$ of the fit to the fiber profile, and the fiber number on the CCD\null. The values (in very small print) to the right of the grid represent (from top to bottom): the fiber center distance to the detection center (in arcsecs), the CCD pixel coordinate of the fiber center, the exposure date, the observation number and exposure number for that date, and the IFU spectrograph ID, amplifier ID, and fiber number on that amplifier.
    \item \textit{Key CCD Region} - $\pm 10$ fibers in the CCD direction and $\pm 40$\,\AA\ in the spectral direction around the detection center for the fiber nearest the detection, shown before and after sky subtraction. 
    \item \textit{1D Line Fit} - the 1D emission line fit to the data. This matches the gold highlighted section in the full 1D spectrum. Values are integrated fluxes in 2\,\AA\ wide bins. 
    \item \textit{1D Spectrum} - the full 1D spectrum as integrated fluxes in 2\,\AA\ wide bins. The gray background gives the estimated. The two vertical gray-hashed bars point out the two strongest sky-lines. The gold highlighted region is the anchor emission line. Any other colored regions, if present, highlight other spectral lines that support the selected multi-line redshift solution. The other red labels ("NV","SiII","SiIV","CIV","HeII") mark the positions of other possible lines in the spectrum, assuming the anchor line is \lya; in this spectrum, none of these confirming lines are detectable.   The colored labels above the spectrum represent the positions of other common lines if the anchor emission line were one of the features listed below the spectrum with the matching color.
    \item \textit{Main Catalog Summary} - displays the name of the catalog with the deepest imaging used in the report, along with the number of potential catalog counterparts (if any) and the P(LAE)/P(OII) found from the continuum estimate of the listed filter. 
    \item \textit{Fiber Positions} - the footprint of all fibers contributing to the detection plotted over a stacked image from the catalog with the deepest imaging. The four colored fibers match those in the $5\times 3$ grid in (4). Fibers with a dashed outer ring are at the edge of the detector. The PSF weighted center of the detection is marked with a red cross.
    \item \textit{Lineflux Map} - wavelength collapsed flux intensity map summing over $\pm3\sigma$ from the emission line center. The values under the image are an estimate of significance based on the flux inside a 1\arcs\ radius aperture and the standard deviation of flux inside a 5\arcs\ to 7\arcs annulus, corrected for area. The lower section of the Lineflux Map in this example is blank as that region happens to fall off the edge of the CCD.
    \item \textit{Imaging Stamps} - postage stamp cutouts of the deepest imaging available to ELiXer, shown in increasing  order from blue (left) to red (right).  Only the bluest five filters are shown, though more may be available. Overplotted are colored 1\arcs\ per side squares corresponding to the positions of possible catalog counterparts. The top three (see (12) are shown in blue, red, and green, with all others displayed in white. In this example, the blue and red squares overlap, so only the red is obviously visible, but they mark the same object. The overplotted ellipses are \textit{SEP} identified sources (\S \ref{elixer_aperture_phot}). A gold ellipse marks the object selected by ELiXer as the most likely counterpart, while all other objects are marked in white. If the bounding ellipse is dashed, then it has been expanded to be a 1\arcs\ radius circle for visibility. The text above each cutout indicates the catalog name, and the approximate imaging depth and the filter. The values under the cutouts correspond to the gold aperture and are: 'm' = aperture magnitude, 're' = the effective radius of the ellipse in arcsecs, 's' = separation between the center of the aperture and the HETDEX PSF weighted center in arcsecs,"EWr" - the equivalent width in the \lya\ rest-frame using the aperture magnitude as the continuum estimate, "PLAE" - P(LAE)/P(OII) using the aperture magnitude as the continuum estimate. All values are computed for $g$ and $r$ (or equivalent) filters, but not always for other bands.  
    \item \textit{Catalog Counterparts} - basic information on up to the top three most likely catalog counterparts, based on magnitude and distance, which correspond to colored squares on the Imaging Stamps. In this example, the blue, red, and green objects are actually the same source, but reported from different catalogs, and their corresponding squares in the Imaging Stamps overlap. The "Separation" is the distance in arcsec between the HETDEX detection position and the catalog reported position. This offset can sometimes be sizeable, especially for extended objects where the catalog reports a surface brightness center and the HETDEX detection is more toward the object's edge. The reported P(LAE)/P(OII) value uses the catalog's reported bandpass magnitude as the continuum estimate, not the aperture magnitude from the Imaging Stamps.
    \item \textit{Catalog z PDFs} - if available, shows the photometric redshift PDFs, color coded to match the top three catalog counterparts. In this example, there is no PDF for the red counterpart (from the CFHTLS catalog), so only blue and green PDFs are shown. Circles, again with a matching color, mark the reported spectroscopic redshift, if available. The green dashed line represents the redshift if the emission line is \OII, while the red dashed line shows the same for \lya. Since the anchor line in this example is \lya, there is a precise match with the spec-$z$ and a close match with the phot-$z$ for the object marked in blue.
\end{enumerate}

\end{document}